\documentclass[fleqn,usenatbib]{mnras}

\usepackage{newtxtext,newtxmath}

\usepackage[T1]{fontenc}

\DeclareRobustCommand{\VAN}[3]{#2}
\let\VANthebibliography\thebibliography
\def\thebibliography{\DeclareRobustCommand{\VAN}[3]{##3}\VANthebibliography}

\usepackage{graphicx}	
\usepackage{amsmath}	

\usepackage{ulem}
\usepackage{xcolor}
\usepackage{url}

\newcommand{\pO}{\hbox{I007}}
\newcommand{\pOc}{\hbox{I007 catalog}}

\newcommand{\GG}{\hbox{$G_{Gaia}$}}
\newcommand{\GS}{\hbox{$G_{SDSS}$}}

\title[SDSS Stripe 82 Standard Stars catalogue: New and Improved]{Photometric Cross-Calibration
of the SDSS Stripe 82 Standard Stars catalogue with Gaia EDR3, and Comparison with Pan-STARRS1, DES, 
CFIS and GALEX catalogues}

\author[K. Thanjavur et al.]{
Karun Thanjavur,$^{1}$\thanks{E-mail: karun@uvic.ca (KT)}
\v{Z}eljko Ivezi\'{c},$^{2}$
Sahar S. Allam$^{3}$
Douglas L. Tucker$^{3}$
J. Allyn Smith$^{4}$
and Stephen Gwyn$^{5}$
\\
$^{1}$Department of Physics \& Astronomy, University of Victoria, 3800 Finnerty Road, Victoria, BC V8P 5C2, Canada\\
$^{2}$Department of Astronomy and the DiRAC Institute, University of Washington, 3910 15th Avenue NE, Seattle, WA 98195, USA\\
$^{3}$Fermi National Accelerator Laboratory, Batavia, Il 60510, USA\\
$^{4}$Dept. of Physics, Engineering \& Astronomy, Austin Peay State University, 601 College St., Clarksville, TN 37044, USA\\
$^{5}$National Research Council, Canadian Astronomy Data Centre, Victoria, BC, V9E 2E7, Canada
}

\date{Accepted XXX. Received YYY; in original form ZZZ}

\pubyear{2021}

\begin{document}
\label{firstpage}
\pagerange{\pageref{firstpage}--\pageref{lastpage}}
\maketitle

\begin{abstract}

\noindent We extend the SDSS Stripe 82 Standard Stars catalogue with post-2007 SDSS imaging data. This improved version lists averaged SDSS $ugriz$ photometry for nearly a million stars brighter than $r\sim22$ mag. With 2-3$\times$ more measurements per star, random errors are 1.4-1.7$\times$ smaller than in the original catalogue, and about 3$\times$ smaller than for individual SDSS runs. Random errors in the new catalogue are $\la$0.01 mag for stars brighter than 20.0, 21.0, 21.0, 20.5, and 19.0 mag in $u,g,r,i,$ and $z$-bands, respectively. We achieve this error threshold by using the Gaia Early Data Release 3 (EDR3) Gmag photometry to derive grey photometric zeropoint corrections, as functions of R.A. and Declination, for the SDSS catalogue, and use the Gaia BP-RP colour to derive corrections in the $ugiz$ bands, relative to the $r$-band. The quality of the recalibrated photometry, tested against Pan-STARRS1, DES, CFIS and GALEX surveys, indicates spatial variations of photometric zeropoints $\lid$0.01 mag (RMS), with typical values of 3-7 millimag in the R.A., and 1-2 millimag in the Declination directions, except for $\sol$6 millimag scatter in the $u$-band. We also report a few minor photometric problems with other surveys considered here, including a magnitude-dependent $\sim$0.01 mag bias between $16 \leq \GG \leq 20$ in the Gaia EDR3. Our new, publicly available catalogue offers robust calibration of $ugriz$ photometry below 1\% level, and will be helpful during the commissioning of the Vera C. Rubin Observatory Legacy Survey of Space and Time.
\end{abstract}




\section{Introduction} \label{sec:intro}

Modern multi-band photometric sky surveys aim to deliver measurements uniform at or better than the 1\% (0.01 mag) level, to enable cosmological and other high-precision measurements \cite[e.g., the Vera C. Rubin Observatory Legacy Survey of Space and Time,][]{LSSToverview}. Photometric data are usually calibrated using sets of standard stars whose brightness is known from previous work. One of the largest catalogues with sub-percent measurement precision and optical multi-band $ugriz$ photometry was constructed by averaging multi-epoch data for about a million stars collected by the Sloan Digital Sky Survey \citep[SDSS,][]{York2000}  in a 300 deg$^2$ region known as SDSS Stripe 82 \citep[][hereafter \pO]{Ivez07}. The SDSS $ugriz$ photometric system is now in use at many observatories worldwide and the \pOc\ has been used both for calibration and testing of other surveys. 

After completion of the \pOc, SDSS obtained additional imaging data, about 2-3 times more measurements 
per star depending on its sky position within Stripe 82. This increased number of data points results in averaged photometry with 1.4-1.7 times smaller random errors (precision) than in the original catalogue (and about three times as small as for individual SDSS runs). In addition, the availability of photometric data from recent wide-field surveys such as the Dark Energy Survey \citep[DES,][]{2016MNRAS.460.1270D}, Pan-STARRS \citep[PS1,][]{2010SPIE.7733E..0EK} and Gaia \citep{GaiaCollab2018b}, enables a much more detailed and robust cross-calibration, including correcting for residual photometric zeropoint errors in SDSS flat-fielding. For example, \cite{2013A&A...552A.124B} reported a saw-tooth pattern in photometric residuals between the SDSS and the Supernova Legacy Survey (SNLS) catalogues, as a function of Declination (see their Fig.~23). Such a pattern is most likely due to errors in SDSS zeropoint calibration because of the flat-field correction of Stripe 82 being only a function of Declination (due to drift-scanning in R.A. direction). Given that systematic errors in other catalogues are much smaller (Gaia), or are expected to display different spatial patterns (DES and PS1), it is likely that a cross-comparison of several catalogues can result in significant improvements. These are the main reasons that motivated our decision to construct an updated version of the \pOc. 
 
We describe datasets used in our analysis in \S2, and the construction of the new catalogue and its analysis in \S3. 
Our results are summarized and discussed in \S4. 


\section{Datasets} \label{sec:data}

\subsection{SDSS Stripe 82 Imaging Data} \label{ssec:s82}

In the SDSS survey, Stripe 82 is a contiguous, 300 deg$^2$ equatorial region, which stretches between $-60^{\circ}\;\leq\;RA\;\leq\;60^{\circ}$ [20h to 4h], and $-1.266^{\circ}\;\leq\;Dec\;\leq\;1.266^{\circ}$. Following the initial, concerted effort by the SDSS collaboration between 2001 and 2008 to map this region repeatedly to a forecast imaging depth, $r \leq 22$, several other surveys in various wavebands have also targeted this same patch of sky to provide a rich multi-wavelength dataset suitable for a variety of investigations. SDSS observations have also continued in this region \citep[e.g., the SDSS-II search for supernovae,][]{2008AJ....135..338F}, resulting in an imaging depth greater than what was initially planned.  

Data from the SDSS imaging camera \citep{1998AJ....116.3040G} were collected in drift-scan mode. The images which correspond to the same sky location in each of the five photometric bandpasses (these five images are collected over $\sim$5 minutes, with an exposure time of 54 seconds for each band) are grouped together for simultaneous processing as a {\it field}. A field is defined as a 36 second (1361 rows) stretch of drift-scanning data from a single column of CCDs (sometimes called a {\it scan line}; for more details, see \pO\ and references therein). 

\subsubsection{The 2007 SDSS Standard Star catalogue}

The SDSS standard star catalogue, \pO\ (version 2.6) \footnote{This public catalogue is available by following the link provided in the data availability section.} was constructed by averaging multiple SDSS photometric observations (at least four per band, with a median of 10) in the $ugriz$ system. The catalogue includes 1.01 million non-variable stars. The measurements for individual sources have random photometric errors below 0.01 mag for stars brighter than 19.5, 20.5, 20.5, 20, and 18.5 in $ugriz$, respectively (about twice as good as for individual SDSS runs). Several independent tests of the internal consistency suggested that the spatial variation of photometric zero points is not larger than $\sim$0.01 mag (RMS).  

\subsubsection{Post-2007 SDSS data \label{ssec:DR15}}

In this work, we used the SDSS Data Release 15 (DR15) as available in April 2019 \citep{Blan17}. In DR15, the Stripe 82 region is covered by 118 {\it runs}, which include 32,292 fields, each with observations in the five   $ugriz$ SDSS filters. Using our programmatic query tool, we obtained the processed data for all these runs from the DR15 public database\footnote{https://dr15.sdss.org/sas/dr15/eboss/photoObj/}. In the database, the data are presented as individual FITS tables, named \verb photoObj_<run>_<camcol>_<field>.fits. From each fits table, we extracted photometric and astrometric quantities, time of observation, and several ancillary data for all the objects into a formatted master file for further processing.

The objects in each of these data files were then matched with the standard stars in the \pO\ catalogue using their mean sky positions (R.A. and Declination) and a matching radius of 0.5 arcsec. For matching, only deblended objects ({\it nchild}=0), lying between rows $64 < objc\_rowc \leq 1425$ in each field, were selected to avoid poor photometry due to blending or lying close to edges of the CCD. From these matched objects, only those with photometric error $<$0.1 mag were selected to compute photometric zeropoint offsets between the \pO\ catalogue and DR15. These offsets were obtained independently for all runs and fields, in all five filters, and applied to bring our DR15 based catalogue to the same photometric scale as the \pO\ catalogue -- in essence, we have re-calibrated photometry for all Stripe 82 runs in DR15 using the \pO\ catalogue. In addition, the MJD and fractional MJD of observation were computed using the median of the TAI values (the GPS based time reported by the SDSS Apache Point Observatory) for these matched objects. In the final step, the photometric, astrometric and other details for each of these matched standard stars were written to independent (one per star) light curve files. Further processing of these light curves is described in Section~\ref{sec:averaging}. 

The final dataset, consisting of all the light curves in the five $ugriz$ filters for the 1,006,849 standard stars in the \pO\ catalogue resulted in $\sim$20 GB of tabular data. To make file search and access fast, the data were chunked into sub-directories, each spanning 1 $\deg$ in RA, and 0.1 $\deg$ in Dec (a ``poor-man's'' two-dimensional tree structure). These light curve data files can be made available as a single tarball by emailing the contact author. 

\subsection{Gaia Early Data Release 3 Data} \label{ssec:gaia}
 
For our primary astrometric and photometric cross calibration we use the Gaia Early Data Release 3 (EDR3) catalogue.  Gaia is a European Space Agency (ESA) mission designed to map over a billion stars in the Milky Way and the local group in three dimensions, providing accurate proper motion and radial velocity measurements. \citet{GaiaCollab2016} provide a detailed overview of the Gaia mission (spacecraft, instruments, survey and measurement principles, and operations), while technical details for specific topics relevant to our work may be found in the following citations: pre-processing and source-list creation \citep{Fabr2016}, astrometric solution \citep{LInd2018}, processing the photometric data \citep{Riel2018}, photometric content and validation \citep{Evan2018}, and full catalogue validation \citep{Aren2018}. A detailed description of Gaia data products may be found in \citet{GaiaCollab2018b}. Here we summarize only the pertinent details.

EDR3 includes astrometry, photometry, radial velocities, and information on astrophysical parameters and variability, for approximately 1.8 billion sources\footnote{See https://www.cosmos.esa.int/web/gaia/earlydr3.}. This dataset is based on the first 34 months of the mission and includes celestial positions and the apparent brightness in the broad-band G (\GG\ hereafter) for sources brighter than \GG$\sim$21.  This data release also contains two additional broad-band magnitudes, the BP (330-680 nm) and RP (630-1050 nm). Gaia EDR3 photometry is generally superior to ground-based photometry for sources with sufficient signal-to-noise ratio, and we use it to derive zeropoint corrections for SDSS photometry, as described in Section~\ref{sec:GaiaCorr}. 

\subsection{Dark Energy Survey (DES) Data} \label{ssec:des}

The Dark Energy Survey (DES; \citealt{2016MNRAS.460.1270D}) is an imaging survey of the Southern Galactic Cap in 5 filter passbands ($grizY_{DES}$) that was conducted from 2013 to 2019 with the 570 mega-pixel Dark Energy Camera (DECam; \citealt{2008arXiv0810.3600H,2015AJ....150..150F}) on the Victor M.\ Blanco 4-m telescope at the Cerro Tololo Interamerican Observatory.

For this paper, we made use of the DES Data Release 1 (DES DR1; \citealt{Morg2018}, \citealt{2018ApJS..239...18A}) public data set, which is based on the first 3 years of DES observations.  The DES DR1 object catalogue consists of $\sim$ 400 million objects to a depth of 24.33, 24.08, 23.44, 22.69, 21.44 mag in $grizY_{DES}$ bands ($S/N=10\sigma$).  The DES DR1's photometric calibration is uniform at the sub-percent level (RMS) for each of the five filter passbands over the entirety of the survey footprint.  Its astrometric precision is quoted to be 151~milli-arcsec (RMS).  We downloaded DES DR1 data via the NOAO Data Lab\footnote{http://datalab.noao.edu} Table Access Protocol (TAP) service, selecting stars in the general area of the SDSS Stripe~82 region of the DES footprint.  For the purposes of our analysis, we downloaded from the co-added catalogue, the weighted mean PSF magnitude ({\tt WAVG\_MAG\_PSF}) and magnitude error ({\tt WAVG\_MAGERR\_PSF}) as well as the number of observations that went into the weighted catalogue-coadd weighted mean PSF magnitude ({\tt N\_EPOCHS}) in each filter band for each downloaded DES star.  In total, 3,585,229 stars were downloaded in the region (RA $>$ $270^{\circ}$ or $<$  $105^{\circ}$; DEC $=$ $-3.5^{\circ}$ to $+3.5^{\circ}$), of which 619,741 were matched to our SDSS catalogue using a match radius of 0.5~arcsec.


\subsection{Pan-STARRS (PS1) Data} \label{ssec:ps1}
 
The Panoramic Survey Telescope And Rapid Response System, Telescope 1 (Pan-STARRS-1, or PS1), commissioned in 2010, is the first of four planned 1.8m telescopes, designed to map three quarters of the entire sky visible from Hawaii \citep{2010SPIE.7733E..0EK}. This panchromatic, synoptic survey, called the 3$\pi$ Steradian Survey was carried out in five bands, $grizy_{P1}$, with limiting magnitudes of 23.3, 23.2, 23.1, 22.3, and 21.4 mag. \citet{2016arXiv161205560C} provide full details about the PS1 observatory, the surveys being carried out, and an overview of the resulting data products. More detailed technical information on the following specific topics may be found in the following series of papers: the Pan-STARRS data processing system \citep{Euge20a}, pixel analysis, source detection and characterization \citep{Euge20b}, detrending, warping and stacking \citep{Wate2020}, photometric and astrometric calibration \citep{Euge20c}, and the database and data products \citep{Flew2020}.

Here we used the 2019 data release PS1-DR2 from $MAST$\footnote{https://dx.doi.org/10.17909/T9RP4V}. The catalogue overlapping the Stripe 82 region contains over 7 million point sources taken from the stacked 3$\pi$ Steradian Survey. From the large set of measured and derived quantities available for these objects, we downloaded $grizy_{P1}$ PSF photometry and various quality flags. For this region, the mean positional uncertainties are 12 and 11 milli-arcsec in the RA and Dec directions. Based on this positional precision and that of our SDSS standard stars catalogue, we compared both catalogues using a matching radius of 0.5 arcsec. The resulting matched catalogue used in our analysis presented in \S \ref{sec:DESPS1} contains 909,000 objects. Taking the $r$ band as being representative, the mean number of visits used to obtain the mean PSF magnitudes for these matched stars is 14 visits. 


\subsection{Canada-France Imaging Survey (CFIS) Data} \label{ssec:cfis}

The Canada-France Imaging Survey (CFIS) \citep{2017ApJ...848..128I} is a large observing program being carried out at the Canada-France-Hawaii Telescope (CFHT) to map the northern sky in the MegaCam $u$ and $r$ bands. With a broad range of science goals, including providing ground-based optical photometry to complement the Euclid space mission, CFIS aims to cover 10,000 deg$^2$ in the $u$ band to a depth of 24.4 mag. For our analysis, we were provided the $u$ band data which overlaps the Stripe 82 footprint from the CFIS Data Release 2 (DR2, January 2020). In this region, CFIS data are available only for Declination $>+0.45$ degree, thus covering only about a quarter of the Stripe 82 survey footprint. 

The available CFIS DR2-Stripe 82 catalogue contains close to 965,000 sources, of which $\sim$315,000 matched within a link radius of 0.5 arcsec with stars in our SDSS standard stars catalogue. The astrometry over the complete CFIS survey region has been calibrated against Gaia DR2.  The residuals of the astrometric solution are typically 10 mas.  From the matched objects, we select only bright stars, $r \leq 20$ mag,  with colours  $ 1 \leq (u-g)_{SDSS} \leq 2.1$, following \cite{2017ApJ...848..128I}. This colour-magnitude cut yields 150,114 stars, which are used in the comparison described in \S \ref{sec:CFIStest}. 


\subsection{{\it GALEX} Data} \label{ssec:galex}

The Galaxy Evolution Explorer ($GALEX$) all-sky survey catalogue in the NUV (1771-2831 \AA) and FUV (1344-1786 \AA) bands provides a unique photometric dataset to crosscheck the u-band of a multiband survey such as the SDSS. Processed and calibrated archival data from the eight year, All-sky Imaging (AIS) and Medium depth Imaging (MIS) surveys are available at $MAST$\footnote{https://galex.stsci.edu/GR7/}. Due to the failure of the FUV detector midway through the survey, the available imaging depth in this band is shallower by nearly 1 ABmag than the NUV depth of 20.8 ABmag in the AIS. Overall, the source number count in the FUV is, on average, only a tenth of that in the NUV. Given these limitations in the FUV survey, we extracted only sources in the NUV catalogue corresponding to the Stripe 82 footprint.  
 
\citet{2017ApJS..230...24B} provide full details of the GALEX survey, and we summarize only the NUV details, which are pertinent to our analysis here. The coverage over the survey region led to total exposure times of $\sim$150s in the AIS, and $\sim$1500s in the MIS. The corresponding NUV survey depths are 20.8 and 22.7 ABmag in the two surveys respectively. Objects in the Stripe 82 region were extracted and matched with our SDSS DR15 catalogue, described in \S \ref{ssec:DR15} using a matching aperture radius of 3'', which corresponds to the matching radius used internally by GALEX to match their NUV and FUV sources. The GALEX catalogues do not list the RA, Dec positional errors for their sources, and we found their listed sky positions generally carry a greater uncertainty than the other catalogues used in this work.

Our matched GALEX catalogue consists of 150,945 NUV sources in the Stripe 82 footprint for which we obtained aperture magnitudes and uncertainties, as well as various fitted geometric measures for each of them. The astrometry has greater uncertainty than that of SDSS, with mean positional uncertainties of 1'' RMS. We also note that there is an overall mean shift in GALEX astrometry relative to that of the SDSS by -0.1 arcsec in RA, and +0.3 arcsec in Dec; our GALEX-SDSS positional offset results are comparable to what others have reported in the past (\citealt{Ague2005, Morr2007, Hein2009, Bian2014}). In addition, we find that the available depth in the Stripe 82 region is limited, with $>$80\% of the matched sample being brighter than r=18 mag. 
  
 
\section{The Construction and Analysis of the New v4.2 catalogue \label{sec:v42}}

We first describe the construction of the new SDSS catalogue, v4.2, and the derivation of photometric zeropoint corrections using Gaia EDR3 data, and then compare the resulting photometry to Gaia EDR3, DES, Pan-STARRS, CFIS and GALEX catalogues. 


\subsection{The construction of raw SDSS catalogue from light curves \label{sec:averaging}} 

Given the light curve data files described in Section~\ref{ssec:DR15}, we computed the median and mean magnitudes, their formal uncertainties and $\chi^2$ (assuming constant brightness) for all stars, in all five bands. Due to more observational epochs in DR15, the new data are more sensitive to variability; following \pO, we applied $\chi^2<3$ in the $gri$ bands, as well as requirements for at least 4 epochs in the same three bands and the formal uncertainty of the mean $r$ band magnitude below 0.05 mag. These selection criteria recovered 98.5\% stars from the original catalogue, resulting in a new catalogue with 991,472 stars. 

\begin{figure*}
\centering
\includegraphics[width=0.48\textwidth, keepaspectratio]{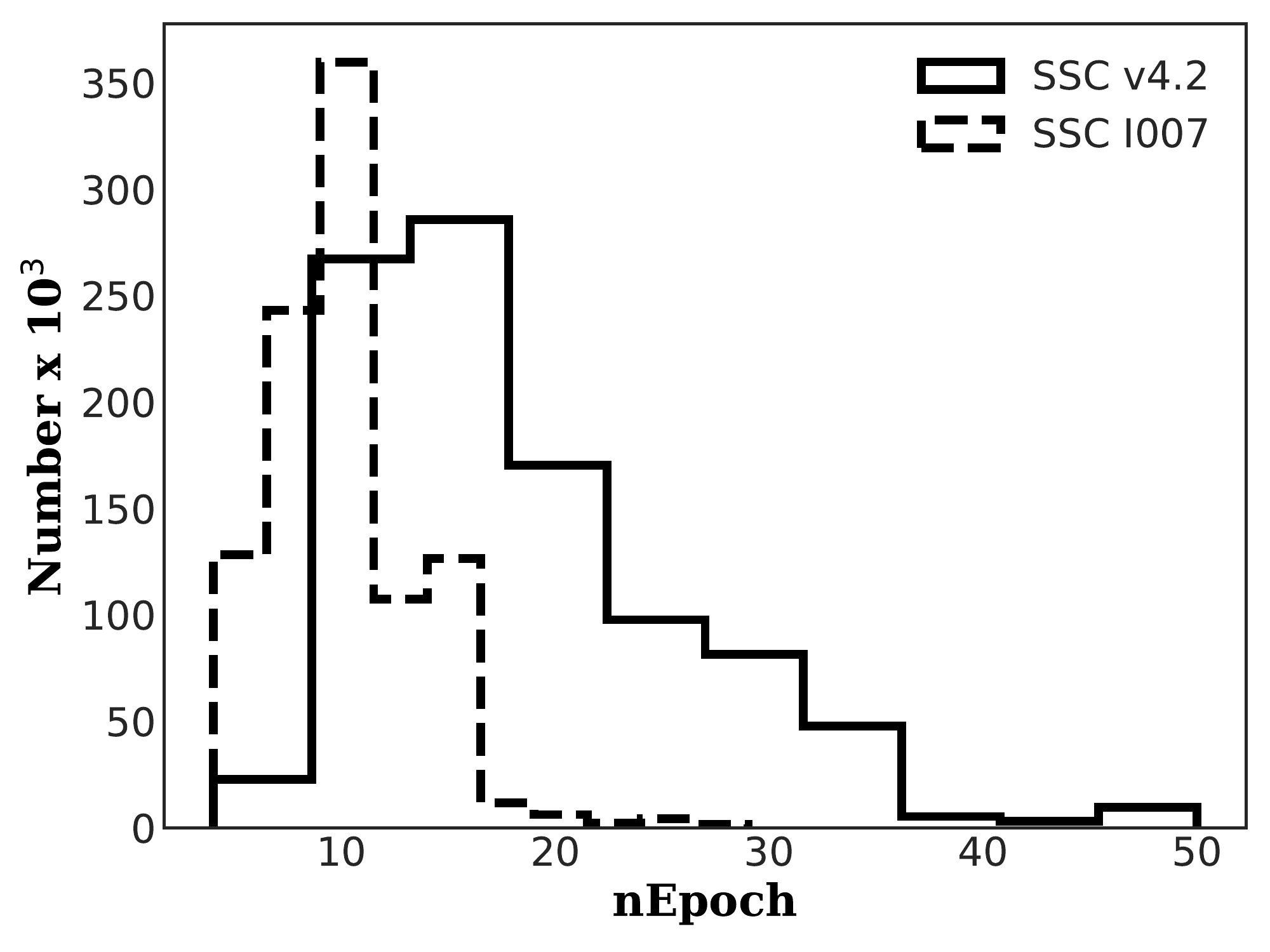}
\includegraphics[width=0.48\textwidth, keepaspectratio]{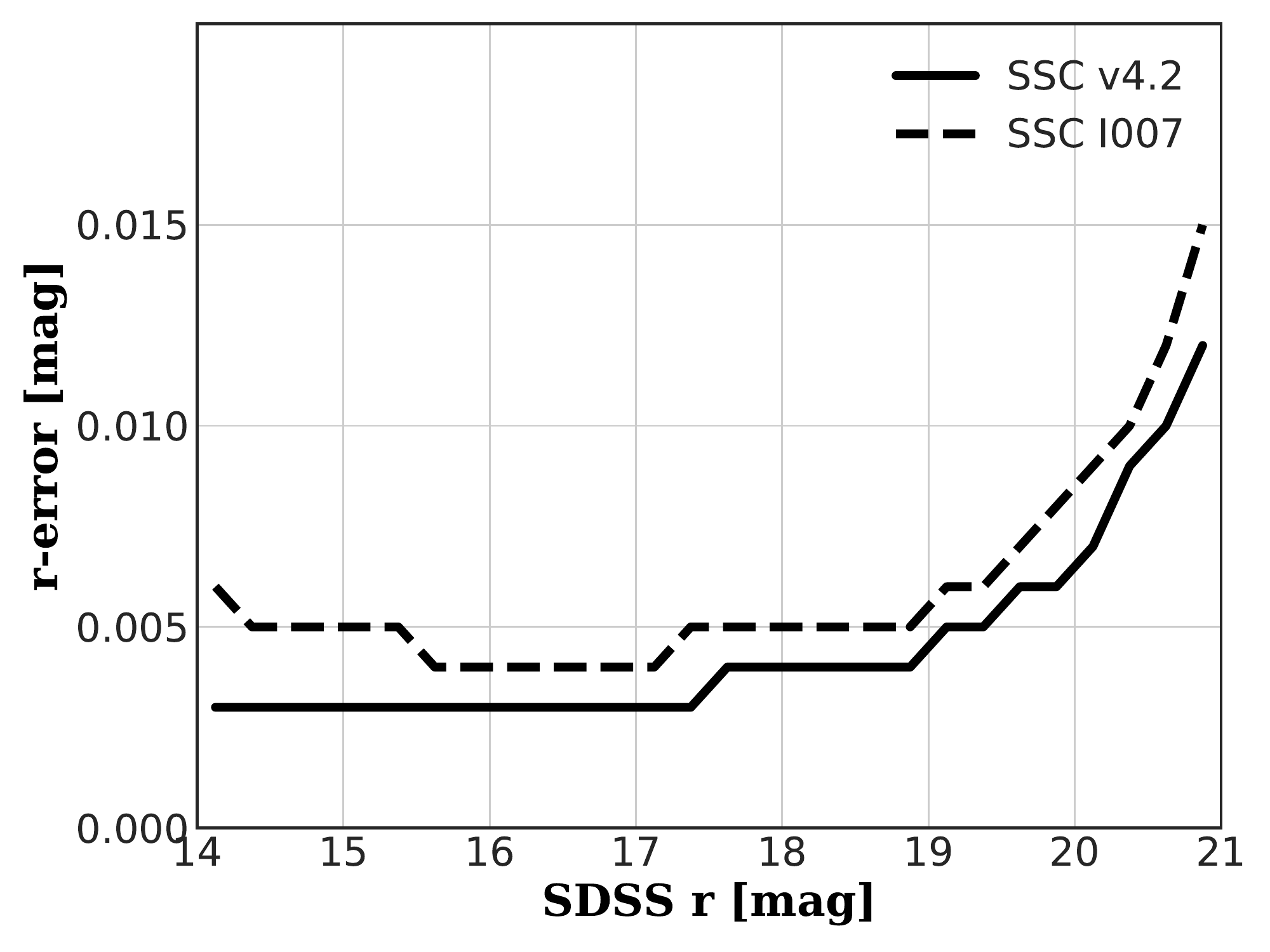}
\caption{({\it Left}) A comparison of the number of observational epochs for matched stars in the new standard star catalogue (SSC) v4.2 versus the \pOc. ({\it Right}) A comparison of the median formal $r$ band photometric uncertainties of matched objects in the v4.2 versus \pOc, as a function of their mean $r$ magnitudes.
\label{fig:rerr_nvso}}
\end{figure*}

Figure~\ref{fig:rerr_nvso} compares the numbers of epochs for matched stars (left panel) and their formal uncertainties of the mean $r$ band magnitude (right panel). The new catalogue has about 2-3 times more measurements per star, depending on its sky position within Stripe 82. Consequently,  formal photometric uncertainties (``random errors'') are about 1.4-1.7 times smaller. Using Gaia photometry, we recalibrate the zeropoints in this catalogue, as explained in the following subsection. This new, recalibrated version of the SDSS standard star catalogue, v4.2,  is publicly available by following the link provided in the data accessibility section. A detailed comparison of the photometry between the old and new catalogues is discussed in Section~\ref{sec:v26v42}. 

\subsection{The derivation of  photometric zeropoint corrections using Gaia EDR3 data\label{sec:GaiaCorr}} 

The variation of photometric zeropoints with position on the sky in the \pOc\ (see their eq.~4) was constrained using a combination of stellar colours \citep[the principal axes in colour-colour diagrams, for details see][]{2004AN....325..583I} and a standard star network \citep{2002AJ....123.2121S,2006AN....327..821T}. It is likely that residual errors in zeropoint calibration (e.g., a saw-tooth pattern as a function of Declination, which was reported by \citealt{2013A&A...552A.124B}; see their Fig.~23) could be further minimized using uniformly calibrated space-based photometry from Gaia EDR3, as shown in this work. 

\subsubsection{Positional matching of the SDSS and Gaia catalogues}

\begin{figure*}
\centering \includegraphics[width=0.45\textwidth, keepaspectratio]{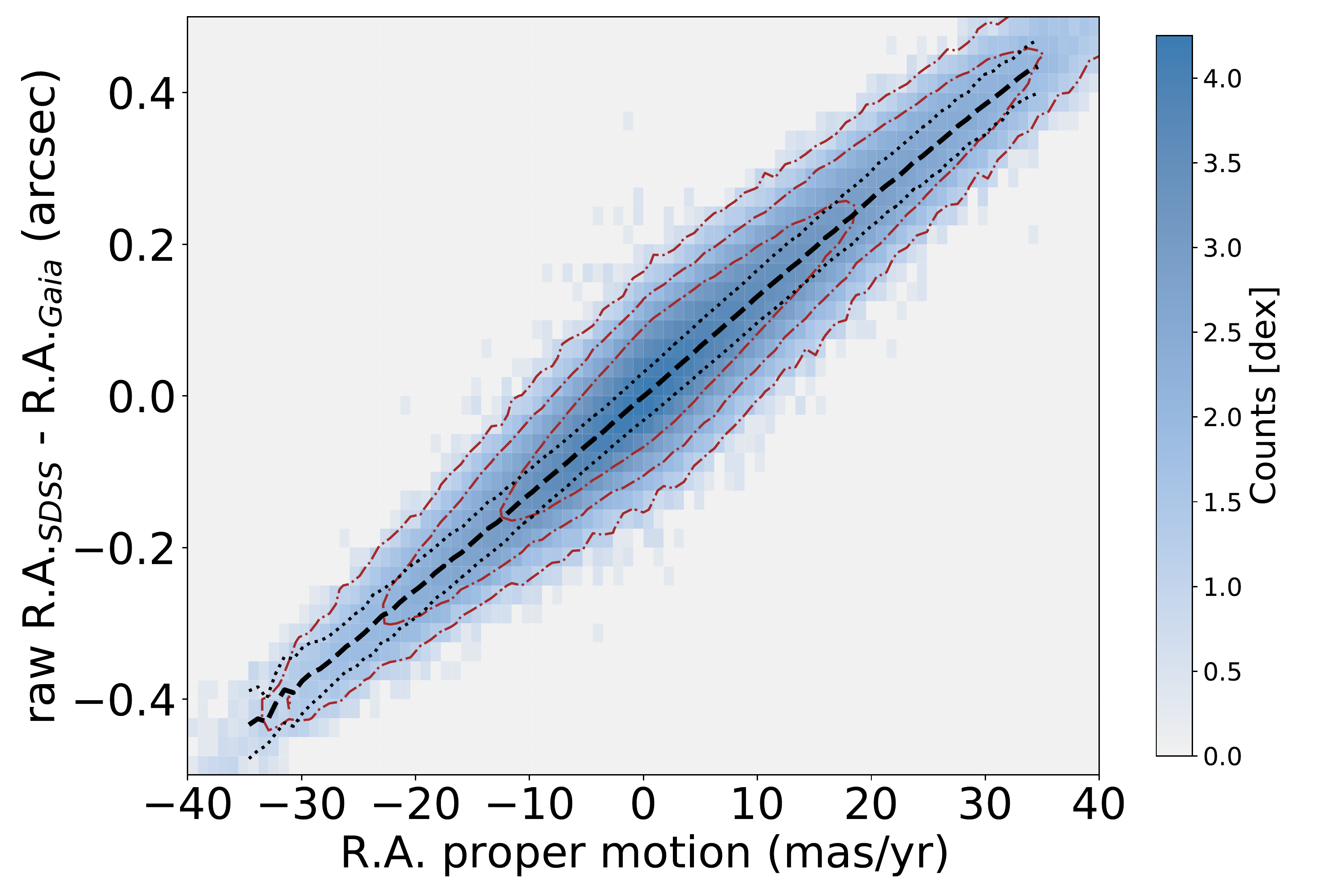}
\centering \includegraphics[width=0.45\textwidth, keepaspectratio]{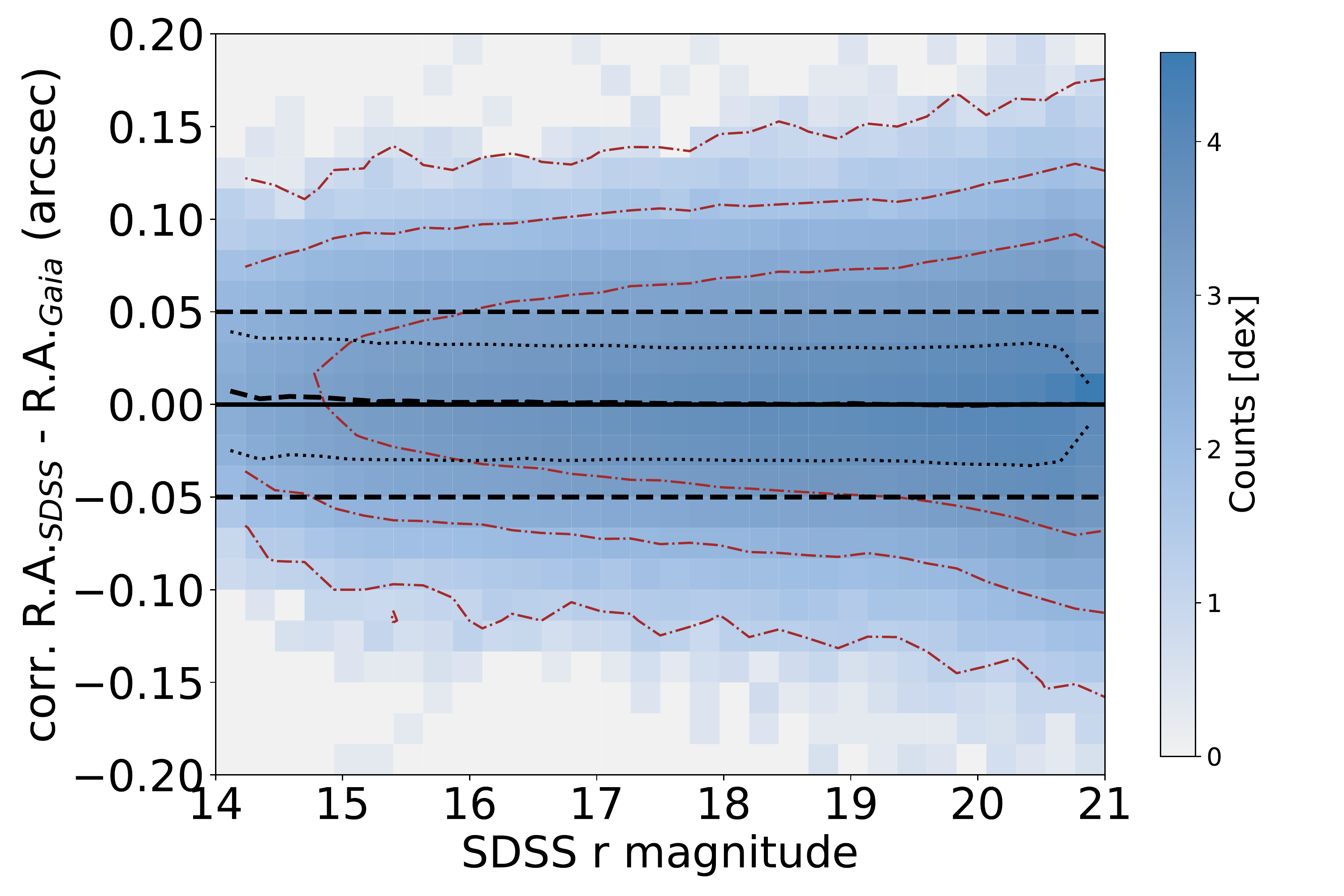}

\caption{({\it Left}) The R.A. difference between SDSS and Gaia vs. R.A. proper motion reported by Gaia EDR3. The solid line shows the median difference in bins of proper motion and the dashed lines mark the $\pm \sigma_G$ envelope around the medians, where $\sigma_G$ is the robust standard deviation. The density of the distribution (in dex) is indicated by the colour bar, and the overplotted contours represent 1, 2 and 3 dex ({\it Right}) The R.A. difference after correcting using the best-fit R.A. difference vs. proper motion curve, as a function of the SDSS $r$ magnitude. Overplotted contours represent 1, 2 and 3 dex. The residual differences are dominated by systematic errors in SDSS astrometry at the level of $\sim28$ milliarcsec (note that there is no increase with magnitude). In an analogous plot for Declination, the residuals are similar. 
\label{fig:GaiaRApm}}
\end{figure*}

Naively, one would positionally match the SDSS and Gaia EDR3 catalogues using a matching radius of about 0.3 arcsec because SDSS positions are accurate to better than 0.1 arcsec per coordinate (RMS) for sources with $r < 20.5$ mag \citep{2003AJ....125.1559P}.  However, observational epochs are sufficiently different that stellar proper motions need to be accounted for; indeed, we find a very strong correlation between the SDSS-Gaia positional differences and proper motions published in the Gaia EDR3 catalogue (see the left panel in  Figure~\ref{fig:GaiaRApm}). After accounting for proper motions,  the positions agree at the level of $\sim28$ milliarcsec (robust\footnote{We use a robust estimator of the standard deviation computed as $\sigma_G = 0.741*(q_{75}-q_{25})$, where $q_{25}$ and $q_{75}$ are the 25\% and 75\% quantiles, and the normalization factor 0.741 assures that $\sigma_G$ is equal to the standard deviation for a normal (Gaussian) distribution.} RMS, per coordinate). The residual differences are dominated by systematic errors in SDSS astrometry because there is no increase of this RMS with magnitude (see the right panel in Figure~\ref{fig:GaiaRApm}), and because the contribution of Gaia's astrometric measurement uncertainties is negligible. The implied SDSS astrometric accuracy of $\sim28$ milliarcsec is substantially better than the 0.1 arcsec reported by \cite{2003AJ....125.1559P}, but note that here we used positions ``averaged'' over typically $\sim20$ SDSS runs (see the left panel in Figure~\ref{fig:rerr_nvso}). 

\subsubsection{Gaia-based photometric zeropoint corrections  \label{sec:GaiaCorr2}}

Gaia EDR3 reported \GG\ magnitudes, which approximately span the SDSS $griz$ bandpasses, and BP and RP magnitudes, which approximately correspond to the blue and red halves of the \GG\ bandpass. We first use \GG\ data to derive ``grey'' zeropoint corrections (applied to all five SDSS bands), and then use the BP-RP colour to derive zeropoint corrections for the $ugiz$ bands, relative to the $r$ band. 

\begin{figure}
  \centering\includegraphics[width=0.98\columnwidth]{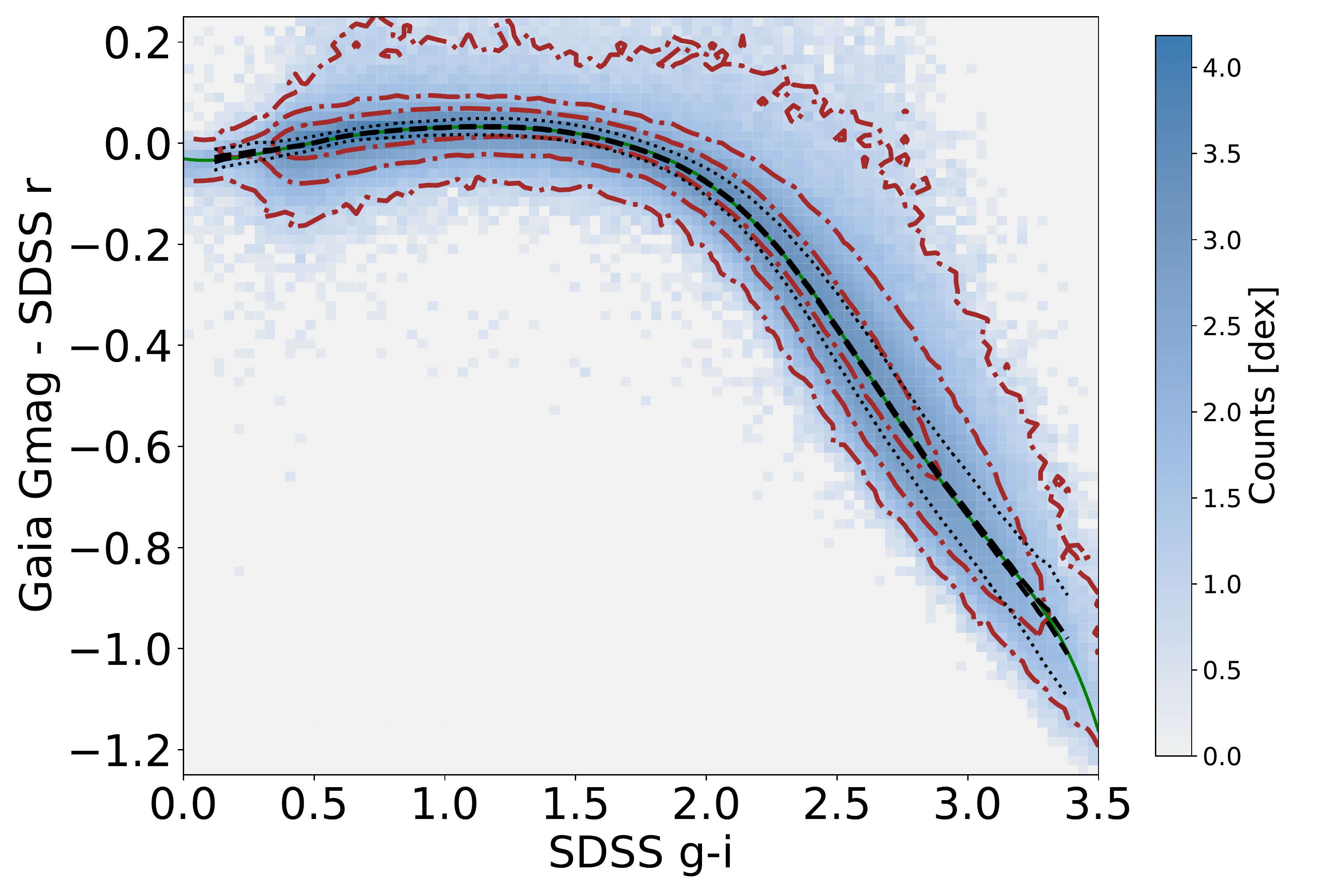} 
\caption{The variation of the difference between Gaia's \GG\ magnitude from Early Data Release 3 and SDSS $r$ magnitude with the SDSS $g-i$ colour. The  colour map and contours illustrate the distribution of $\sim 393,000$ matched stars with $16<$~\GG~$<19.5$. The thick dashed line represents the median values for 0.05 mag wide $g-i$ bins. These values are used to derive the grey zeropoint correction, as a function of R.A. and Declination. The thin dotted lines show the median values $\pm$ the robust standard deviation for each bin. The green solid line is a polynomial fit to the median values, given by Equation \ref{eq:poly} and discussed in \S \ref{sec:GaiaCorr2}. The fitted polynomial coefficients are listed in Table \ref{tab:poly}. }
\label{fig:GrVSgi}
\end{figure}

The basic idea is simple: first, use Gaia's G magnitude, \GG, and the SDSS $gri$ magnitudes to derive a synthetic G magnitude, \GS\ based on SDSS data; see Figure~\ref{fig:GrVSgi}. Next, bin the $\Delta$G = (\GS~-~ \GG) residuals by R.A. and Dec, and use the median residuals per bin as the grey correction for SDSS photometry (as functions of R.A. and Dec). Similarly, use Gaia's BP-RP colour to derive synthetic $u-r$, $g-r$, $r-i$ and $r-z$ colours, and use the median residuals per bin as zeropoint corrections for the $ugiz$ bands. 

Given a large number of matched stars ($\sim 400,000$), and a large number of colour combinations, we do not attempt to derive analytic fits for synthetic magnitudes and colours but instead use narrow colour bins of 0.05 mag, and linearly interpolate between the bins. We have verified that even sixth-order polynomial fits do not provide better results than this simple numerical approach. For example, we had to use two piece-wise polynomial fits to be able to fit the \GG-$r$ vs.$g-i$ relation with residuals smaller than 0.01 mag separately for $0.0 < g-i < 2.0$ and $2.0 < g-i < 3.4$ ($g-i$=2.0 approximately corresponds to M0 MK spectral type. These fit residuals are dominated by systematic errors, mainly the inability of even the sixth order polynomial to follow all the variations at the millimag level).

\begin{equation}
\label{eq:poly}
         \GG-r = \sum_{k=0}^{k=6} \, a_k\,(g-i)^k,
\end{equation}
The best fit polynomial coefficients for both colour ranges are listed in Table~\ref{tab:poly} and the fit is shown by the green solid line in Figure~\ref{fig:GrVSgi}.

\begin{table*}
	\centering
	\caption{The best-fit coefficients for eq.~\ref{eq:poly}.}
	\label{tab:poly}
	\begin{tabular}{r|r|r|r|r|r|r|r} %
		\hline
		Colour range & $a_0$ & $a_1$ & $a_2$ & $a_3$ & $a_4$ & $a_5$ & $a_6$    \\
		\hline
        $0.0 < g-i < 2.0$ &  $-$0.0307 &  $-$0.0885 &   0.6632 &  $-$1.0179 &   0.7262 &   $-$0.2547 &   0.0337  \\
        $2.0 < g-i < 3.4$ &    9.6488 &  $-$8.4115 &  $-$7.9834 &  13.757 &  $-$7.0973 &  1.5993 &  $-$0.1351   \\
		\hline
	\end{tabular} 
\end{table*}

\begin{figure}
    \centering\includegraphics[width=0.95\columnwidth]{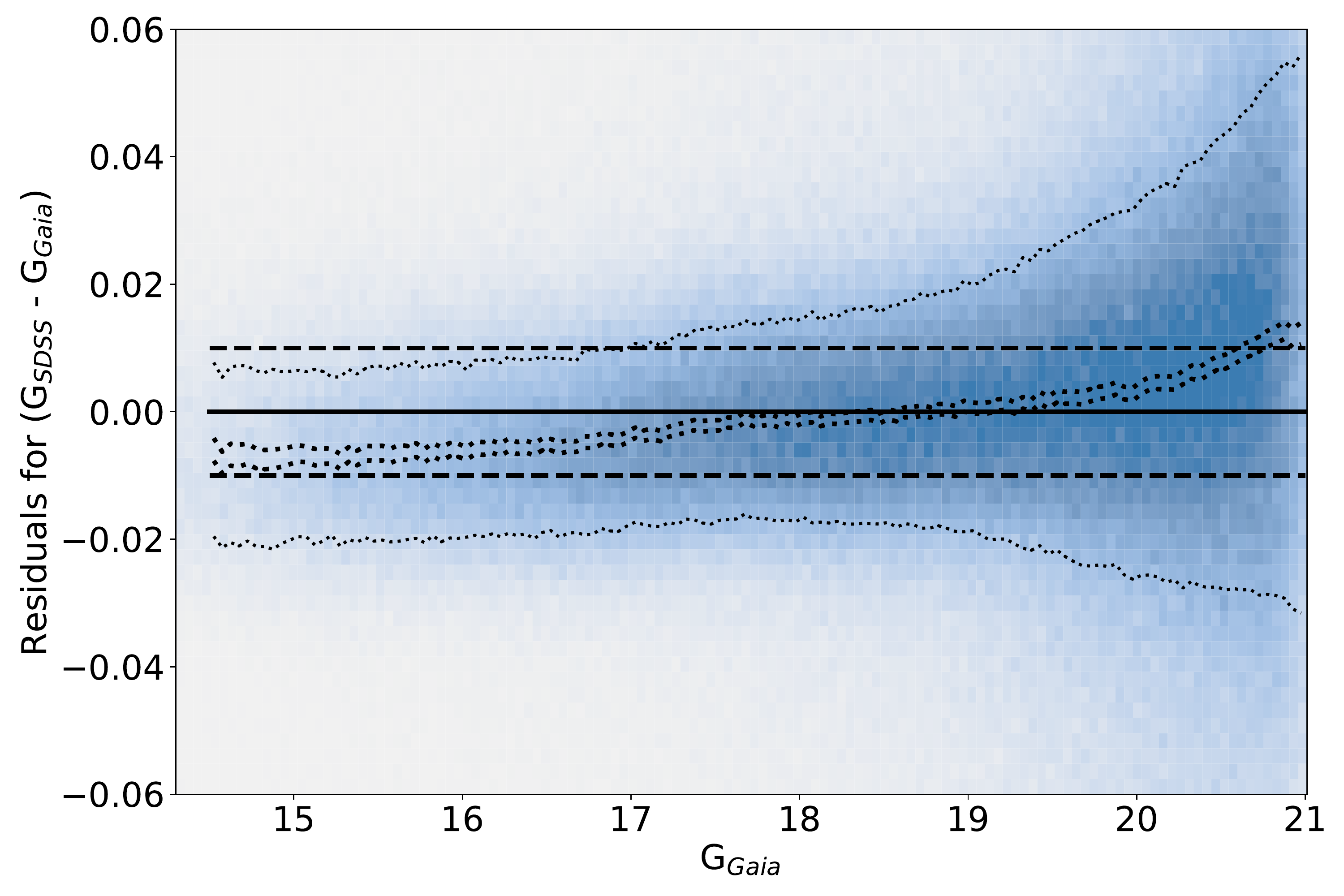} 
\caption{The variation of the residuals between Gaia's G magnitude from Early Data Release 3, \GG, and the synthetic G magnitude values, \GS\ generated using SDSS $gri$ photometry. The two solid lines represent the median values $\pm$ uncertainty of the median for each 0.05 mag wide \GG\ bin. The short-dashed lines show the median values $\pm$ the robust standard deviation for each bin. The horizontal solid and long-dashed lines at zero and $\pm$0.01 mag, respectively, are added to guide the eye. Note an overall gradient of $\sim0.01$ mag from bright (G=16) to faint (G=20) end -- a comparison of the SDSS catalogue with Pan-STARRS and DES catalogues (see Figure~\ref{fig:drVSr}) suggests that its origin is a magnitude-dependent bias in Gaia's EDR3 photometry rather than a problem with SDSS photometry.}
\label{fig:gaiaJump}
\end{figure}

The variation of G magnitude residuals, $\Delta G$, with \GG\ (see Figure~\ref{fig:gaiaJump}) shows an overall gradient of $\sim10$ millimag from bright (G=16) to faint (G=20) end. A comparison of the SDSS catalogue with Pan-STARRS and DES catalogues (see Section~\ref{sec:DESPS1} and Figures~\ref{fig:DESPSRA} and \ref{fig:drVSr}) strongly suggests that the origin of this discrepancy is a magnitude-dependent bias in Gaia's \GG\ photometry, rather than a problem with the SDSS catalogue (offsets between the SDSS and DES photometry,  are $<2$ millimag at \GG$\sim$20, and $<5$ millimag when compared to Pan-STARRS).

\begin{figure}
    \centering\includegraphics[width=0.99\columnwidth]{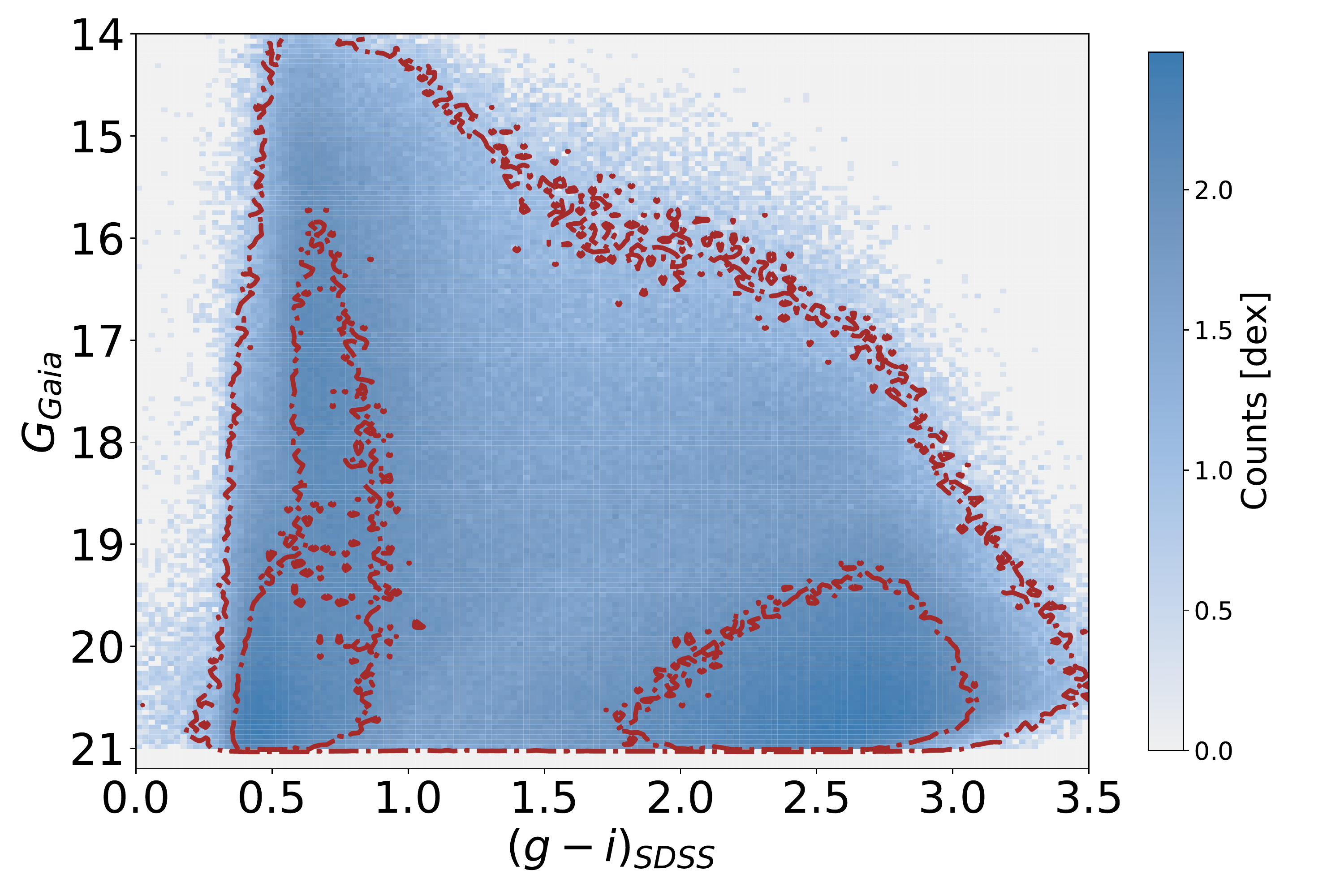} 
\caption{A Hess diagram constructed with Gaia's G magnitude and SDSS $g-i$ colour. The counts are shown on logarithmic scale, according to the colourbar. The 1-$\sigma$ and 2-$\sigma$ contours are also shown.}
\label{fig:Ggi}
\end{figure}

\begin{figure*}
  \centering\includegraphics[width=0.49\textwidth]{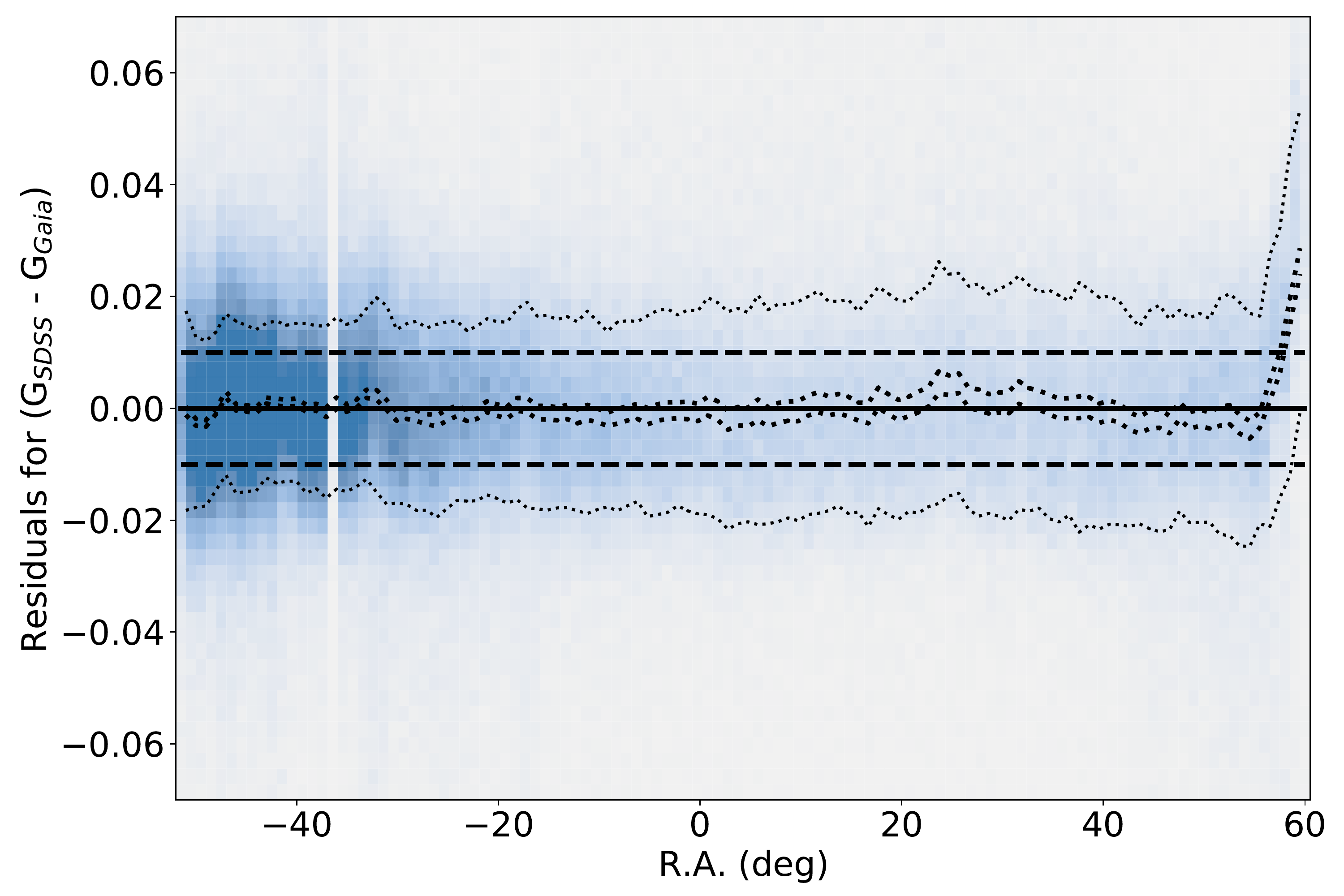} 
  \centering\includegraphics[width=0.49\textwidth]{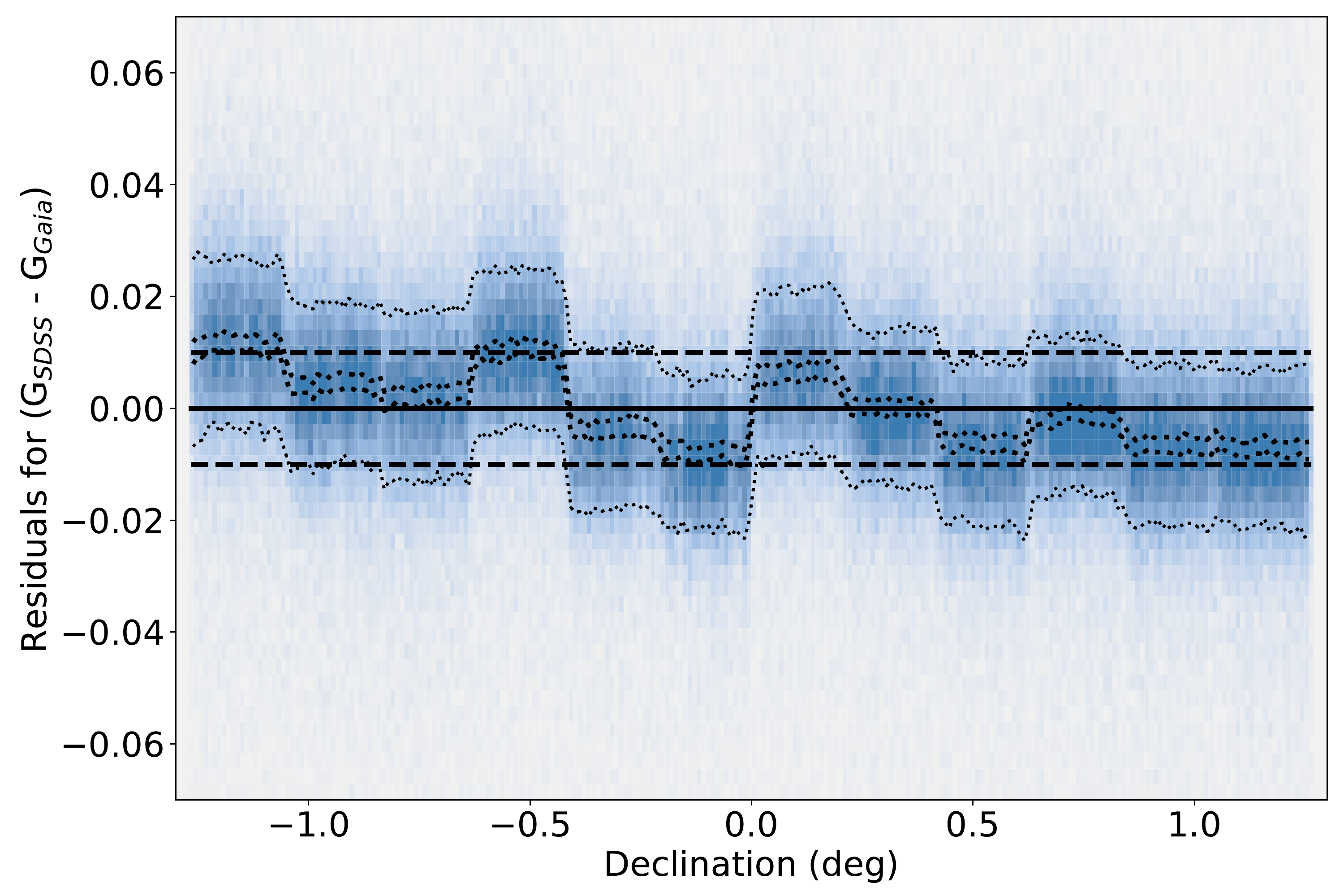} 
\caption{({\it Left}) The R.A. variation of the residuals between Gaia's G magnitude from EDR3, \GG\ and synthetic G magnitude values, \GS, generated using SDSS $gri$ photometry. The  colour map illustrates the distribution of $\sim 372,000$ matched stars within $16<$ \GG $<19.5$ and $0.4 < g-i < 3.0$. The two solid lines represent the median values $\pm$ uncertainty of the median for 1 degree wide R.A. bins. The short-dashed lines show the median values $\pm$ the robust standard deviation for each bin. The horizontal solid and long-dashed lines at zero and $\pm$0.01 mag, respectively, are added to guide the eye. The mean of the two solid lines is the grey correction, as a function of R.A., applied to the SDSS $ugriz$ magnitudes. The standard deviation for the applied correction is 3.2 millimag. {\it Note}: The stripe of missing data seen at R.A.$\sim$ -37 deg is due to the presence of the globular cluster, $M2$. The resulting high stellar density region was masked off during data processing. ({\it Right}) Analogous to the left panel, except that here results are shown for 0.01 degree wide Declination bins. The 12 clearly visible regions correspond to two SDSS scans (in R.A. direction) and six CCD columns in the SDSS camera. The standard deviation for the applied correction is 6.5 millimag, with a maximum absolute value of $\sim0.01$ mag.}
\label{fig:graycorrRA}
\end{figure*}

Given these two features, we limit the calibration sample to the $16<$~\GG~$<19.5$ magnitude range. We further restrict calibration stars to the $0.4 < (g-i)_{SDSS} < 3.0$ colour range (approximately A0 to M5 spectral range), yielding a sample of $\sim372,000$ stars (see Figure~\ref{fig:Ggi} for the \GG\ vs. $g-i$ colour-magnitude diagram). The behaviour of the median G magnitude residuals per R.A. and Declination bin are shown in the left and right panels respectively of Figure~\ref{fig:graycorrRA}.

Except for a few degrees long region at the edge of Stripe 82 (R.A.$>$55 deg), the SDSS photometric zeropoints are remarkably stable with respect to R.A.; the scatter is only 3.5 millimag. On the other hand, there are clear deviations in the Declination direction, which clearly map to the 12 scanning strips that fill Stripe 82. We note that discrepancies never exceed 0.01 mag (with a scatter of 6.2 millimag), which was the reported accuracy of the \pOc. Thanks to a large number of stars in the sample, and well calibrated Gaia's photometric zeropoints across the sky, we can now constrain SDSS zeropoints with a precision of about 1 millimag per 0.01 degree wide Declination bin. 

The residuals shown in the left and right panels of Figures~\ref{fig:graycorrRA} are applied as ``grey'' zeropoint corrections to $ugriz$ magnitudes, as functions of R.A. and Declination, to all 991,472 stars in the catalogue. 

\begin{figure*}
 \centering\includegraphics[width=0.49\textwidth]{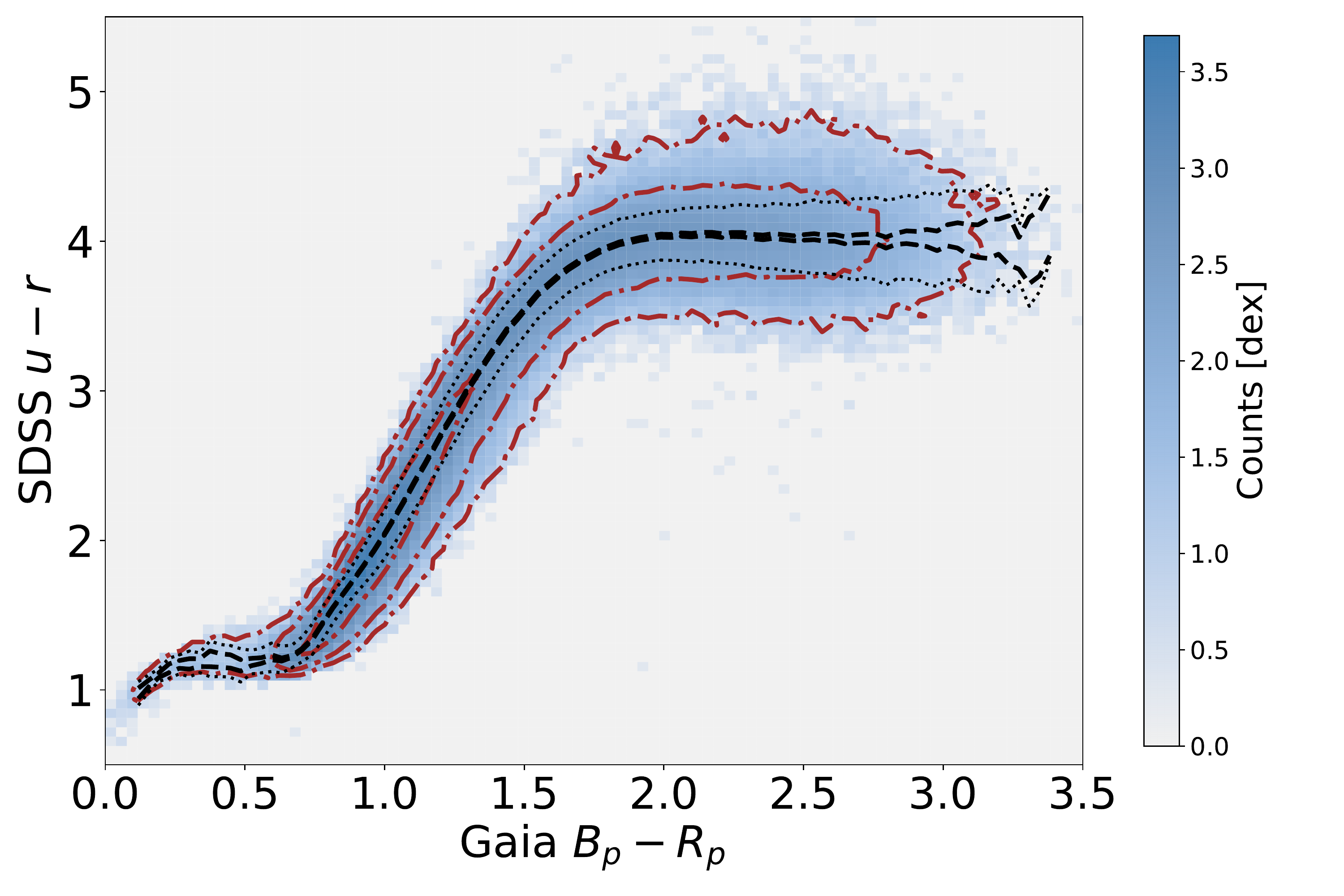} 
 \centering\includegraphics[width=0.49\textwidth]{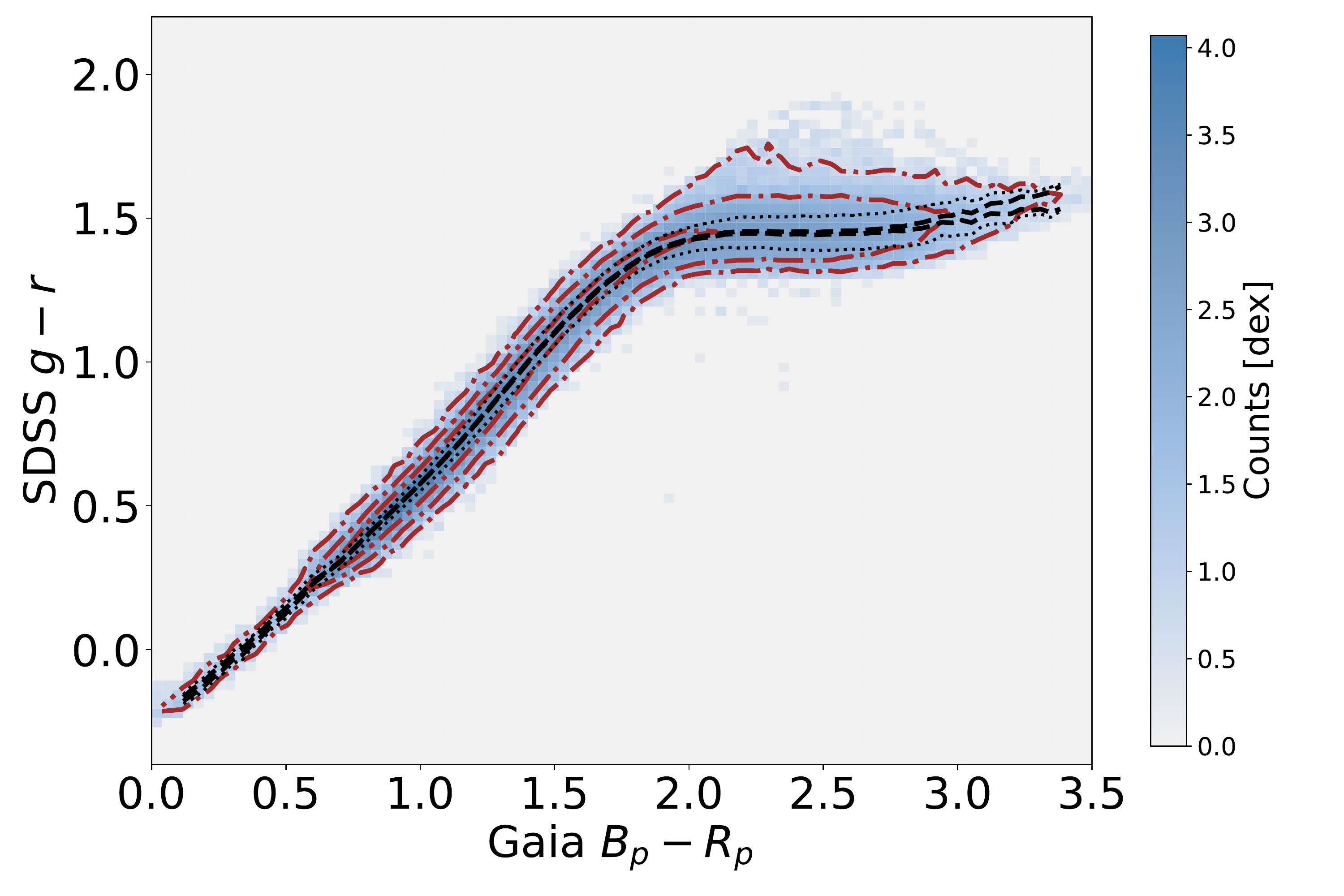} 
 \centering\includegraphics[width=0.49\textwidth]{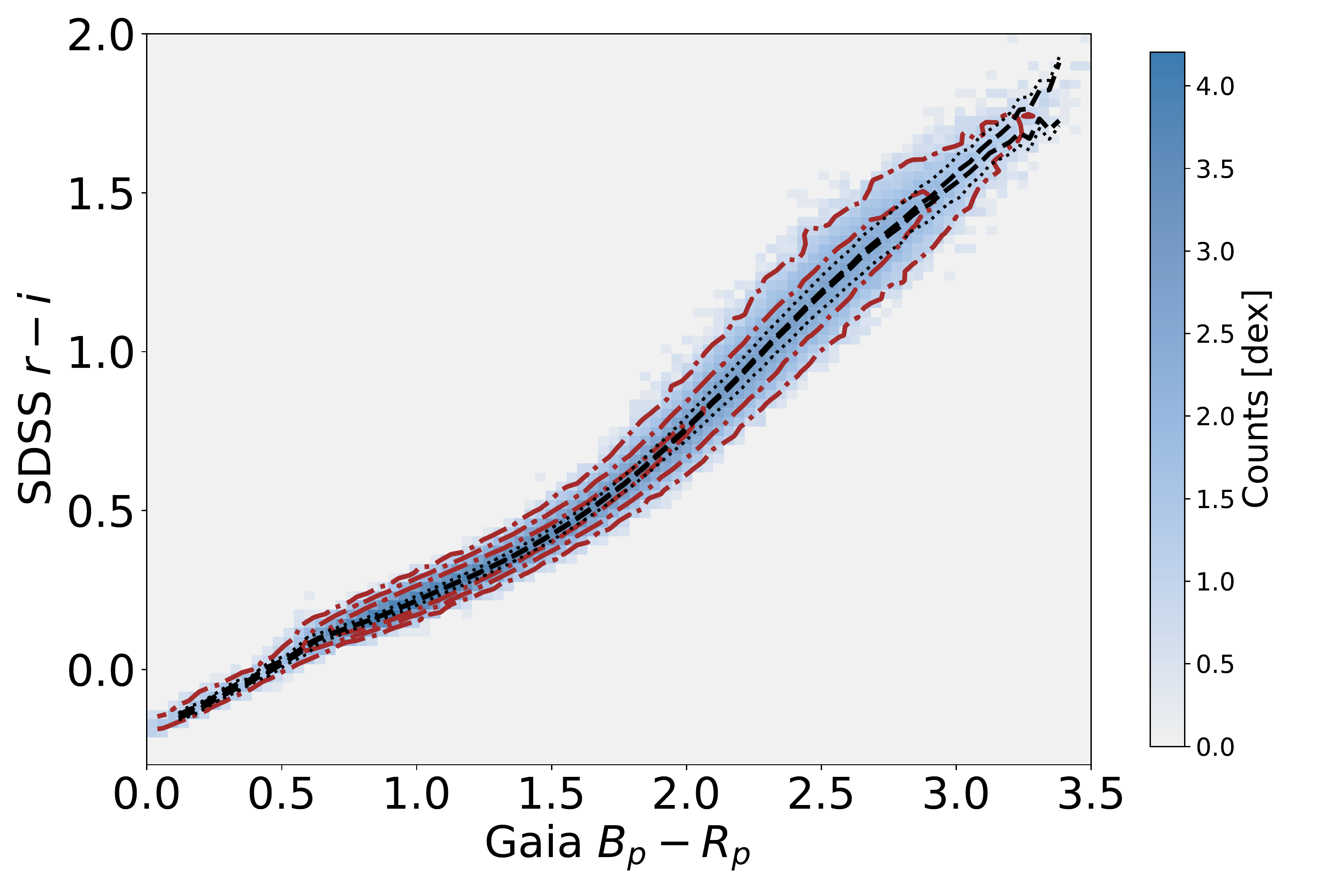} 
 \centering\includegraphics[width=0.49\textwidth]{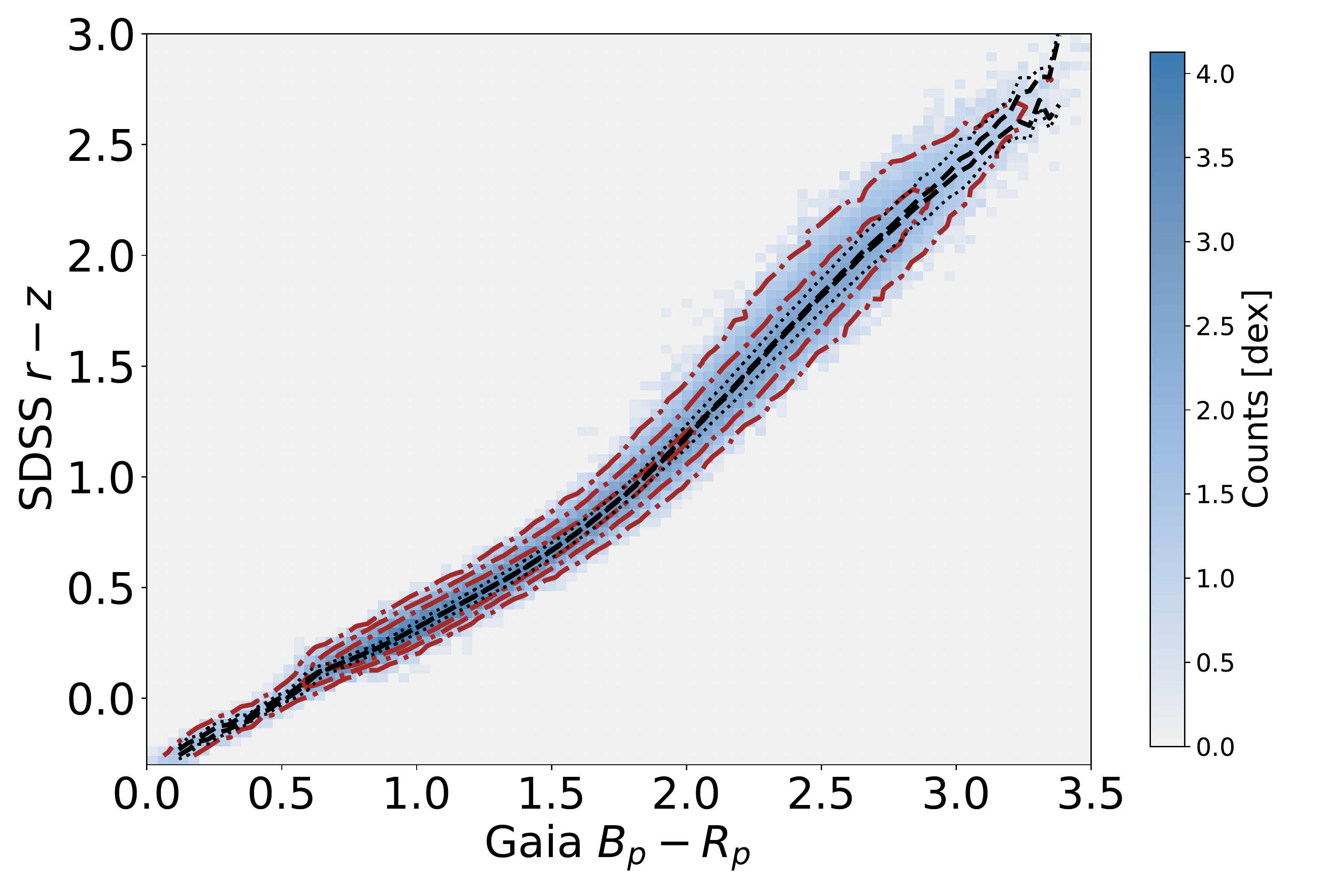} 
\caption{Analogous to Figure~\ref{fig:GrVSgi}, except that the correlations between SDSS colours and Gaia's BP-RP colour are shown.}
\label{fig:VSBpRp}
\end{figure*}

\begin{figure}
    \centering\includegraphics[width=0.98\columnwidth]{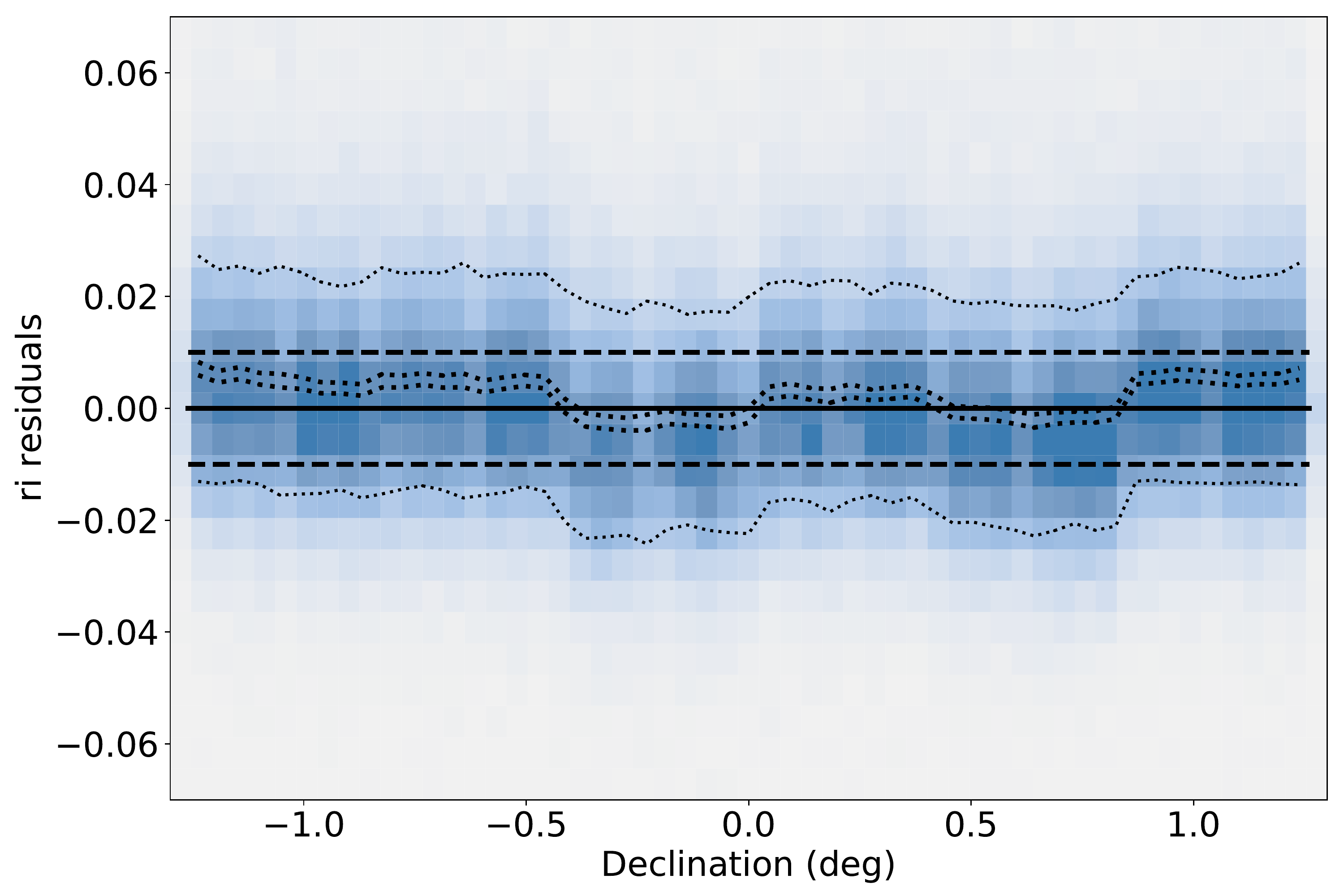} 
\caption{Analogous to Figure~\ref{fig:graycorrRA} ({\it Right}), except that here residuals correspond to differences between the SDSS $r-i$ colour and a synthetic $r-i$ colour generated using Gaia's $BP-RP$ colour. Note the signature of SDSS camera columns at the level of a few millimags. The standard deviation for the binned medians is 3.2 millimag (for other bands, please see Table~\ref{tab:GaiaRMS}).}
\label{fig:riresid}
\end{figure}

In the next re-calibration step, we derive synthetic $u-r$, $g-r$, $r-i$ and $r-z$ colours from Gaia's BP-RP colour, using the same binning procedure as we used above for the \GG~$-r$ vs. $g-i$ variation (see Figure~\ref{fig:VSBpRp}). An example of colour residuals is shown in Figure~\ref{fig:riresid}.  The median residuals per R.A. and Declination bins are then used as zeropoint corrections for the $ugiz$ bands. We required that the median offsets for all stars are vanishing and thus photometry in the new catalogue is on the same AB scale as the old catalogue (for related discussion, see Section~\ref{sec:AB}). The robust standard deviation for all zeropoint corrections is listed in Table~\ref{tab:GaiaRMS}.


\begin{table}
	\centering
	\caption{The robust standard deviation for binned SDSS-based vs. Gaia-based colour residuals$^a$. }
	\label{tab:GaiaRMS}

	\begin{tabular}{l|c|c} %
		\hline
		Colour & RMS for R.A. & RMS for Dec \\
		\hline
 
 grey (Gmag) &    3.2         &    6.5   \\
    $u-r$        &   0.0$^b$  &   20    \\     
    $g-r$        &   3.9         &    4.2    \\
    $r-i$         &   3.7         &    3.2    \\ 
    $r-z$        &   6.8         &    2.9    \\ 
		\hline
	\end{tabular}
     \vspace{1ex}

     {\raggedright {\bf Notes}: (a) The standard deviation for applied Gaia-based zeropoint corrections. The robust standard deviation is estimated using the interquartile range. The units are in millimag. \newline (b) For the $u$ band, we could not confirm the R.A. behaviour of Gaia-based zeropoint correction with the CFIS data and did not apply it. The large $u$ band correction as a function of Declination was validated with the CFIS data (see Section~\ref{sec:CFIStest}). \par}
\end{table}

The largest corrections were derived for the $u$ band. Given that Gaia's BP-RP colour does not strongly constrain the $u$ band flux, we used the CFIS catalogue (see Section~\ref{ssec:cfis}) as an independent cross check. We verified that zeropoint errors in the SDSS catalogue implied by Gaia and CFIS data agree at the level a few millimags in Declination direction, but found larger inconsistencies of $\sim$0.01-0.02 mag for R.A. bins. For this reason, we only applied the $u$ band correction in Declination direction. The plausible $u$ band zeropoint errors in the new catalogue are further discussed in Section~\ref{sec:CFIStest}. 


\subsubsection{Validation of recalibration  \label{sec:SSCvsGaia}} 
  
\begin{figure*}
    \centering\includegraphics[width=0.49\textwidth]{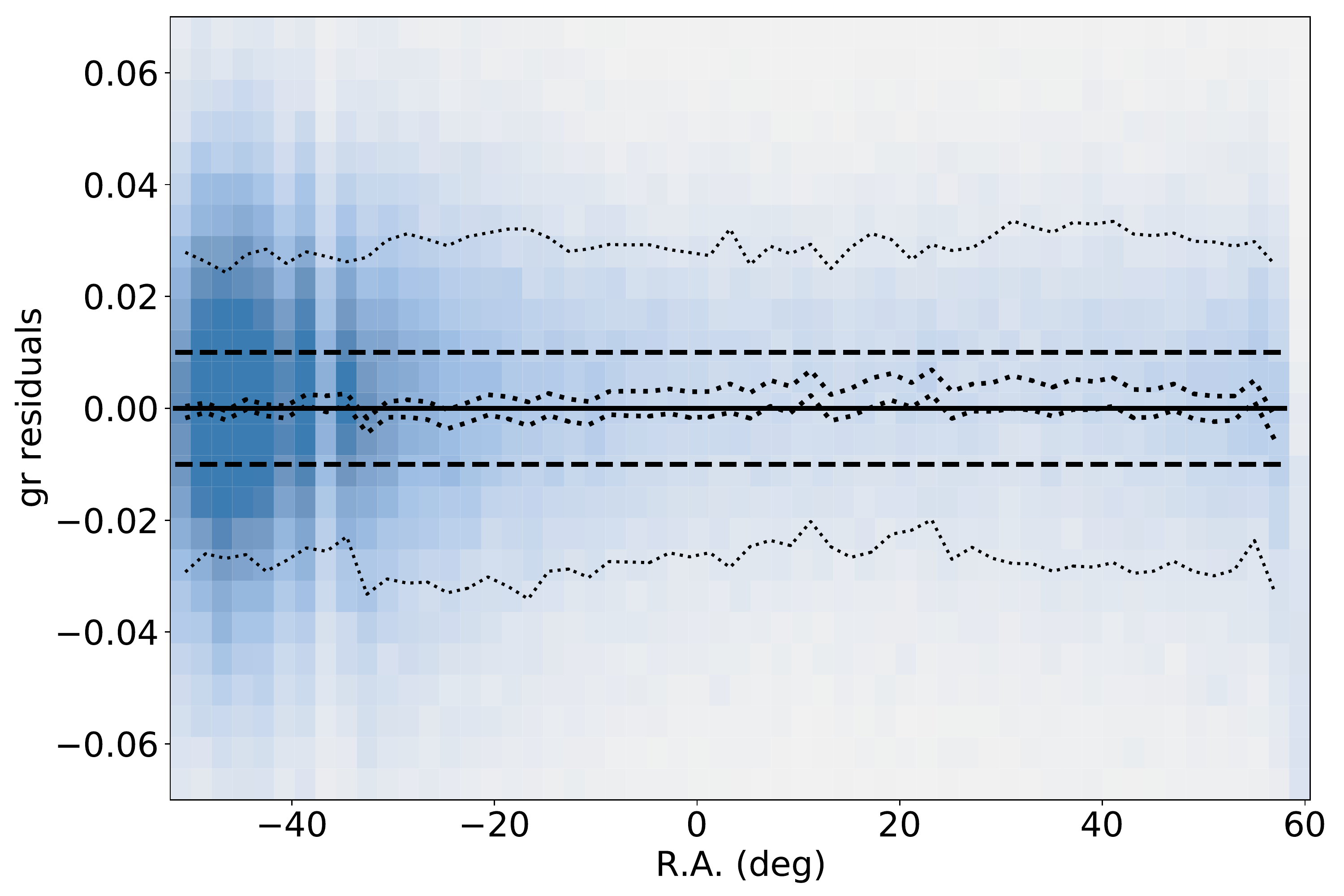} 
    \centering\includegraphics[width=0.49\textwidth]{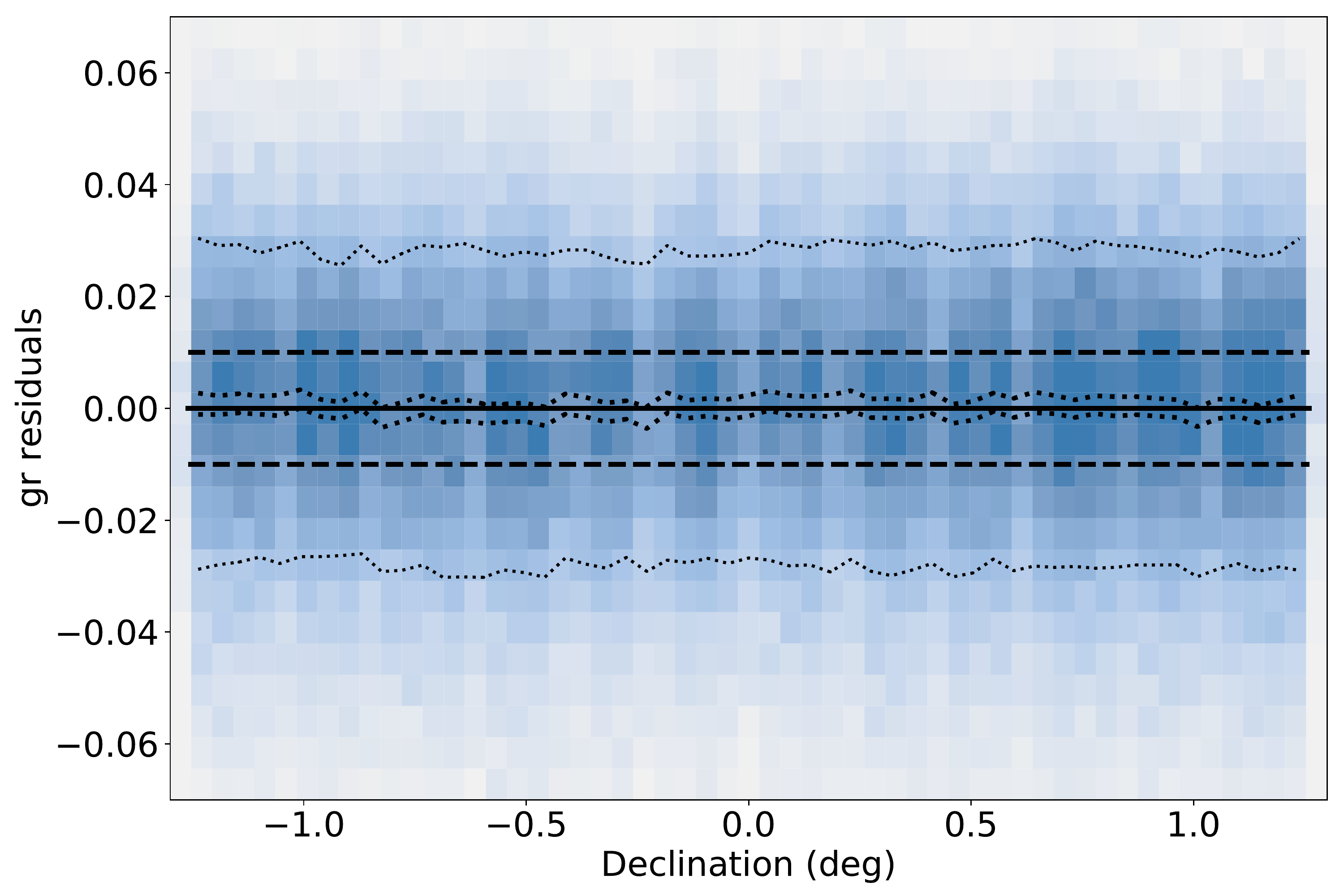} 
\caption{({\it Left}) Analogous to Figure~\ref{fig:graycorrRA}, except that here residuals between the SDSS $g-r$ colour from the v4.2 catalogue and a synthetic $g-r$ colour generated using Gaia's $BP-RP$ colour are shown. The binned median scatter is 1.6 millimag. ({\it Right}) The SDSS $g-r$ residuals are shown as a function of Declination. The binned median scatter is 0.8 millimag.}
\label{fig:grVSgaiaRADec}
\end{figure*}

By construction, the new v4.2 catalogue should not show appreciable zeropoint residuals when binned by R.A. and Declination. We have verified this expectation for all colours used in the recalibration. For illustration, Figure~\ref{fig:grVSgaiaRADec} shows such a test for the $g-r$ colour, with binned median scatter of the order 1 millimag. 


\subsection{Comparison of the SDSS \pOc\ and the v4.2 catalogue \label{sec:v26v42}} 

The v2.6 (``old'') SDSS Standard Star \pOc\  has been extensively used \citep[e.g.,][]{2008AJ....135..338F}, and here we briefly analyze differences between the v4.2 (``new'') and v2.6 magnitudes to inform future users about the consistency of the catalogues. In our analysis, we first compare the magnitudes of individual stars in the v2.6 and v4.2 catalogues, and bin the differences by R.A., Declination and magnitude. 

\begin{figure}
    \centering\includegraphics[width=0.95\columnwidth]{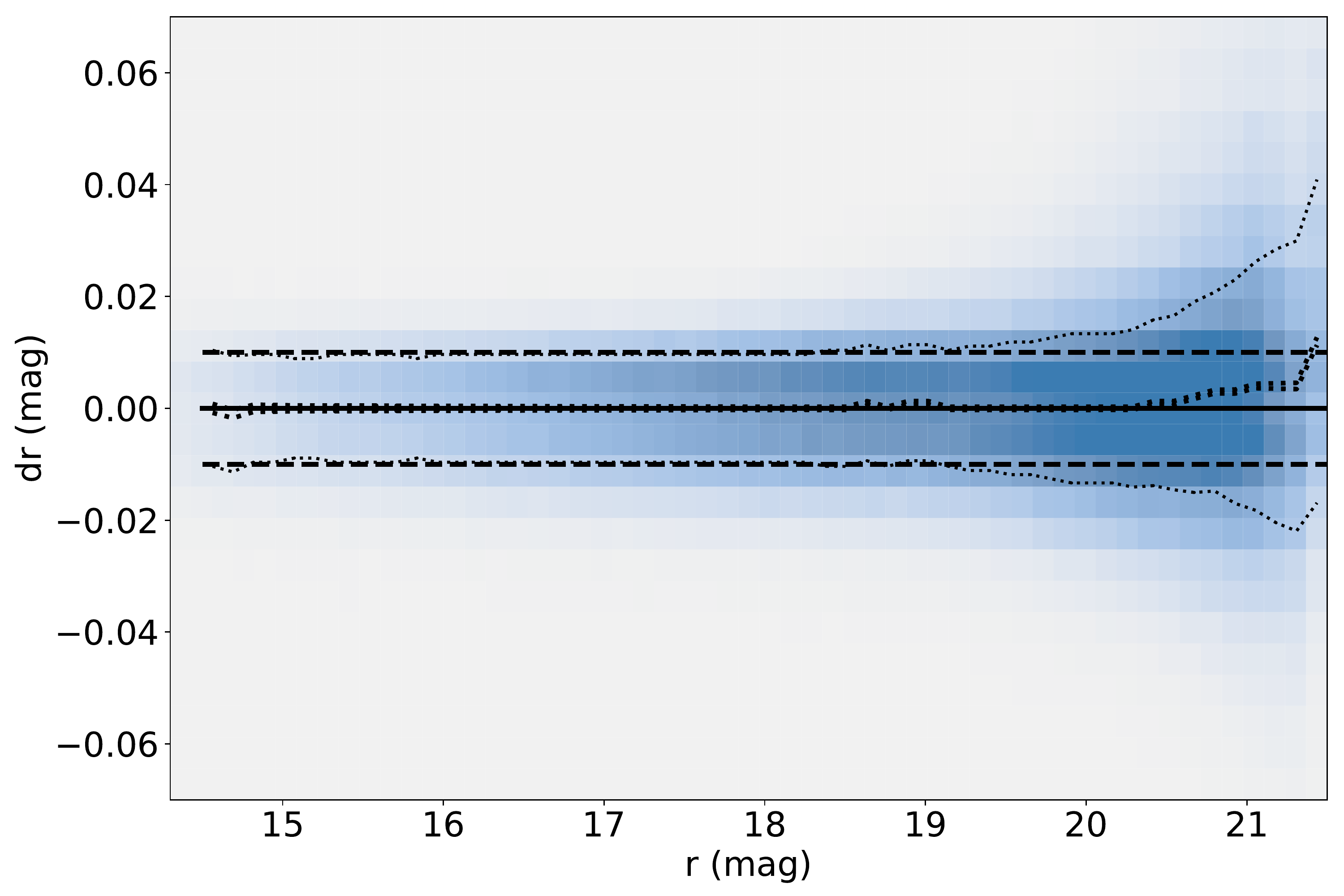} 
\caption{The differences between $r$ band magnitudes listed in the v2.6 and v4.2 SDSS Standard Star catalogues,  shown as a function of the $r$ band magnitude. The scatter of median values per bin is 1.7 millimag. The scatter of individual values is $\sim0.01$ mag for $r<20$, and is due to more data in the new catalogue.} 
\label{fig:v26v34drr}
\end{figure}

\begin{figure}
    \centering\includegraphics[width=0.95\columnwidth]{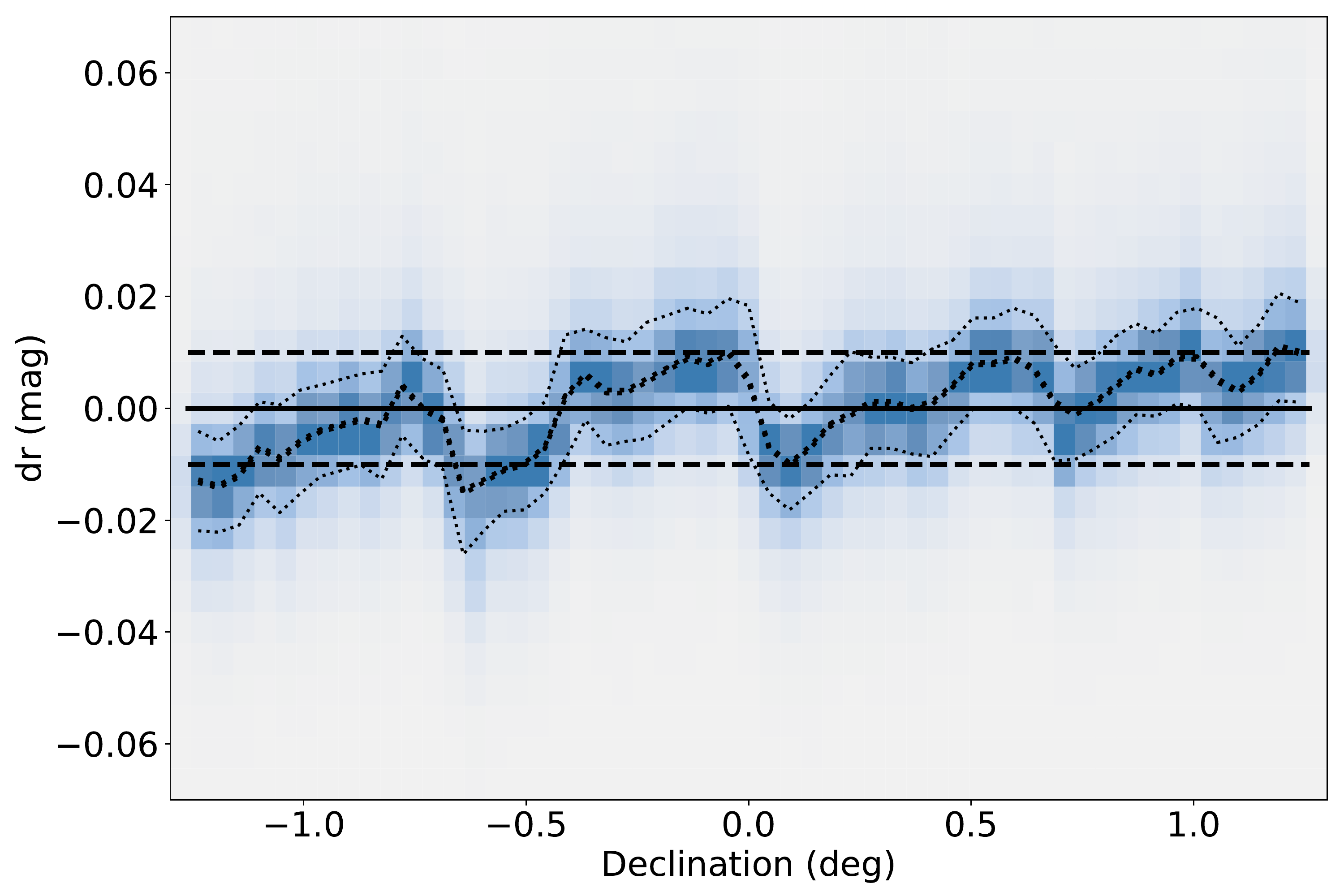} 
\caption{Analogous to Figure~\ref{fig:v26v34drr}, except that here the $r$ band differences are shown as a function of Declination (cross-scan direction). The size of the four regions corresponds to the size of the field-of-view of the SDSS Photometric Telescope \citep{2006AN....327..821T}. The standard deviation for median values per bin is 7.1 millimag, with extreme values of about 0.01 mag. The scatter of binned medians in the R.A. direction is much smaller -- 1.7 millimag. For statistics in other bands, please see Table~\ref{tab:oldnewRMS}.}
\label{fig:v26v34drDec}
\end{figure}

On average, both catalogue versions are on the same magnitude scale (the median $ugriz$ magnitude differences for all stars are zero by construction). There are no systematic offsets when binned by magnitude, as illustrated in Figure~\ref{fig:v26v34drr}. The most obvious differences appear when magnitude differences are binned by Declination. An example is shown in Figure~\ref{fig:v26v34drDec}, where the periodicity of residuals corresponds to the size of the field-of-view of the SDSS Photometric Telescope \citep{2006AN....327..821T}. The standard deviation for median values per bin is 6.8 millimag, with extreme values of about 0.01 mag. It is likely that systematic errors in the photometry of the calibration star network were propagated through ``flat-field corrections'' discussed by \pO\ to the v2.6 catalogue. We note that these errors, now found thanks to the Gaia catalogue, are well within the claimed photometric accuracy by both \pO\ and \cite{2002AJ....123.2121S}. The standard deviation for the median values per bin for the five bands and both coordinates is listed in Table~\ref{tab:oldnewRMS}. 
   
\begin{table}
	\centering
	\caption{The robust standard deviation for magnitude differences between the v2.6 (old) and v4.2 (new) catalogues (in millimag).}
	\label{tab:oldnewRMS}
	\begin{tabular}{l|c|c} %
		\hline
		Band & RMS for R.A. & RMS for Dec \\
		\hline
        
       $u$        &        2.0$^a$    &    25.5      \\
       $g$        &        4.0    &      9.7      \\  
       $r$         &        1.7    &      7.1      \\  
       $i$         &        4.1    &      6.5      \\ 
       $z$        &        7.5    &      8.4      \\ 
		\hline
	\end{tabular}
     \vspace{1ex}

     {\raggedright {\bf Notes}: (a) For the $u$ band, the scatter in the R.A. direction is due to more observations in v4.2 than in v2.6, rather than the zeropoint correction. \par}
\end{table}

\begin{figure*}
    \centering\includegraphics[width=0.49\textwidth]{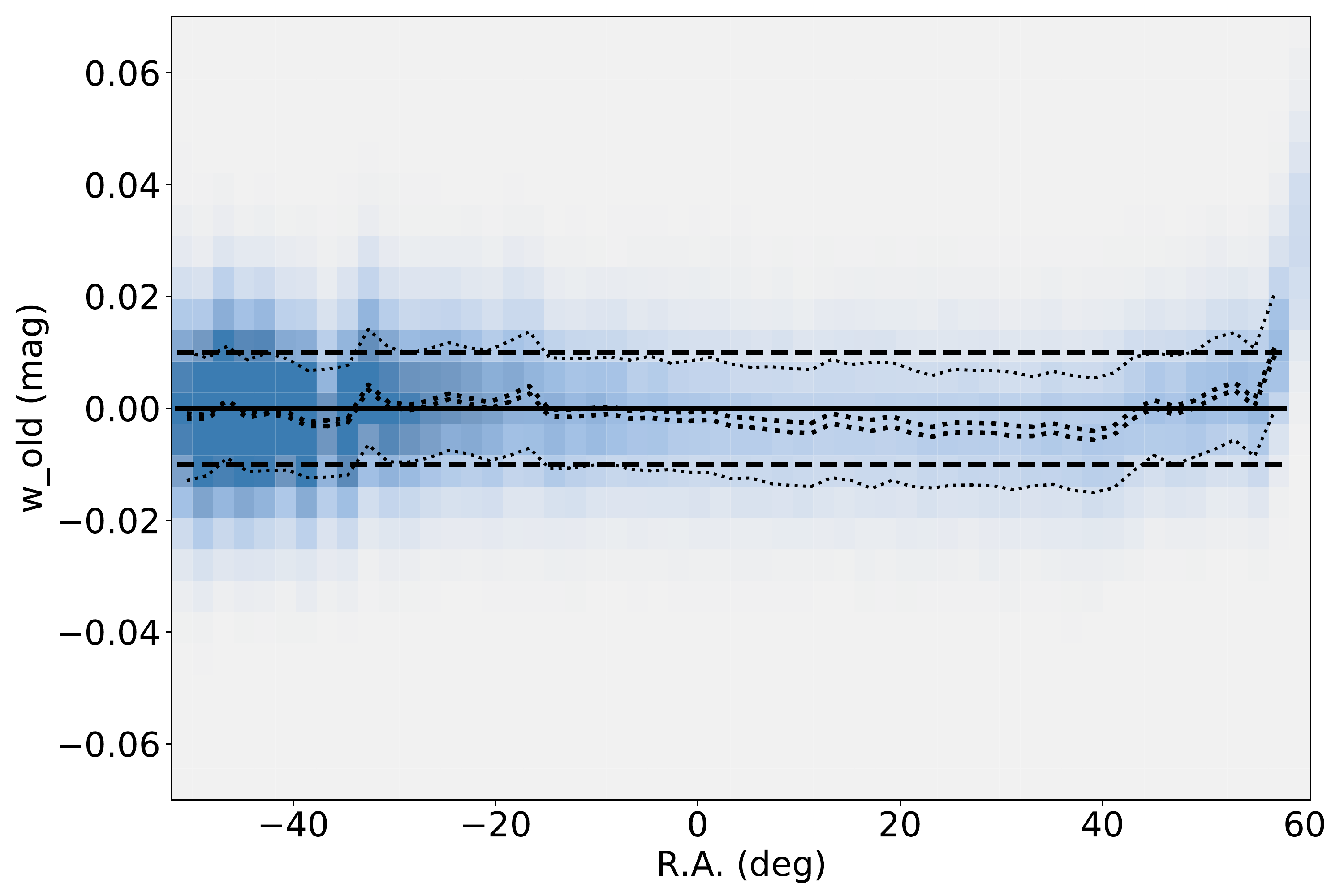}
    \centering\includegraphics[width=0.49\textwidth]{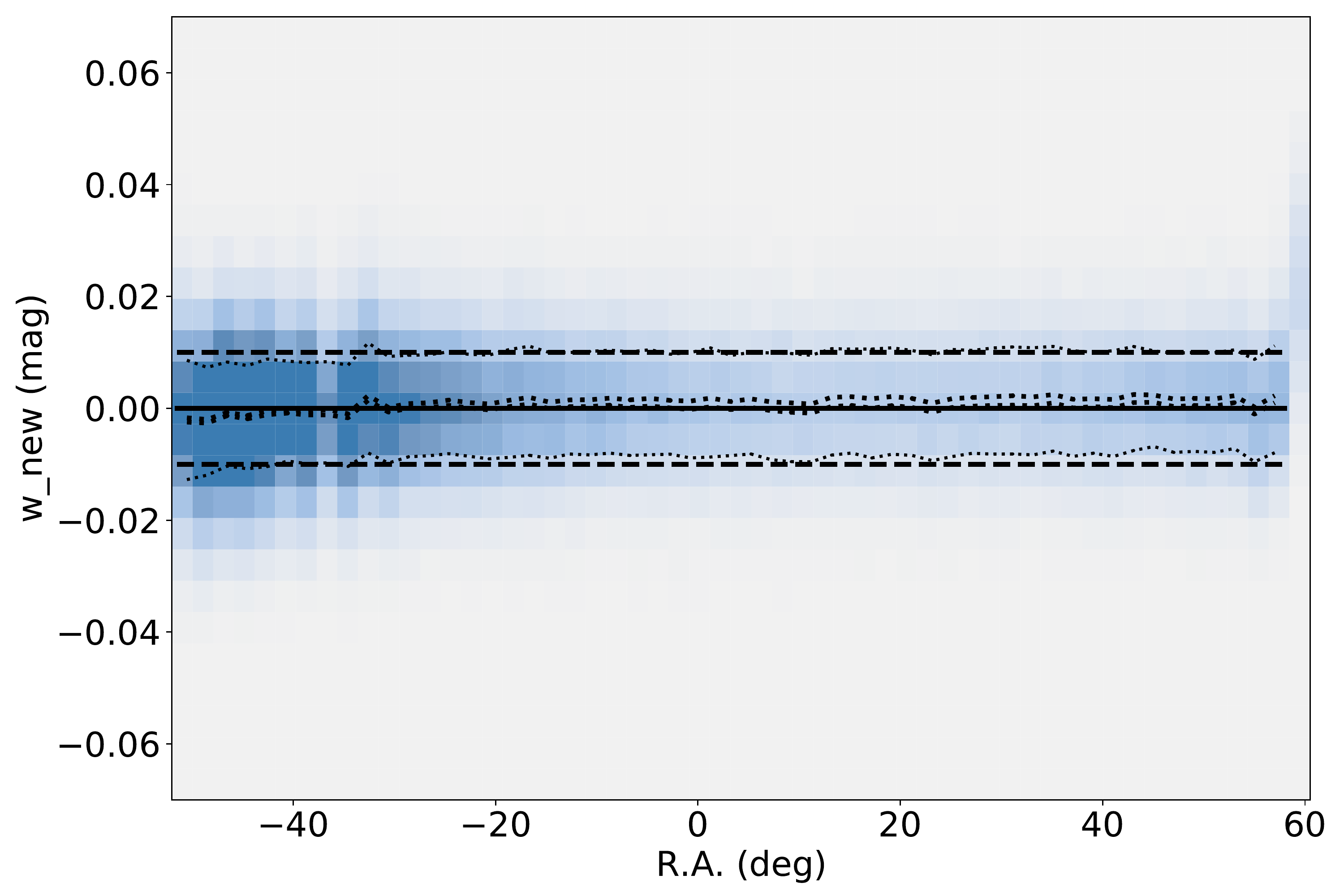}
    \centering\includegraphics[width=0.49\textwidth]{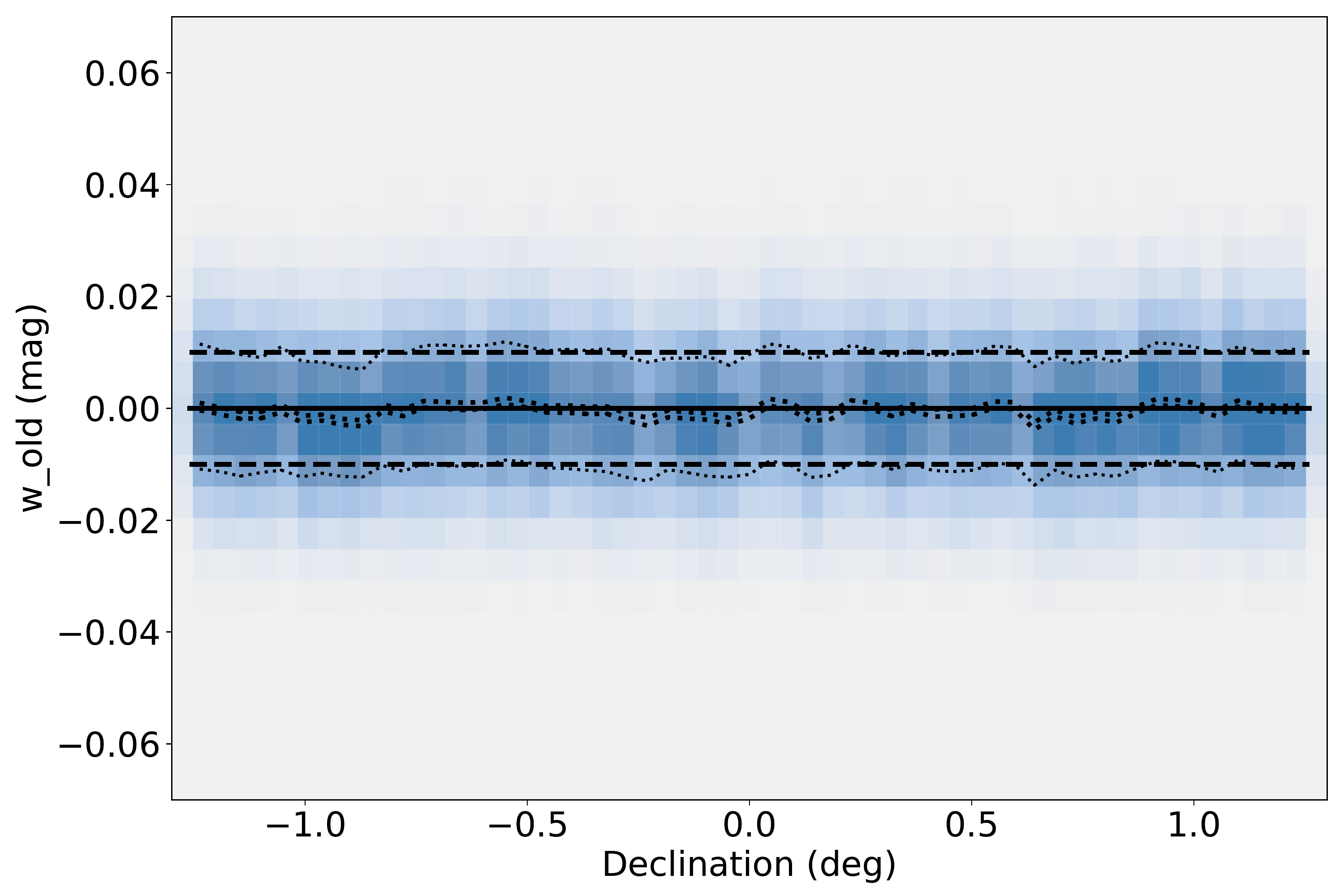}
    \centering\includegraphics[width=0.49\textwidth]{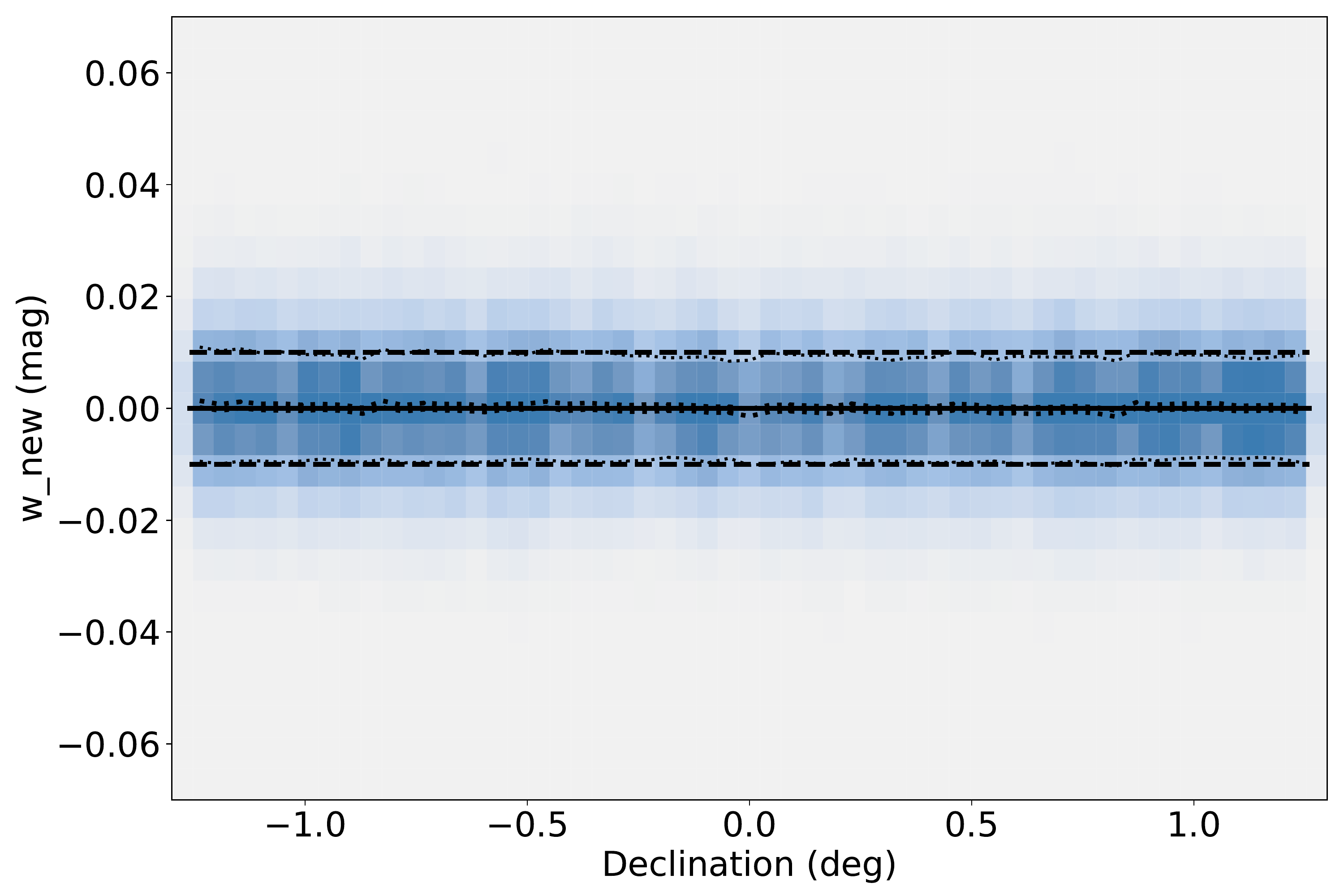}
\caption{A comparison of the $w$ colour, the second principal colour in the SDSS $r-i$ vs. $g-r$ colour-colour diagram, for the v2.6 ({\it Left panels}) and v4.2 ({\it Right panels}) catalogues. The top row shows the behaviour in the R.A. direction, while the bottom row is for the Declination direction. The standard deviation of the median $w$ values binned by R.A. and Dec is 2.6 millimag and 1.1 millimag for v2.6 and 1.0 millimag and 0.3 millimag for v4.2,
respectively.}
\label{fig:comparew} 
\end{figure*}

\begin{figure*}
    \centering\includegraphics[width=0.49\textwidth]{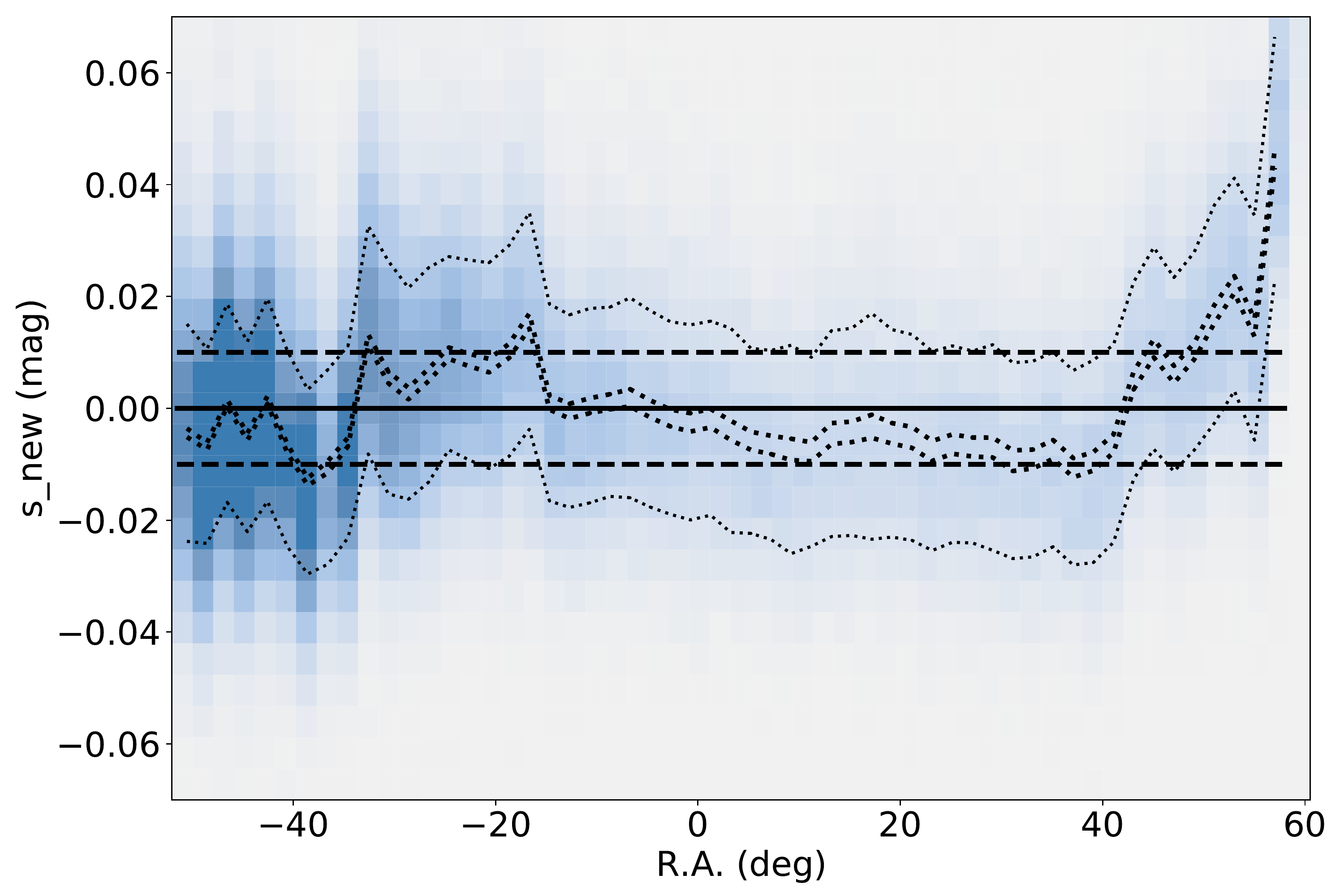}
    \centering\includegraphics[width=0.49\textwidth]{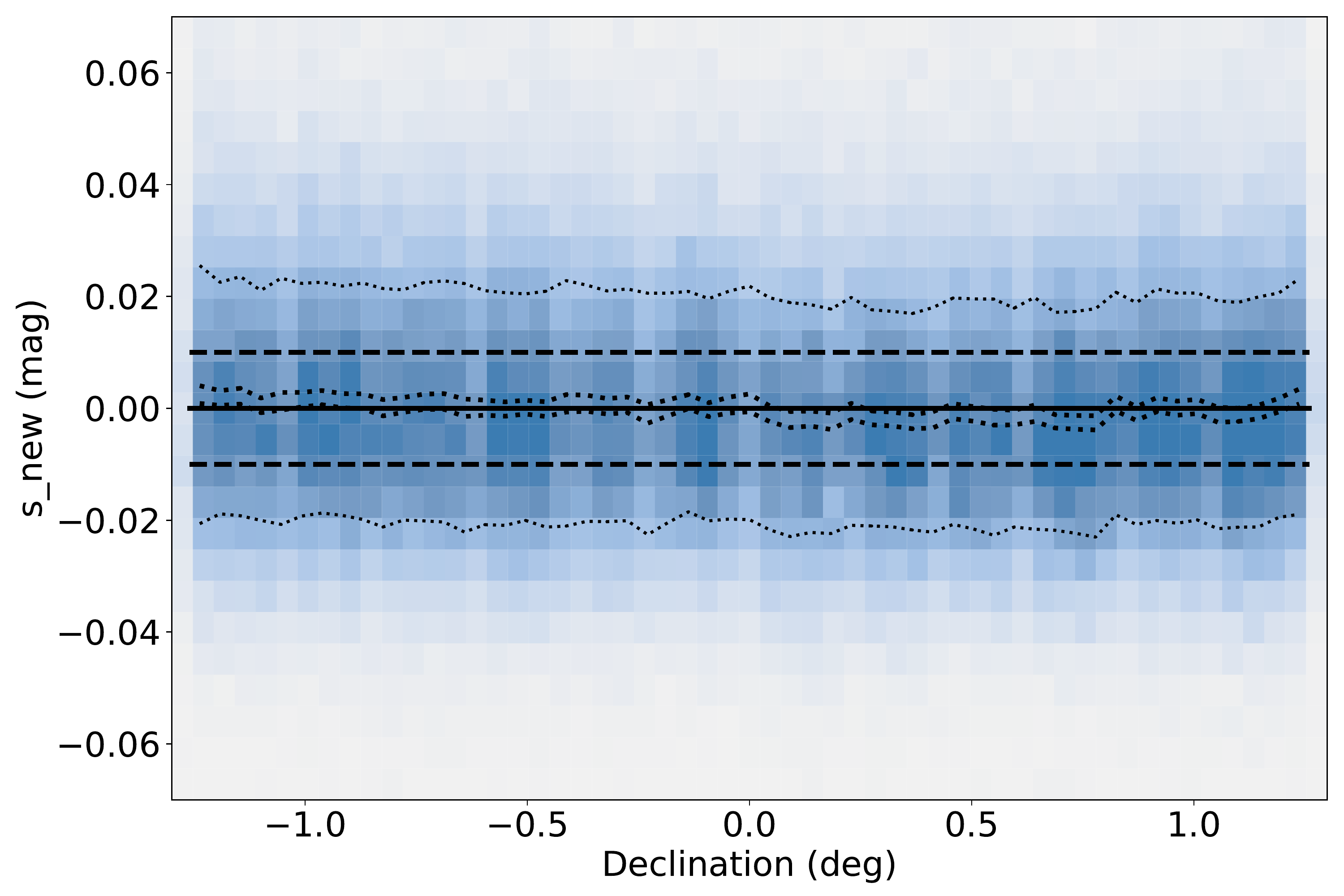} 
    \centering\includegraphics[width=0.49\textwidth]{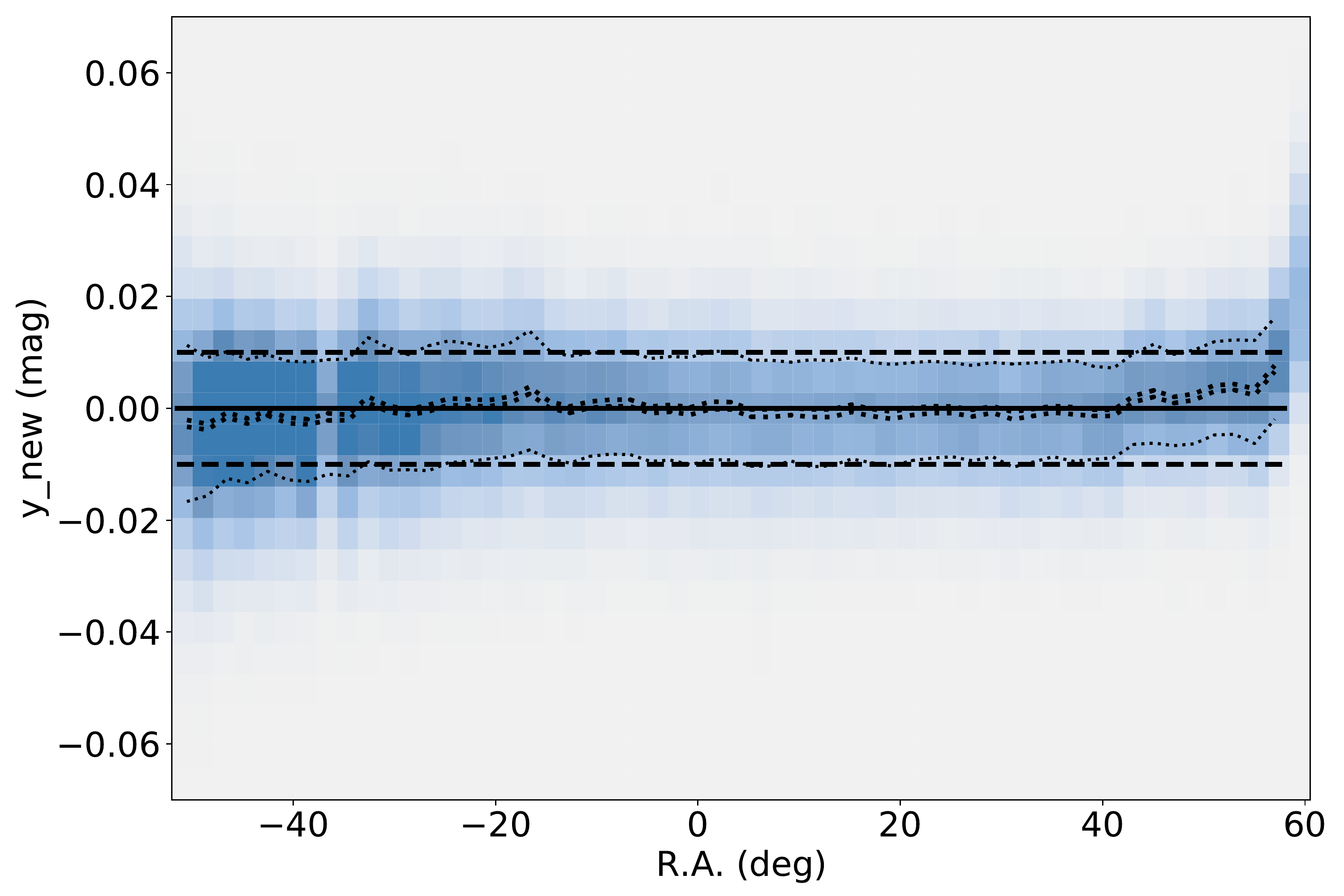} 
    \centering\includegraphics[width=0.49\textwidth]{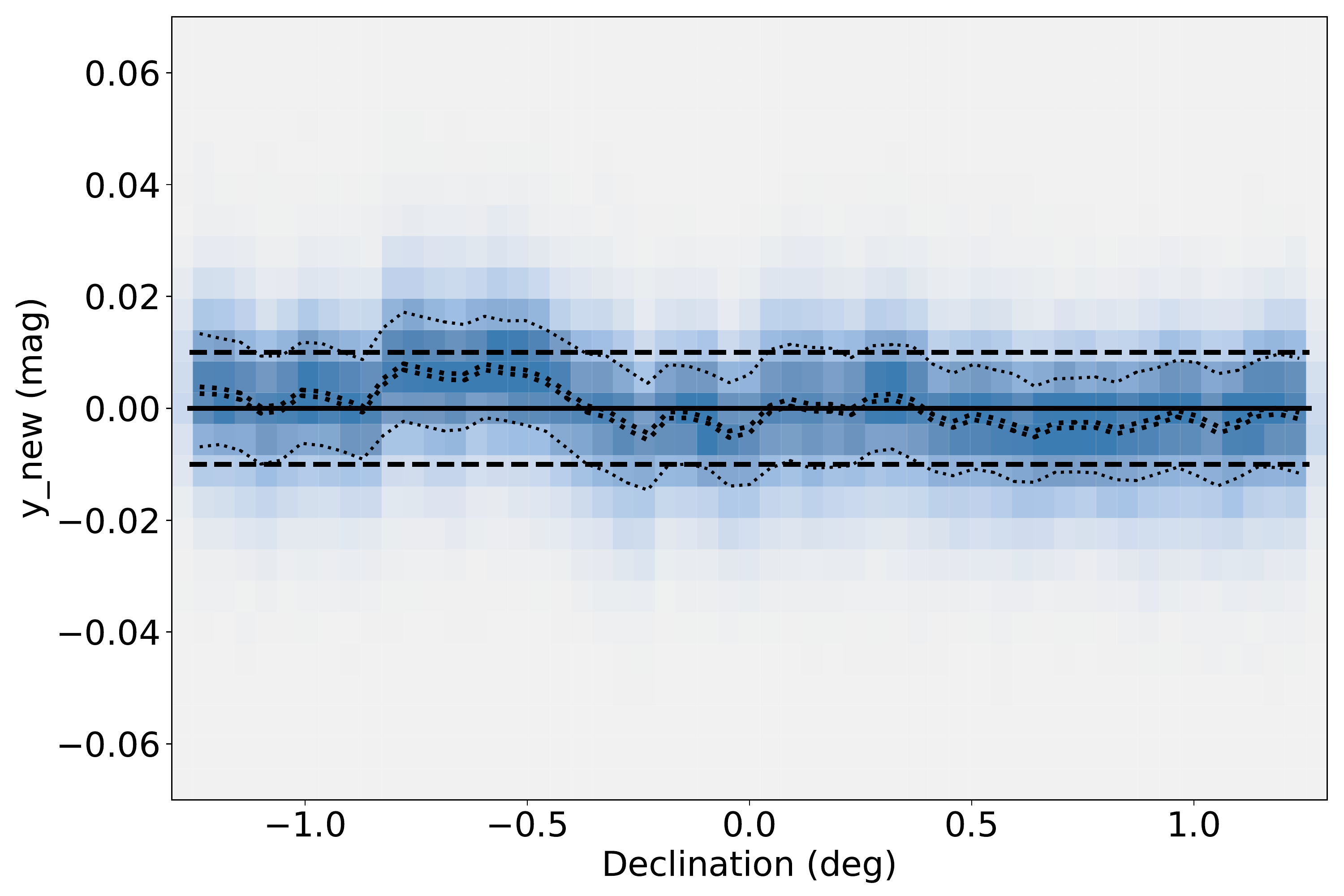}  
\caption{The behaviour of the $s$ colour (top two panels), the second principal colour in the SDSS $g-r$ vs. $u-g$ colour-colour diagram, and the $y$ colour (bottom two panels), the second principal colour in the SDSS $i-z$ vs. $r-i$ colour-colour diagram, for the new v4.2 catalogue. The standard deviation of the median $s$ values binned by R.A. and Declination is 9.8 millimag and 1.3 millimag, respectively, and 1.8 millimag and 3.4 millimag for the $y$ colour.}
\label{fig:comparesy} 
\end{figure*}

Given the quality of the Gaia photometry, there should be no doubt that SDSS $ugriz$ photometry reported in the new v4.2 catalogue is superior to the old v2.6 catalogue. Nevertheless, we perform additional tests, based on the position of the stellar locus in the $g-r$ vs. $u-g$, $r-i$ vs. $g-r$ and $i-z$ vs. $r-i$ colour-colour diagrams  \citep{2004AN....325..583I}. The tests are based on the second principal colour for the blue part of the stellar locus, whose median should not deviate from zero by construction. Figure~\ref{fig:comparew} compares the behaviour of the $w$ colour for the old v2.6 and new v4.2 catalogue and demonstrates that the $gri$ photometry is better calibrated in the latter. The behaviour of the $s$ and $y$ colours for the new catalogue are shown in Figure~\ref{fig:comparesy}. The variation of s colour with R.A. is most likely affected by the variation of interstellar dust properties and stellar metallicity distribution. {\it Based on these tests, we find that the contribution of the zeropoint errors is $<5$ millimag to $gri$ photometry, and $<10$ millimag for the $u$ and $z$ bands.}

\subsection{Comparison of the new v4.2 SDSS catalogue with DES and Pan-STARRS catalogues \label{sec:DESPS1}} 
  
The quality of photometric zeropoint calibration for the new SDSS catalogue can be conveniently tested with the DES (see Section~\ref{ssec:des}) and Pan-STARRS (see Section~\ref{ssec:ps1}) catalogues. Both catalogues list $griz$ photometry of sufficient precision for essentially all stars from Stripe 82. Their photometric calibration procedures are expected to result in different spatial patterns and thus a cross-comparison with the v4.2 catalogue may reveal any residual problems with the zeropoint calibration. They are also deeper than the Gaia EDR3 catalogue and can thus provide further clues about the $\sim$0.01 mag residual Gmag gradient (bright-to-faint end bias) illustrated in Figure~\ref{fig:gaiaJump}. 

\begin{figure*}
    \centering\includegraphics[width=0.49\textwidth]{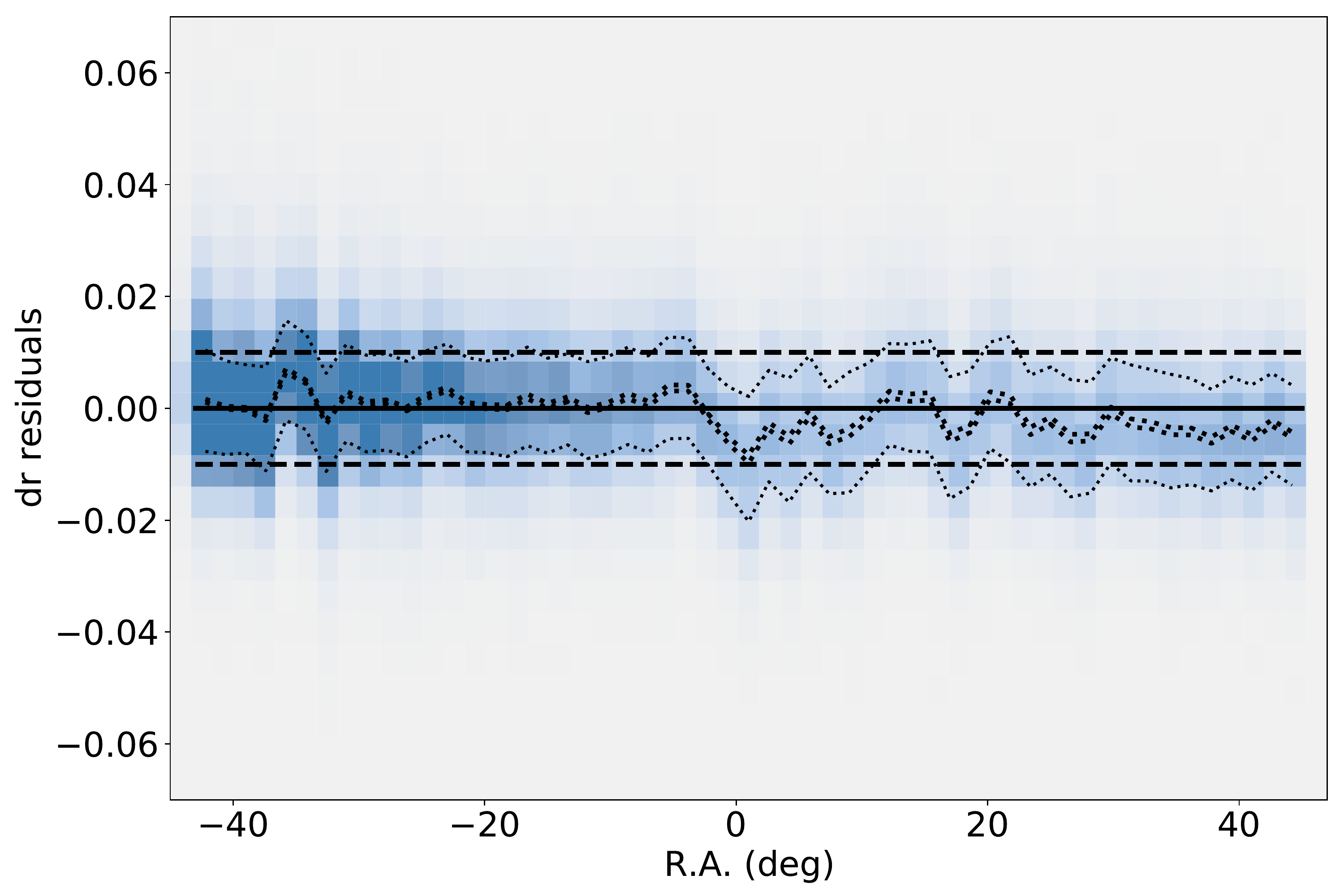}
    \centering\includegraphics[width=0.49\textwidth]{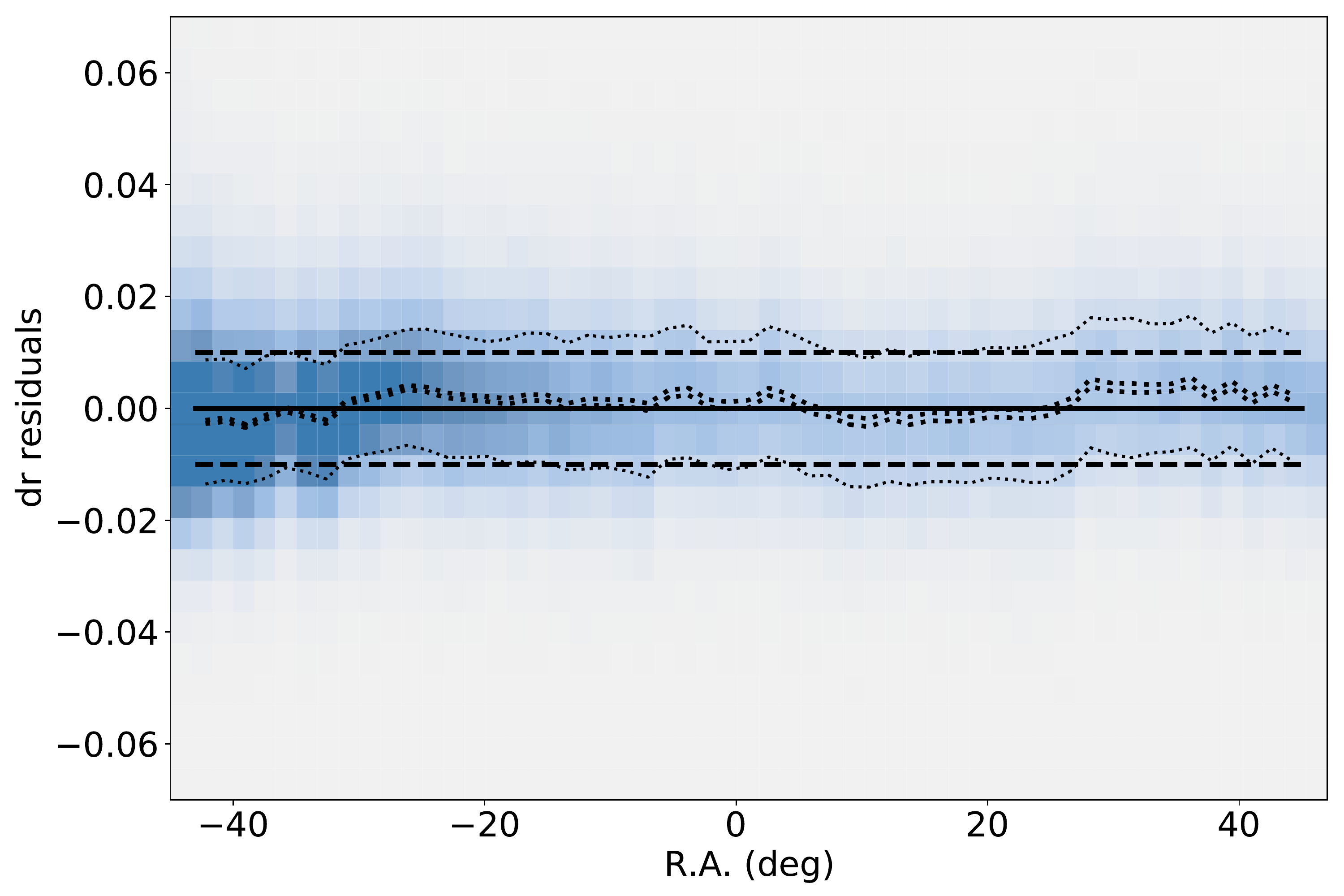}
    \centering\includegraphics[width=0.49\textwidth]{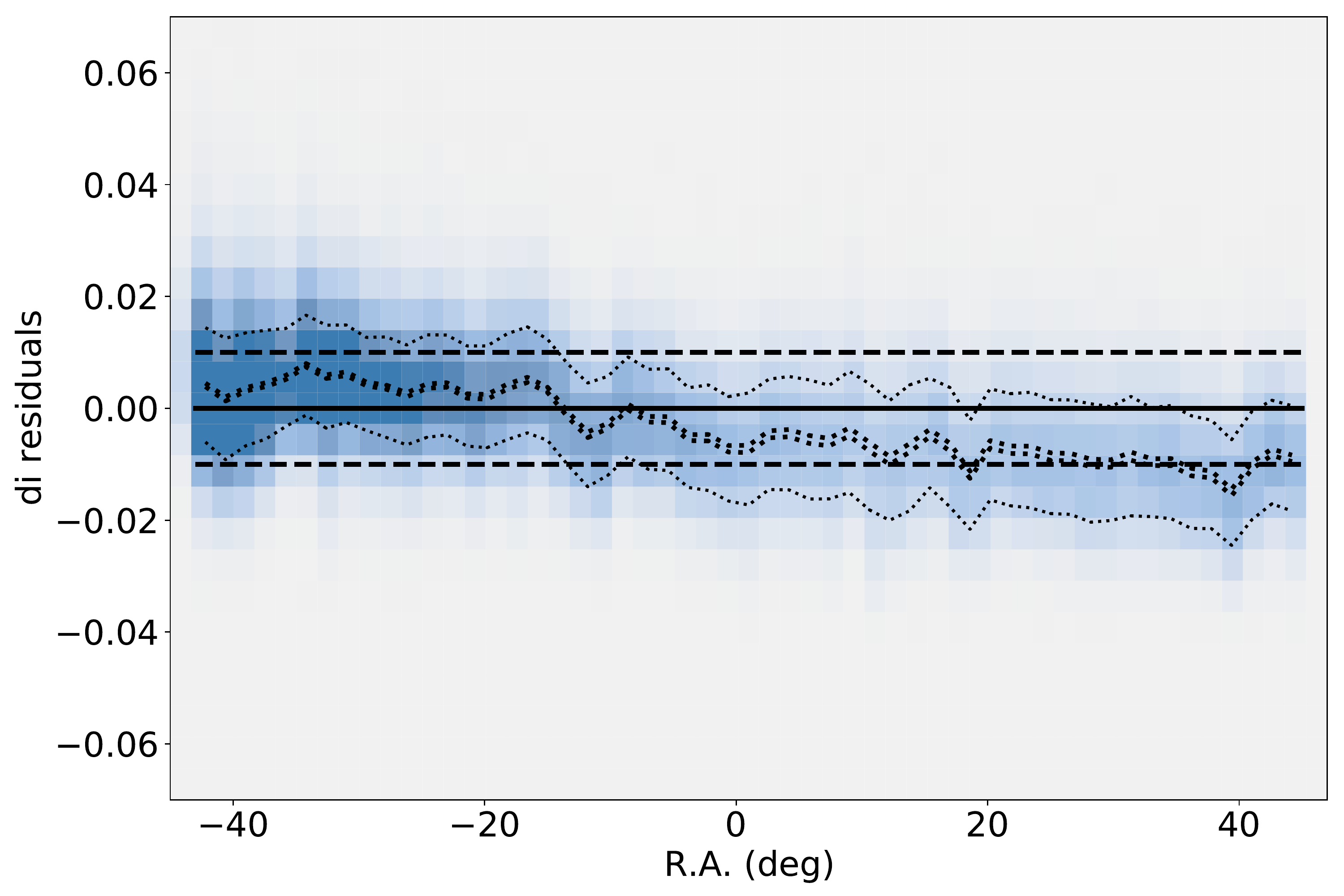}
    \centering\includegraphics[width=0.49\textwidth]{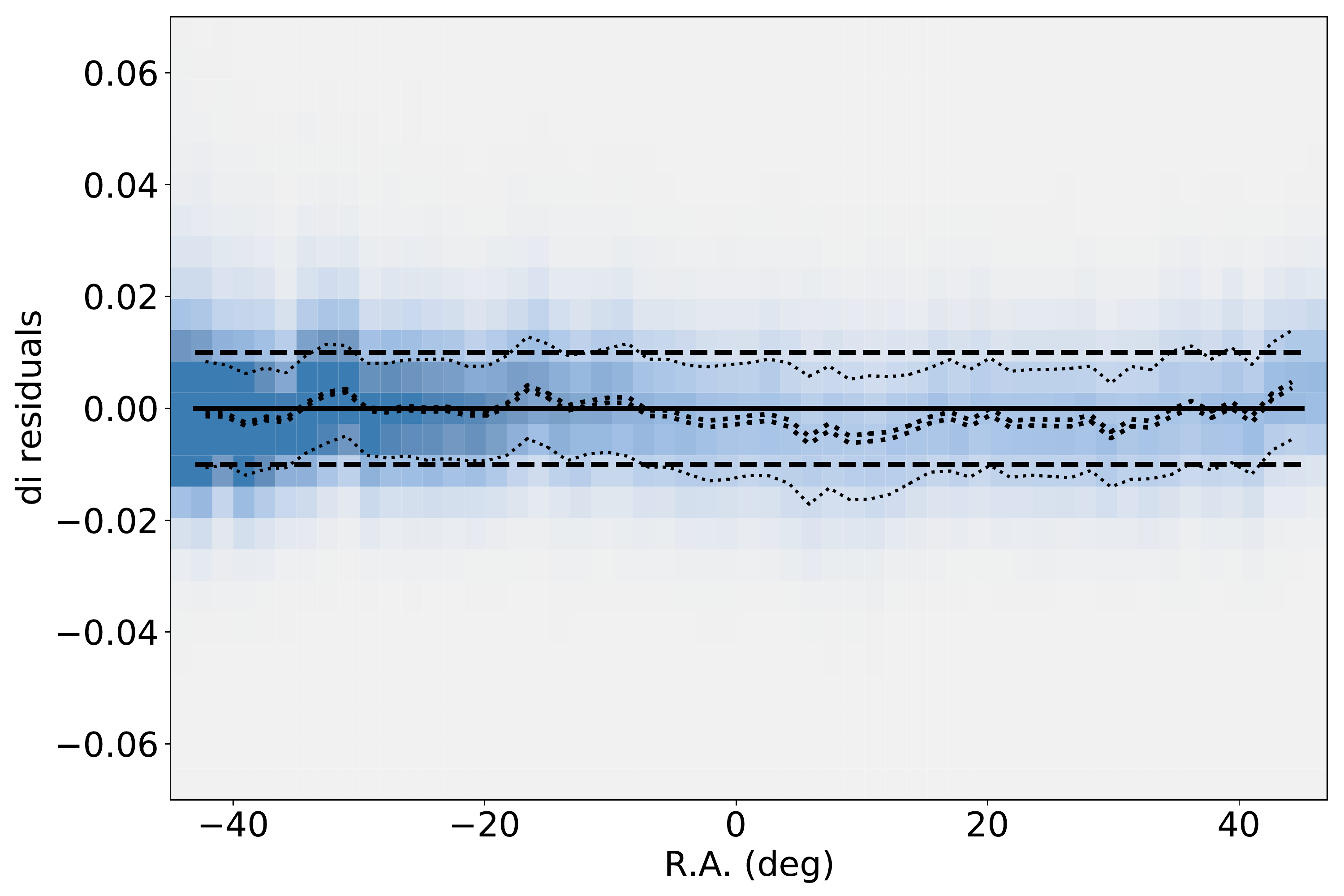}
    \centering\includegraphics[width=0.49\textwidth]{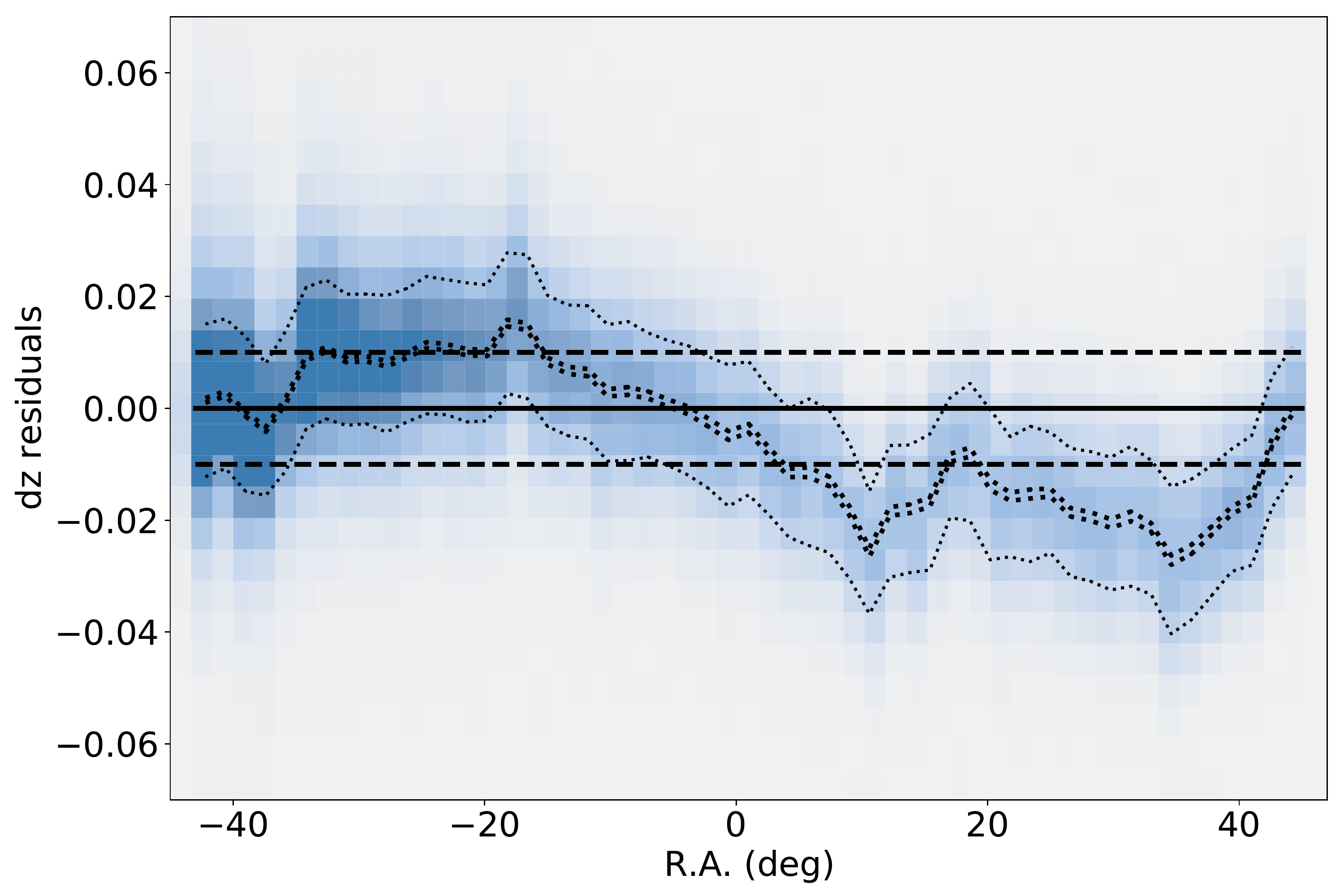}
    \centering\includegraphics[width=0.49\textwidth]{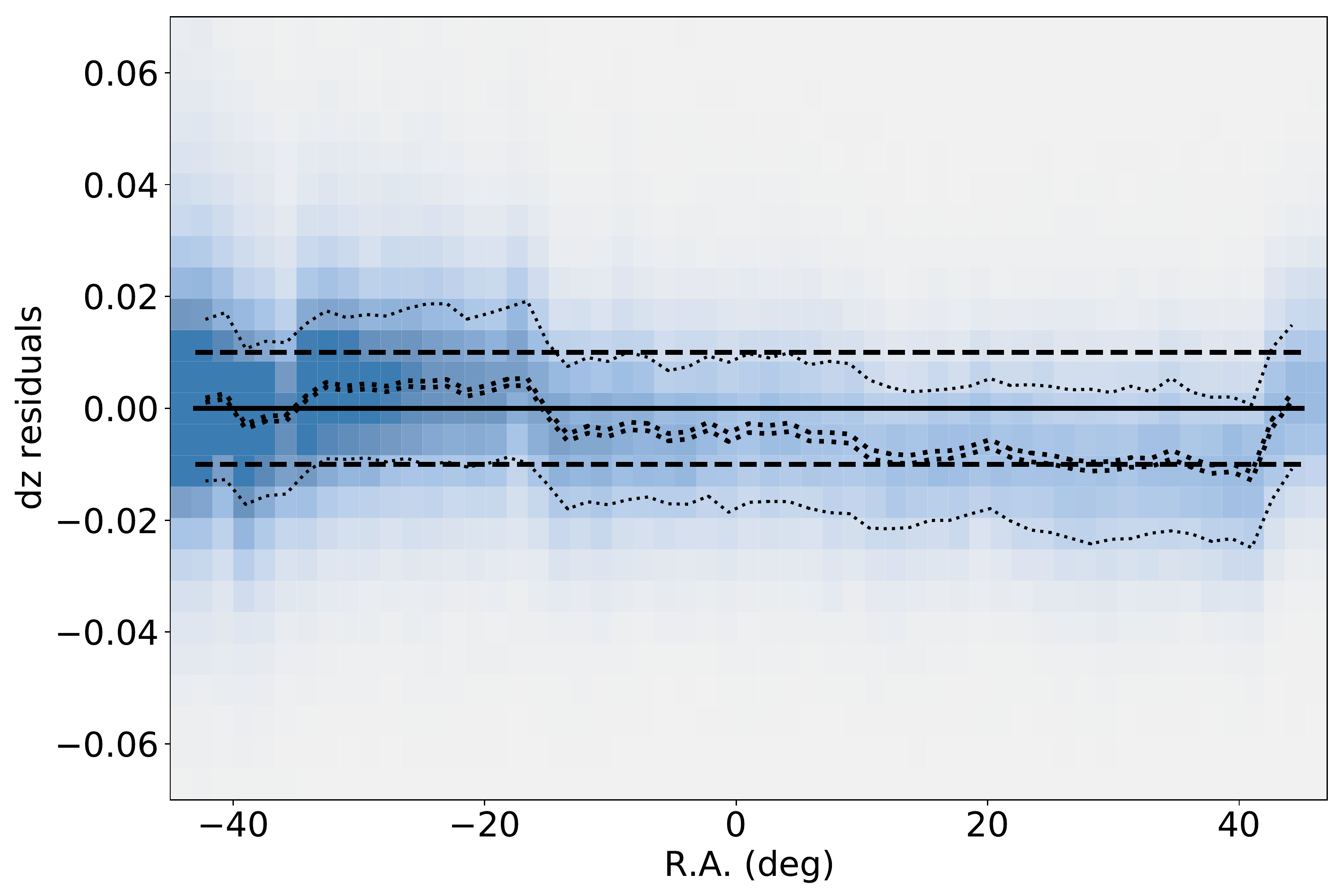}
\caption{A comparison of the magnitude differences between the SDSS v4.2 catalogue and DES (left) and Pan-STARRS (right) catalogues, for the $riz$ bands binned by R.A.}
\label{fig:DESPSRA}
\end{figure*}

\begin{figure*}
    \centering\includegraphics[width=0.49\textwidth]{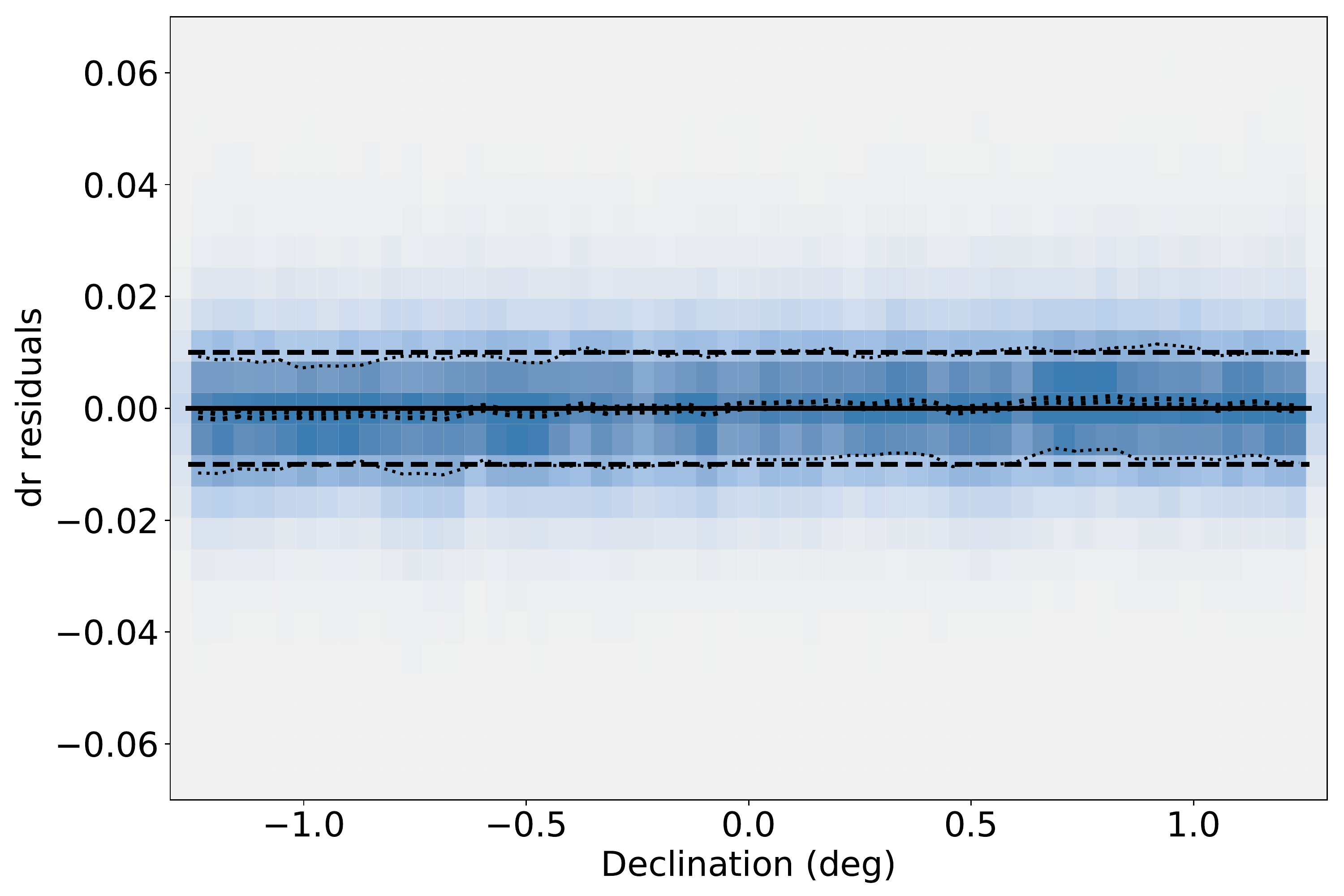}
    \centering\includegraphics[width=0.49\textwidth]{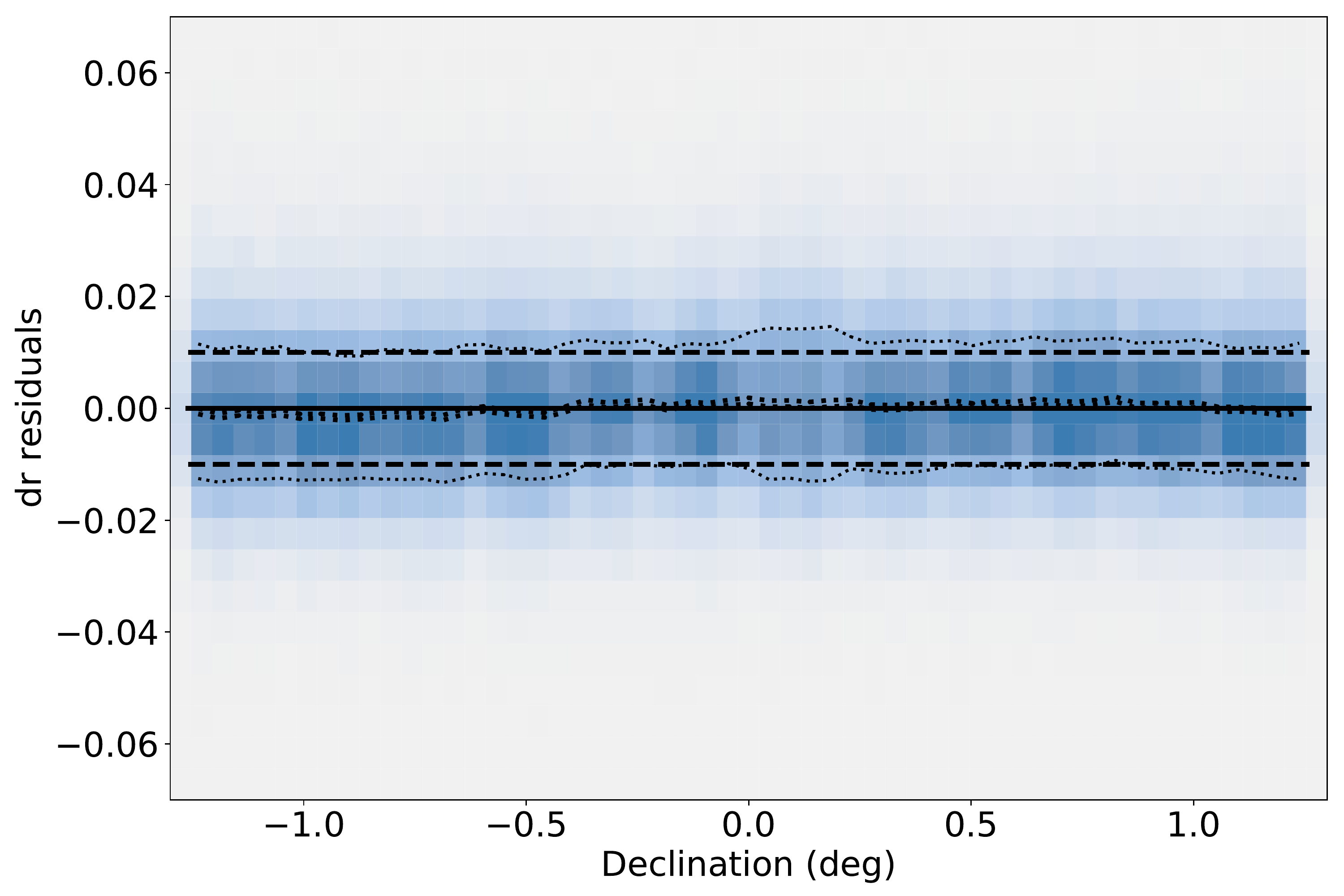}
    \centering\includegraphics[width=0.49\textwidth]{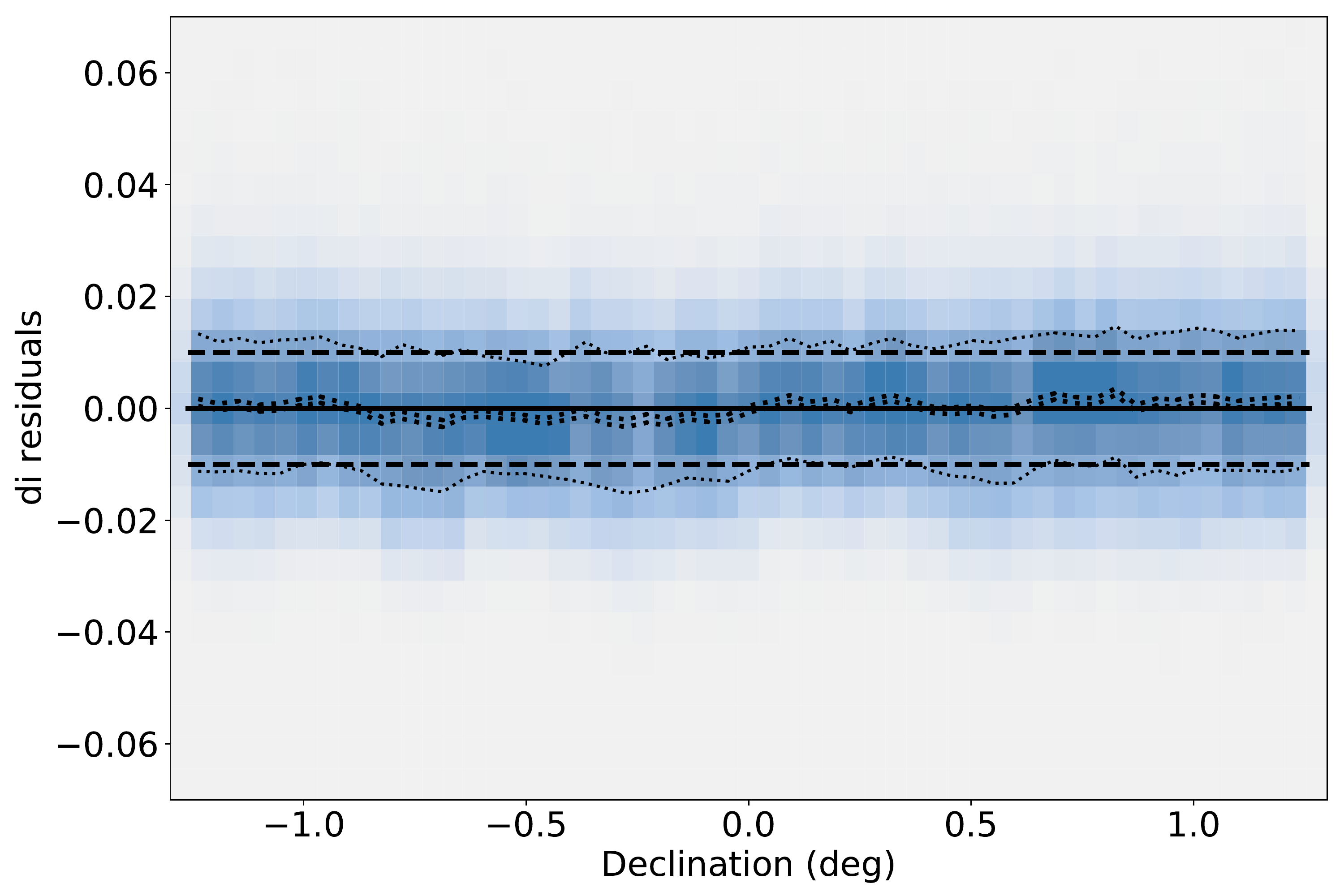}
    \centering\includegraphics[width=0.49\textwidth]{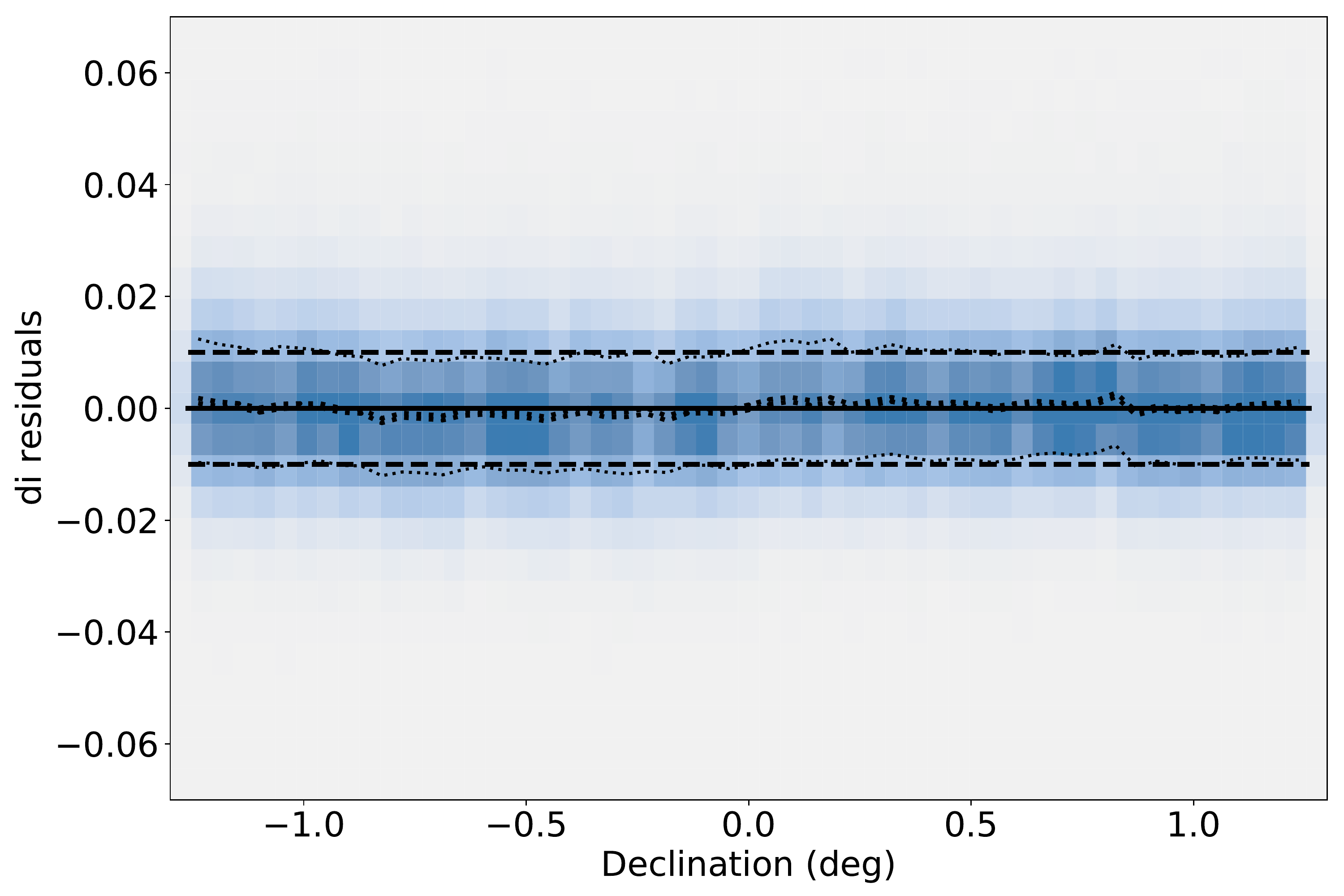}
    \centering\includegraphics[width=0.49\textwidth]{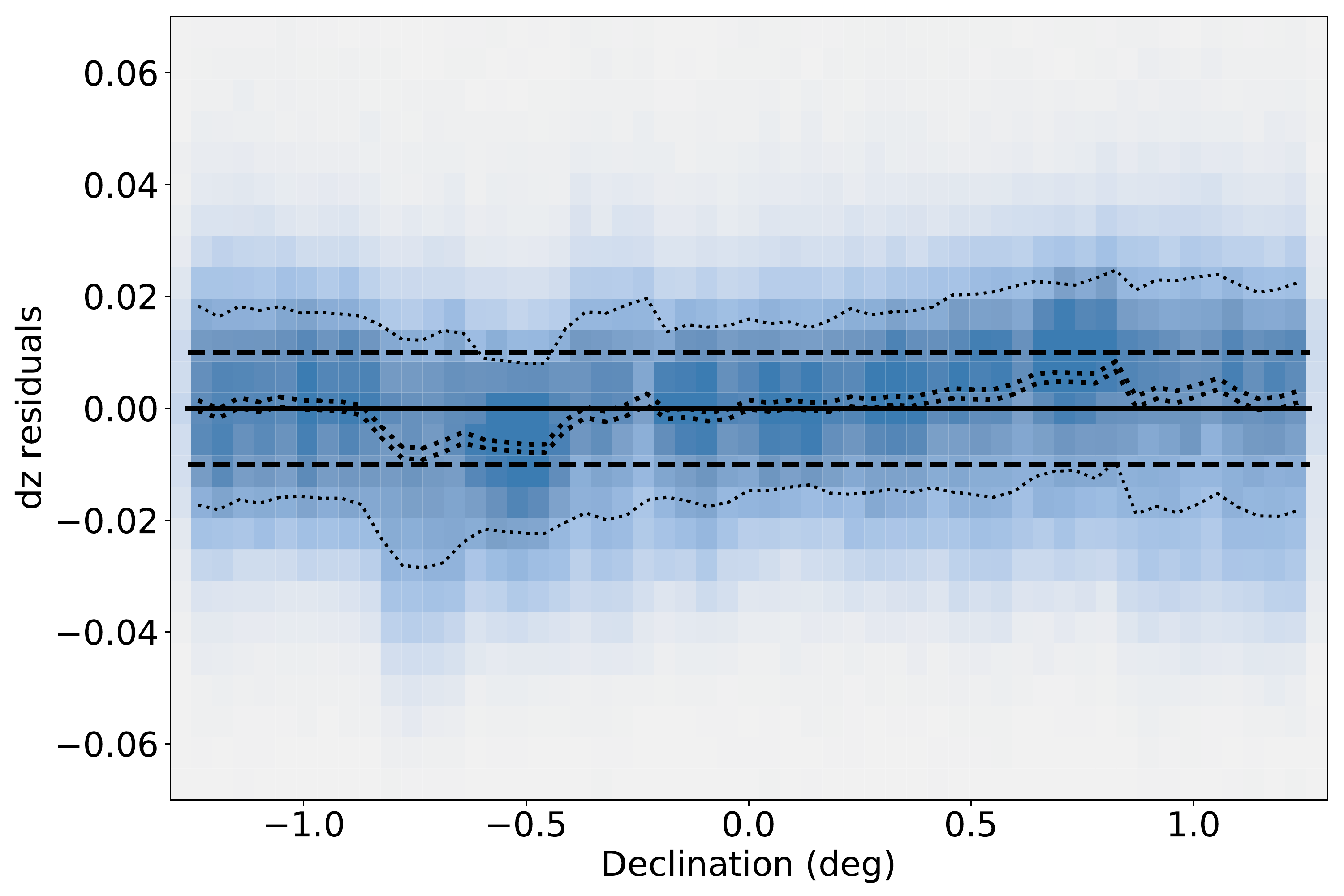}
    \centering\includegraphics[width=0.49\textwidth]{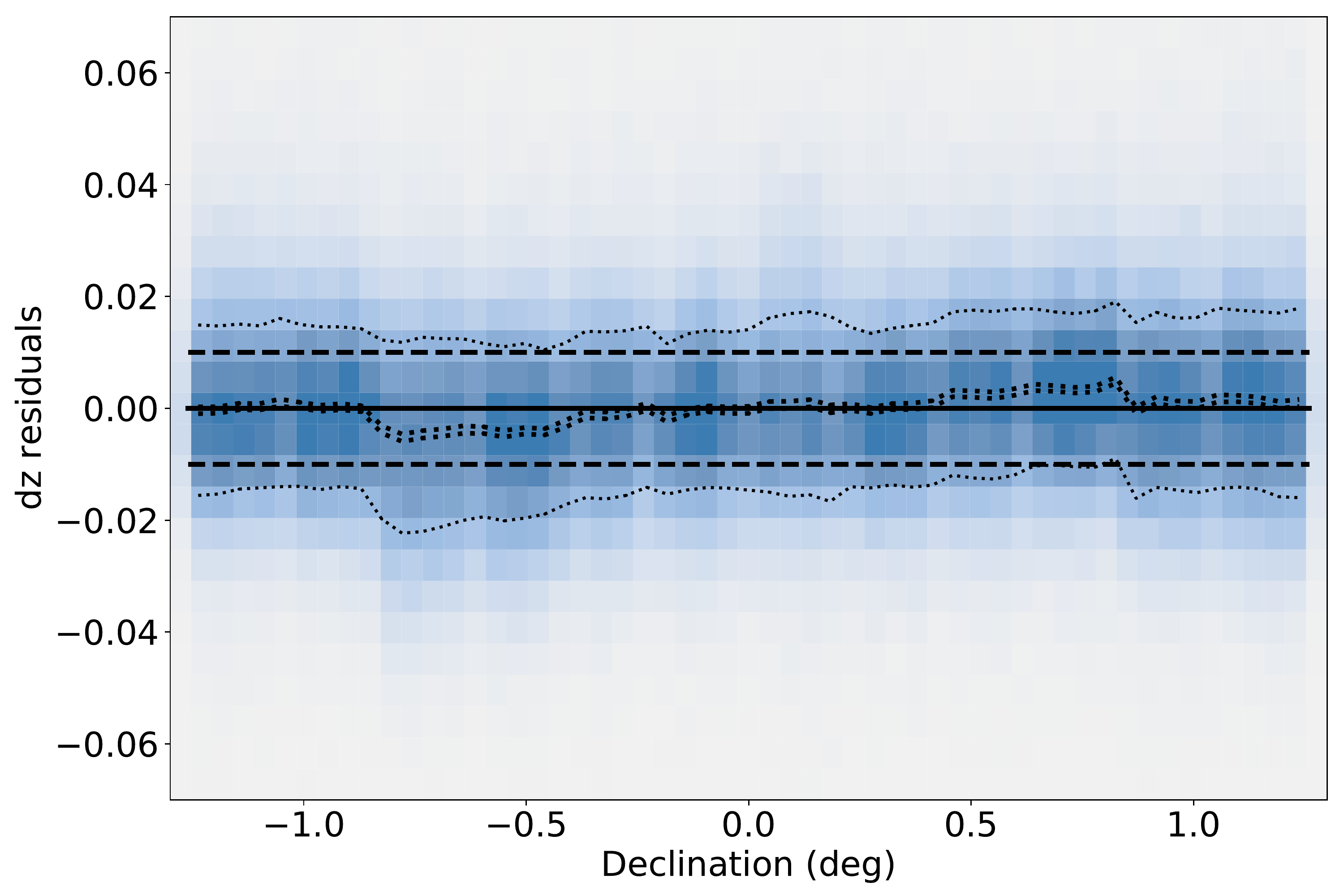}
\caption{Analogous to Figure~\ref{fig:DESPSRA}, except that the magnitude differences are binned by Declination.}
\label{fig:DESPSDec}
\end{figure*}

Our comparison of the magnitude differences binned in the R.A. and Declination directions are illustrated in Figures~\ref{fig:DESPSRA} and \ref{fig:DESPSDec} respectively, and the corresponding robust standard deviations for binned median magnitude differences are listed in Table~\ref{tab:DESPS1}. This multi-survey comparison indicates that the spatial variation of photometric zero points in the updated SDSS catalogue is well below 0.01 mag (RMS), with typical values of 3-7 millimag in the R.A. direction and 1-2 millimag in the Declination direction (excluding the z-band where these values are higher). Note the implied DES $z$ band zeropoint errors as a function of R.A. of up to 0.02 mag (see the bottom left panel in Figure~\ref{fig:DESPSRA}), although a similar trend in the corresponding Pan-STARRS plot (bottom right panel in Figure~\ref{fig:DESPSRA}) indicates that the SDSS catalogue, calibrated using Gaia EDR3, may be at least partially responsible for the observed differences\footnote{For a comparison of DES DR1 and Gaia DR2 calibrations, see e.g., Fig. 9 of \citet{2018ApJS..239...18A}.}

\begin{figure*}
    \centering\includegraphics[width=0.49\textwidth]{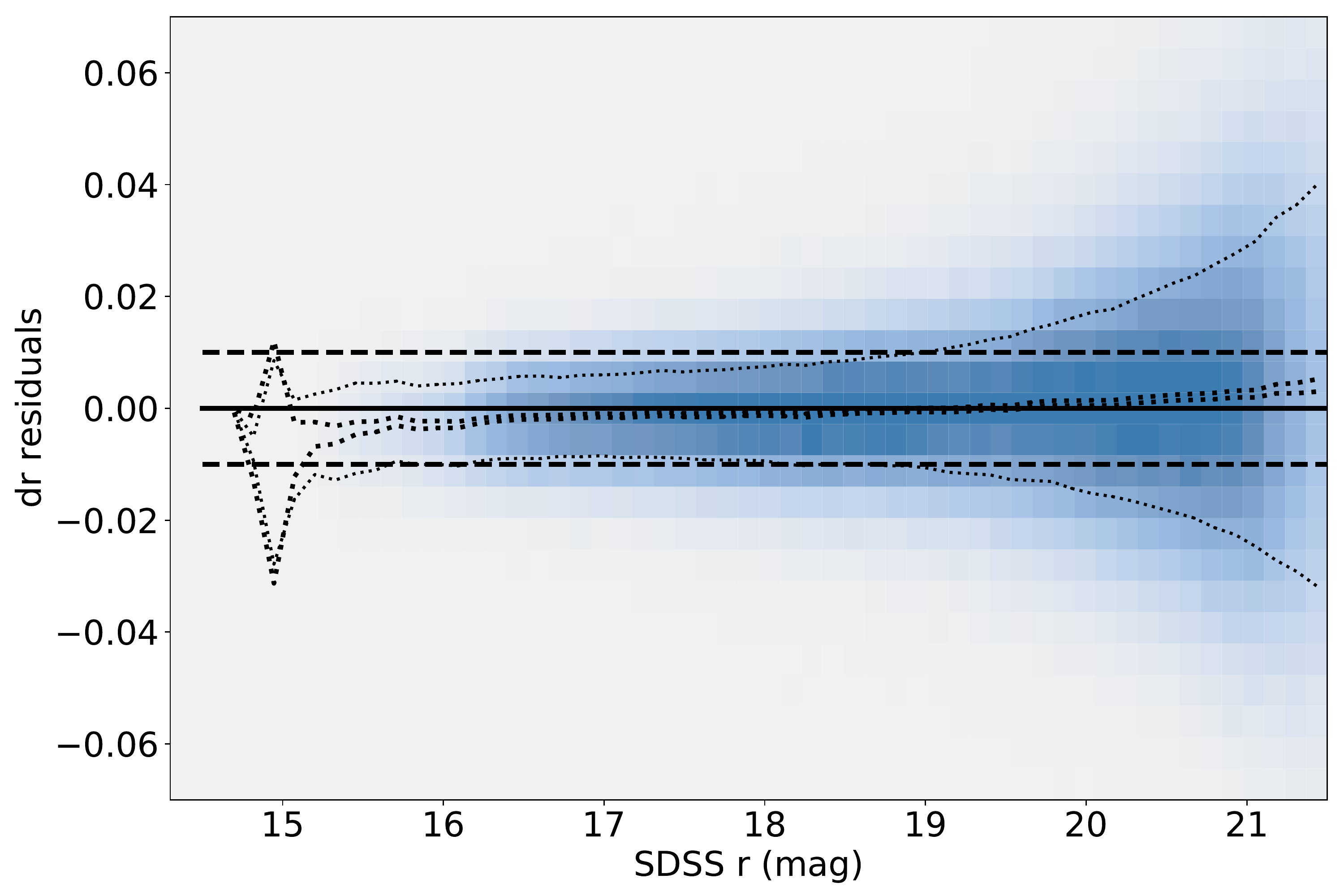}
    \centering\includegraphics[width=0.49\textwidth]{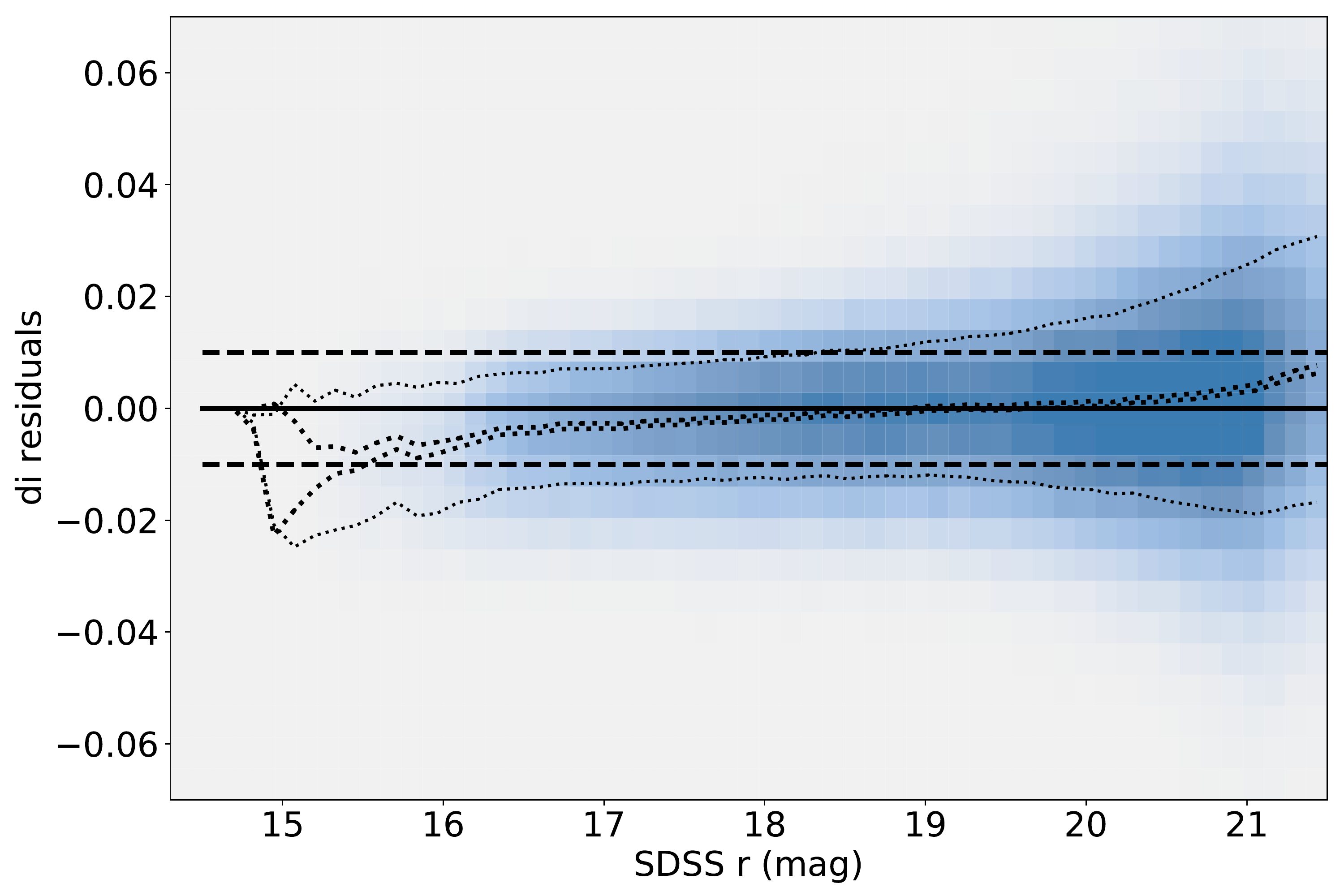} 
\caption{A comparison of the magnitude differences between the SDSS v4.2 catalogue and the DES catalogue, for the $r$ and $i$ bands. Note the very small gradient of the median residuals between the bright and faint ends: about 3-4 millimag and thus much smaller than $\sim$20 millimag when compared to the Gaia G magnitude (see Figure~\ref{fig:gaiaJump}).} 
\label{fig:drVSr}
\end{figure*}

The variation of the residual magnitude differences with magnitude (see Figure~\ref{fig:drVSr} for DES results) is typically flat to within $\sim$3-4 millimag for both DES and Pan-STARRS. This much smaller gradient than the one observed  for Gaia ($\sim$20 millimag) implies a likely problem (a bias between bright and faint ends) with the Gaia EDR3 photometry. The largest discrepancy of about 2 millimag/mag is observed for the Pan-STARRS $r$ band, while the gradient is limited to $<$1 millimag when compared to the DES $r$ band. 
   
\begin{table}
	\centering
	\caption{The robust standard deviation for the binned median magnitude differences between the new v4.2 SDSS catalogue, and DES and Pan-STARRS1 catalogues (in millimag).}
	\label{tab:DESPS1}
	\begin{tabular}{l|c|c|c|c} %
		\hline
		Band & DES R.A. & DES Dec. & PS1 R.A. & PS1 Dec. \\
		\hline
       $g$        &        4.8    &      2.0   &        3.3    &      1.6        \\
       $r$         &        3.3    &      0.9   &        2.1    &      0.9         \\  
       $i$         &        5.9    &      1.4   &        2.2    &      1.0         \\ 
       $z$        &       12.1    &     3.6   &        5.1    &      2.3         \\ 
		\hline
	\end{tabular}
\end{table}

\subsection{Comparison of the new v4.2 SDSS catalogue and $u$ band data from the CFIS catalogue  \label{sec:CFIStest}} 

The comparison of the new SDSS catalogue with the DES and Pan-STARRS catalogues in the previous section did not include the $u$ band. To assess the quality of $u$ band zeropoint calibration, we use the CFIS catalogue (see Section~\ref{ssec:cfis}). The CFIS $u$ band photometry was calibrated using a combination of the SDSS, Pan-STARRS and GALEX UV data. Even given that we recalibrated the new SDSS catalogue using Gaia data, it should not matter for this comparison that SDSS data were used in the calibration of the CFIS catalogue. Nevertheless, this should be kept in mind while interpreting the results presented in this section.  We also note that the transmission curve of the MegaCam $u$ band filter used for the CFIS survey differs from that of the SDSS $u$ band filter\footnote{https://www.cadc-ccda.hia-iha.nrc-cnrc.gc.ca/en/megapipe/docs/filt.html}. However, this does not lead to any noticeable differences in the magnitude comparisons between the two catalogues presented here. 

\begin{figure}
    \centering\includegraphics[width=0.99\columnwidth]{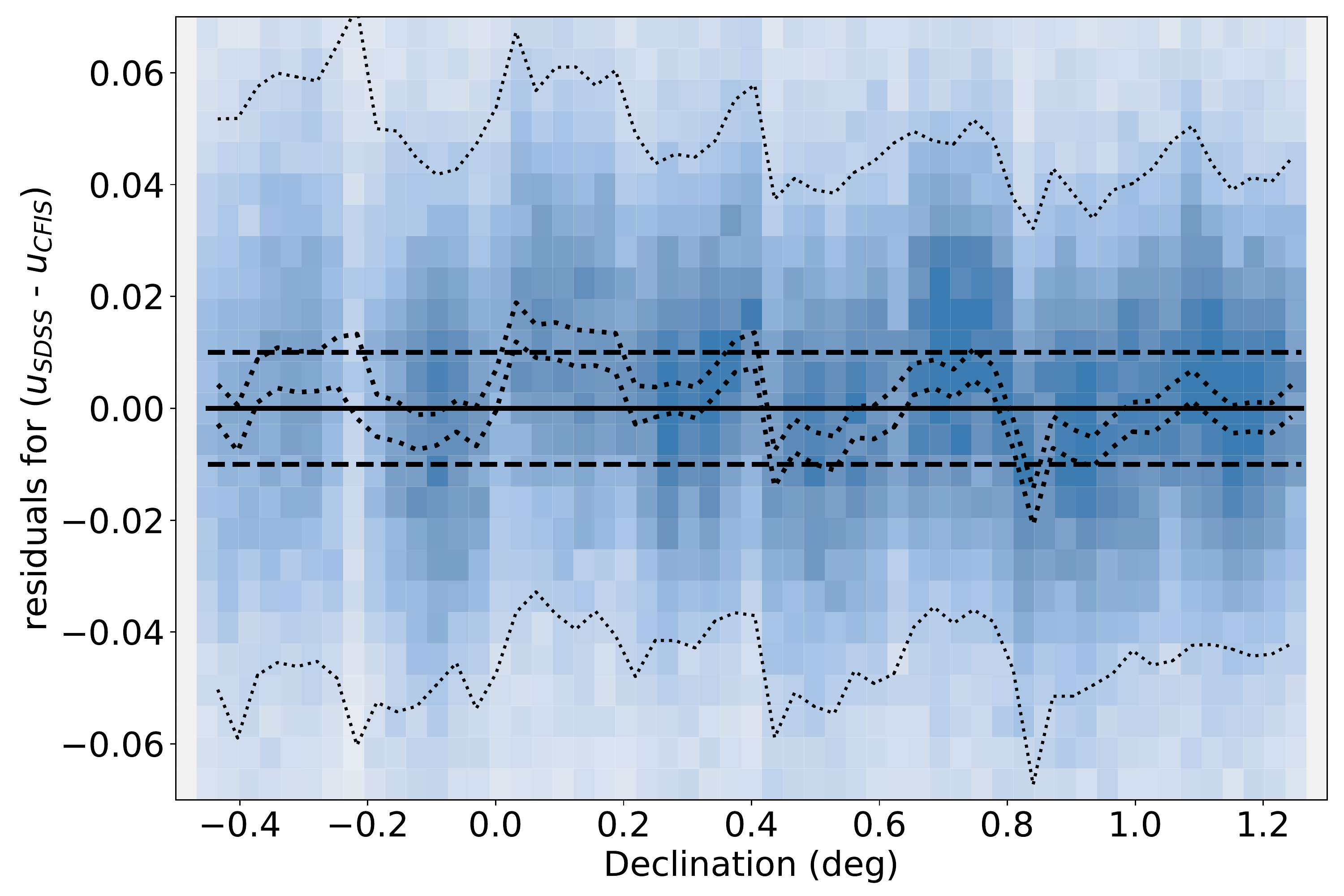} 
\caption{Analogous to Figure~\ref{fig:graycorrRA} ({\it Right}), except that here residuals between the SDSS $u$ band magnitudes and $u$ band magnitudes from the CFIS catalogue (corrected for small colour terms, $\sim0.05$ mag, as a function of the $u-g$ colour), for $\sim$150,000 matched stars with $1.0 <u-g < 2.1$ and $r<20$ are shown. The binned median scatter is 5.7 millimag. Note that the CFIS data are available only for Declination $>$ $+$0.45 degree.}
\label{fig:CFIS}
\end{figure}

A comparison for about 150,000 sufficiently bright stars, $r<20$, with colours matching the main sequence, $1.0 <u-g < 2.1$,  is illustrated in Figure~\ref{fig:CFIS}. The binned median scatter for the Declination direction is 5.7 millimag with systematic differences of up to 0.01 mag. The constraints in the R.A. direction are noisier, with residuals appearing about twice as large as in the Declination direction. These residuals compare favourably to the results of the analysis by \cite{2017ApJ...848..128I}, who showed that some SDSS runs in Data Release 13 have $u$-band zeropoint errors as large
as 0.1 mag. 

\subsection{Comparison of the new v4.2 SDSS catalogue and transformed NUV data from the GALEX catalogue  \label{sec:Galextest}} 

We also use the NUV magnitudes from GALEX (see Section~\ref{ssec:galex}) to provide an independent check on the SDSS u-band magnitudes, following the zeropoint corrections with Gaia photometry described in \S \ref{sec:GaiaCorr}. For this we derive the following GALEX to SDSS u transformation equation:

\begin{equation}
\begin{split}
u_{est} = & NUV - 0.990\times(NUV - g) + 0.014\times(NUV  - g)^2  \\
         & -0.407\times(g - i) + 0.797\times(g - i)^2 + 0.835  \\
         & + T \times [ -1.346\times(E(B-V)-0.15 \\
         & + 3.400\times(E(B-V)-0.15)^{2} ] \\
\end{split}
\end{equation}

where $NUV$ refers to GALEX, and $g$ and $i$ are SDSS magnitudes. $T$ is a step function in the value of the interstellar reddening, E(B-V), from the \citet{1998ApJ...500..525S} reddening map:
\begin{align}
  T = \left\{ \begin{array}{cc} 
    0 & \hspace{5mm} E(B-V) \leq 0.15 \\
    1 & \hspace{5mm} 0.15 < E(B-V) < 0.3 \\
  \end{array} \right.
\end{align}
This transformation relation, which is suitable for stars with $0.5 < (u-g)_{sdss} < 2.0$, $0.2 < (g-r)_{sdss} < 0.8$, $0.3 < (g-i)_{sdss} < 1.0$, and $E(B-V) < 0.3$, has a per star RMS of $\sigma=0.061$ mag (the relatively high RMS is due in part to the relatively high RMS in the GALEX $NUV$ magnitudes).  This relation converts the GALEX $NUV$ magnitudes into very reasonable approximations of the SDSS $u$-band. To refine these approximations, we then applied the same sort of numerical transformation that was used for the SDSS to Gaia transformation in Figure~\ref{fig:GrVSgi}; this improved the overall per star RMS to $\sigma=0.055$, with most of the improvement occurring for the bluest, $(u-g)_{sdss} \la 1.0$, and the reddest, $(u-g)_{sdss} \ga 1.0$ stars in our sample.

\begin{figure}
    \centering\includegraphics[width=0.99\columnwidth]{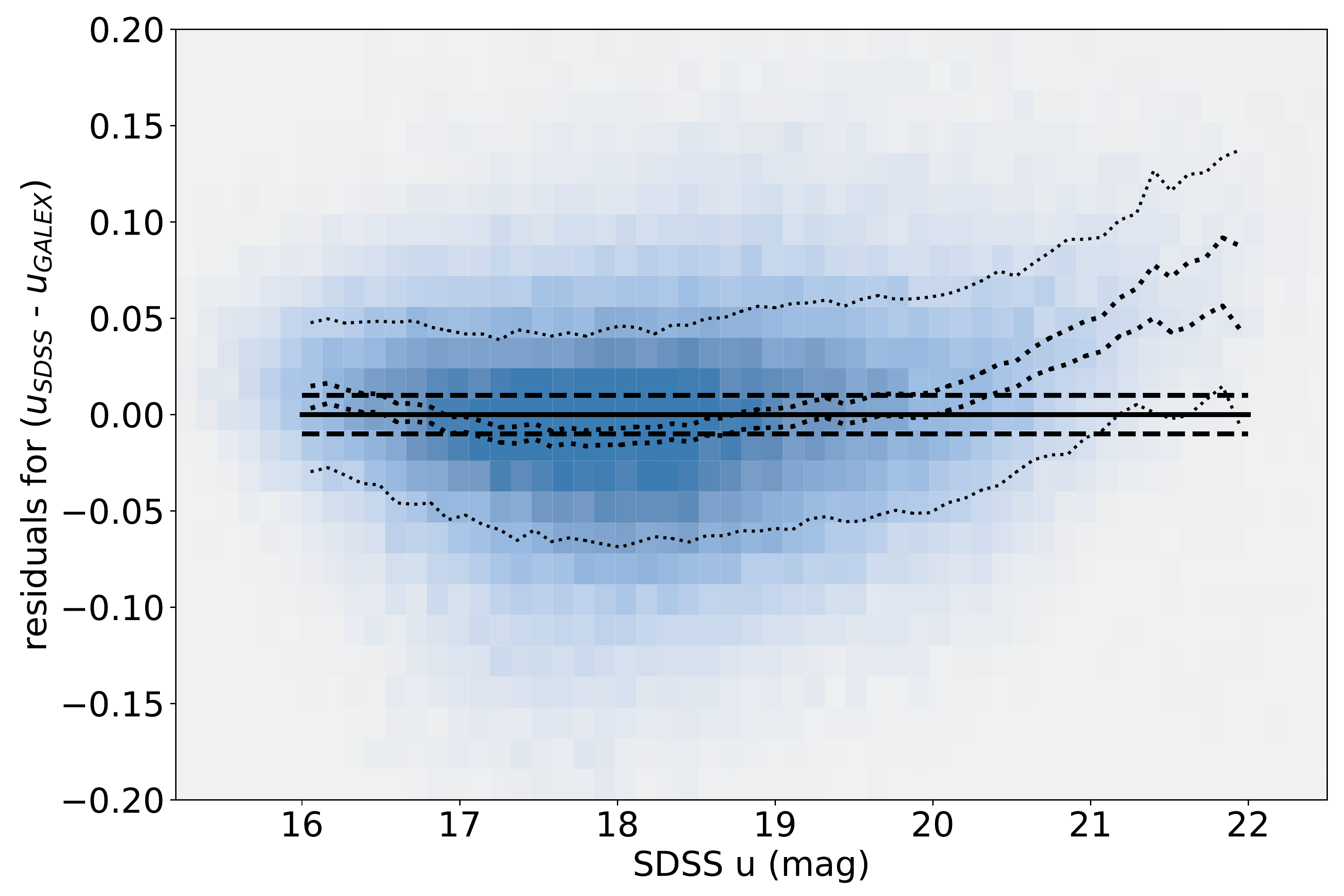}
\caption{Analogous to Figure~\ref{fig:gaiaJump}, except that here the residuals between the SDSS $u$ band magnitudes and predicted $u$ band magnitudes transformed from the GALEX catalogue are shown. The individual star scatter is 54.5 millimag (dot-dashed lines); the binned median scatter is 24.5 millimag solid grey lines).}
\label{fig:GALEX_umag}
\end{figure}

\begin{figure*}
    \centering\includegraphics[width=0.99\columnwidth]{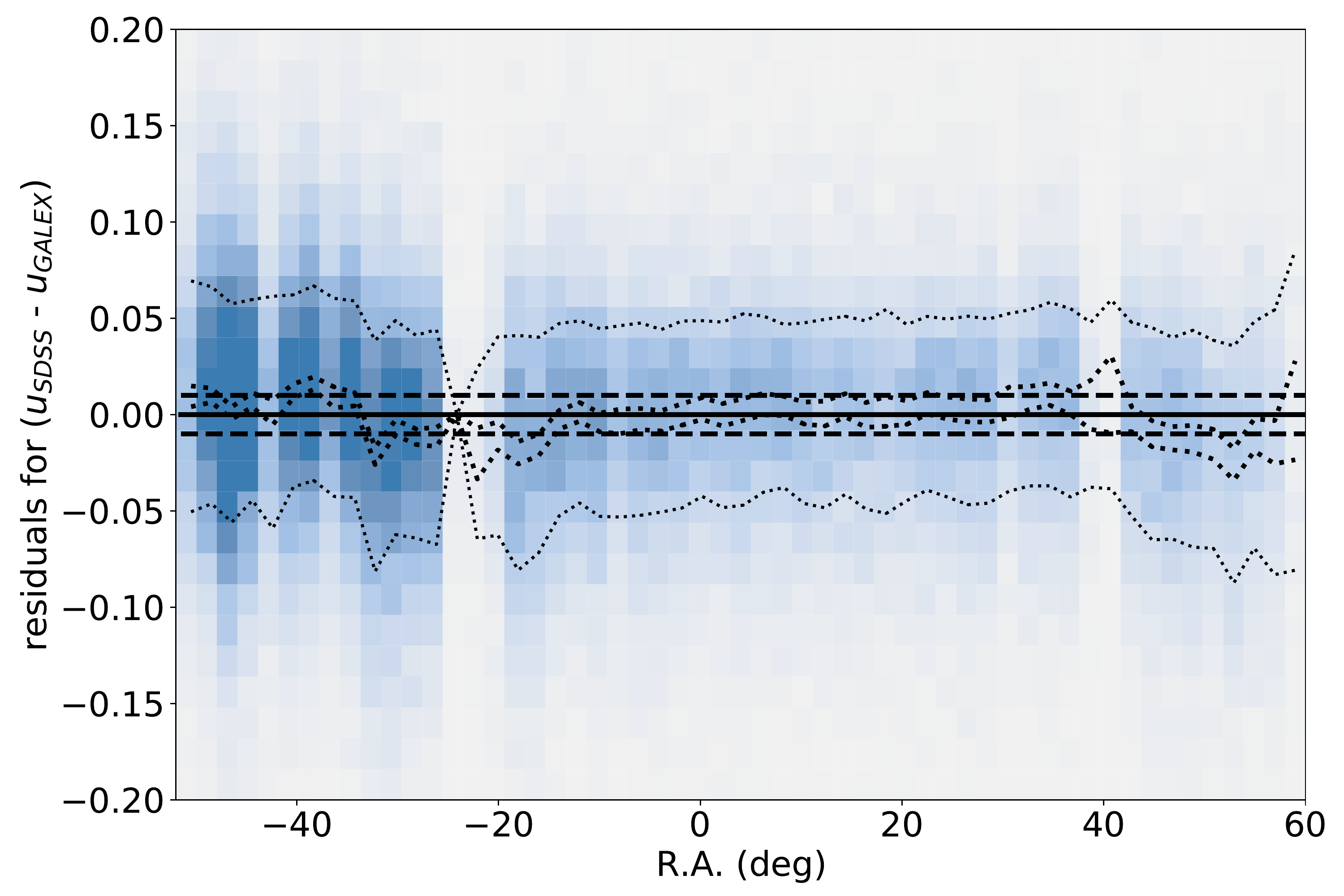}
    \centering\includegraphics[width=0.99\columnwidth]{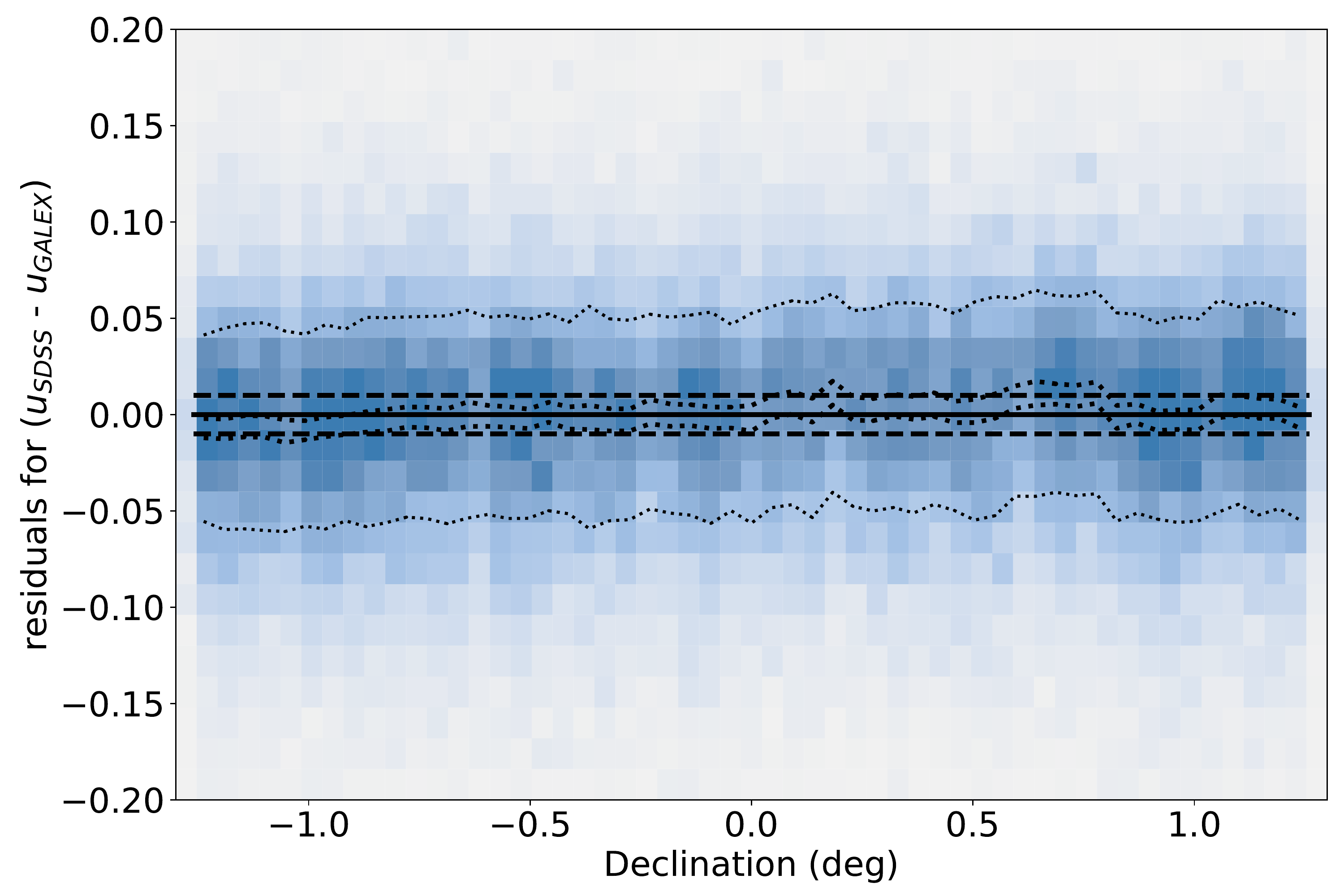}
\caption{({\it Left}) Analogous to Figure~\ref{fig:graycorrRA} ({\it Left}), except that here residuals between the SDSS $u$ band magnitudes and predicted $u$ band magnitudes transformed from the GALEX catalogue are shown.  Unlike in Figure~\ref{fig:GALEX_umag}, we exclude matches for which SDSS $u$$>$20.0 The individual star scatter is 52.9 millimag (dot- dashed lines); the binned median scatter is 9.5 millimag (solid grey lines). ({\it Right}) Analogous to Figure~\ref{fig:graycorrRA} ({\it Right}), except that here the residuals between the SDSS $u$ band magnitudes and predicted $u$ band magnitudes transformed from the GALEX catalogue are shown.  Unlike in Figure~\ref{fig:GALEX_umag}, we exclude matches for which SDSS $u$$>$20.0 The individual star scatter is 52.9 millimag (dot-dashed lines); the binned median scatter is 5.0 millimag (solid grey lines).}
\label{fig:GALEX_RA}
\end{figure*}

Despite the relatively noisy results, we can make several conclusions based upon the binned data consisting of 89,722 matches between the v4.2 SDSS catalogue and the GALEX catalogue. First, in Figure~\ref{fig:GALEX_umag}, we show the difference between the SDSS-measured and GALEX-predicted $u_{sdss}$ vs. SDSS $u_{sdss}$ for the 80,053 matches that lie within the colour and $E(B-V)$ cuts used for the initial transformation equation (Eq.~1).  Although the relation, even when binned like here, is noisy, there is clearly a noticeable trend in the residuals vs. SDSS $u$.  As with the comparison with Gaia (Fig~\ref{fig:gaiaJump}), we expect that this magnitude dependence is associated with the GALEX photometry rather than with SDSS.  Second, in Figure~\ref{fig:GALEX_RA} ({\it Left}), we plot this magnitude difference vs. RA along SDSS Stripe 82.  Here, in addition to the cuts used for Figure~\ref{fig:GALEX_umag}, we also apply a cut of $u_{sdss}\le20.0$, to avoid the worst deviations seen in that figure; this results in a sample of 69,783 matches.  We do note some small but significant large-scale trends with RA, at the 9.5 millimag level (RMS).  Third and finally, in Figure~\ref{fig:GALEX_RA} ({\it Right}), we plot this magnitude difference vs. DEC across SDSS Stripe82 using the same sample of 69,783 matches used for Figure~\ref{fig:GALEX_RA} ({\it Left}).  Unlike for the CFIS comparison in the previous section (see, e.g., Fig.~\ref{fig:CFIS}), the GALEX comparison covers the full DEC range of Stripe 82.  Overall, aside of a possible slight gradient, there appear to be no strong large-scale trends. The hints of smaller scale, $\sim0.4^{\circ}$ variations -- roughly on the scale of a SDSS camera column -- seen in the CFIS comparison (Fig.~\ref{fig:CFIS}), however, are not seen here, perhaps due to the relative ``noisiness'' of the GALEX comparison.

\subsection{Offsets from AB magnitude scale \label{sec:AB}} 

We estimate offsets from the AB magnitude scale using synthetic photometry derived from the spectra of three DA white dwarfs observed by HST (see Table~\ref{tab:HST}): GD50 (specifically, the calibrated spectrum file {\tt gd50\_004.fits}) from the HST CalSpec database of spectrophotometric standard stars \citep{2014PASP..126..711B}\footnote{https://www.stsci.edu/hst/instrumentation/reference-data-for-calibration-and-tools/astronomical-catalogs/calspec}, and SDSSJ~232941.32$+$001107.8 and SDSSJ~010322.19-002047.7 from \citet{2019ApJS..241...20N} (the flux-calibrated modeled spectrum files were provided by G. Narayan, private communication).  These three spectrophotometric calibrators are particularly useful for our purposes since they are faint enough not to be saturated in the SDSS photometry ($r>14$) and they lie within the SDSS Stripe 82 footprint, making direct comparison to the SDSS measurements relatively simple.  Synthetic AB magnitudes for these three DA white dwarfs were derived via numerical integration of Equation~8 of \citet{1996AJ....111.1748F}, which serves to define broad-band AB magnitudes.  This equation requires not only a well-calibrated spectrum, but also the well-calibrated bandpass response curves for the photometric system in question.  Here, we make use of the SDSS bandpass response curves from Table~4 of \citet{2010AJ....139.1628D}.  Our results for the synthetic SDSS AB magnitues for these three stars can be found in Table~\ref{tab:HST}.  The errors on these synthetic AB magnitudes are expected to be $\sim$5--10~millimag (statistical) and $\sim$10~millimag (systematic) (e.g., \citealt{2014PASP..126..711B}).

Table~\ref{tab:AB} presents the numerical summary of the comparison between SDSS magnitudes and HST-based synthetic magnitudes for the three white dwarfs. We used unweighted mean because formal uncertainties for SDSS photometry are subdominant to systematic errors in the HST-based synthetic magnitudes, estimated to be $\sim$0.01 mag. Uncertainties of the mean offsets were computed from the observed scatter of the three values. In the $riz$ bands we detect significant ($>3\sigma$) deviations, ranging from 0.012 mag to 0.033 mag, while in the $u$ and $g$ bands we can only place upper limits (at $2\sigma$: 0.038 mag and 0.028 mag, respectively).

\begin{table*}
	\centering
	\caption{Synthetic SDSS magnitudes for three WD dwarfs with HST photometry$^a$.}
	\label{tab:HST}
	\begin{tabular}{r|r|r|l|l|l|l|l} %
		\hline
		Name & R.A. & Dec. & u & g & r & i & z \\
		\hline
     GD50                     &  57.2091083 &  $-$0.975636 &      13.409 &       13.784 &    14.295 &     14.655 &     {\it no data}$^b$ \\
     SDSSJ~232941.32+001107.8 & 352.4221875 &     0.185500 &      18.154 &       18.145 &    18.466 &     18.754 &     19.042 \\  
     SDSSJ~010322.19-002047.7 &  15.8424625 &  $-$0.346592 &      18.627 &       19.057 &    19.558 &     19.923 &     20.258 \\
		\hline
	\end{tabular}
     \vspace{1ex}

     {\raggedright {\bf Notes}: (a) Sourced from \citet{2014PASP..126..711B} and \citet{2019ApJS..241...20N}. \newline (b) The CalSpec spectrum stopped at about 9000 \AA\ and thus did not cover all of SDSS $z$ band.\par}
\end{table*}

\begin{table}
	\centering
	\caption{AB offsets implied by three WD dwarfs with HST synthetic photometry$^a$.}
	\label{tab:AB}
	\begin{tabular}{l|r|r|r|r|r} %
		\hline
		& $\Delta u$ & $\Delta g$ &$\Delta r$ &$\Delta i$ &$\Delta z$ \\
		\hline
  		 {\rm mean}            &    --     &        --     &      0.015    &       0.016    &      0.035     \\ 
  		$\sigma_{\rm mean}$ &         0.019    &        0.014     &      0.004    &       0.004    &      0.011     \\
		\hline
	\end{tabular}
     \vspace{1ex}

     {\raggedright {\bf Notes}: (a) Offsets are defined as additive corrections to SDSS photometry to place it on AB scale (in mag). Listed values are unweighted mean and its uncertainty. \par}
\end{table}

\section{Discussion and Conclusions} \label{sec:disc}

To enable further progress in cosmological and other high-precision photometric measurements, modern multi-band photometric sky surveys aim to deliver measurements internally consistent to at least the 1\% (0.01 mag) level. Over the last decade a number of such large-scale surveys approached, and often surpassed this photometric accuracy requirement. For ground-based surveys, which are affected by variable atmospheric effects and hardware responses to changes in local environment (e.g. temperature), significant improvements are achieved by averaging multiple observations. 

In this paper, we have described the construction, calibration and testing of an updated version of the SDSS Stripe 82 Standard Star catalogue, \pO\  \citep{Ivez07} that lists averaged SDSS photometry for about a million non-variable stars. Additional post-2007 SDSS data include about 2-3 times more measurements per star than in the original catalogue, resulting in 1.4-1.7 times smaller random photometric errors (precision) than in the original catalogue, and about three times smaller than individual SDSS runs.

Thanks to the availability of photometric data from recent wide-field surveys (Gaia, DES, Pan-STARRS and CFIS), we were able to derive robust zeropoint corrections and establish that this new catalogue is superior to the original catalogue. Using a combination of comparisons to other catalogues and astrophysical constraints, we find that the contribution of the zeropoint errors to photometric uncertainties is $<5$ millimag for the $gri$ bands, and $<10$ millimag for the $u$ and $z$ bands. 

Various catalogue cross-comparisons have revealed minor problems with all the analyzed catalogues. For example, we detected DES $z$ band zeropoint errors of up to 0.01-0.02 mag, as a function of R.A., and demonstrated that the Gaia EDR3 G magnitude, \GG\ has an overall gradient of $\sim$0.01mag from the bright to faint end ($ 16 \leq$ ~\GG~ $\leq 19$), and therefore appears too faint by about 0.02 mag at \GG~$\sim$20.
 
We constrained offsets from the absolute AB magnitude scale using three white dwarfs with the HST CalSpec absolute photometry data. In the $riz$ bands we measured significant ($>3\sigma$) deviations, ranging from 0.015 mag to 0.035 mag (see Table~\ref{tab:AB}), while in the $u$ and $g$ bands we only placed upper limits (at $2\sigma$: 0.038 mag and 0.028 mag, respectively). These constraints on absolute AB magnitude scale could be improved by increasing the number of such calibrators.

Thanks to its high stellar density, about 1 star per square arcmin, and demonstrated sub-percent internal photometric precision, this catalogue is a good resource for both calibrating and testing other surveys. In particular, it will enable high-precision photometric testing of data collected during the commissioning phase of the Vera C. Rubin Observatory Legacy Survey of Space and Time. 

\newpage

\section*{Acknowledgements}
The authors sincerely thank the referee, Dr. Carlos Gonzalez-Fernandez for all the insightful comments and corrections, which greatly augmented the validity and usefulness of our results, as well as improved the readability of our manuscript. \v{Z}.~Ivezi\'{c} acknowledges support from the University of Washington College of Arts and Sciences, Department of Astronomy, and the DIRAC Institute. The DIRAC Institute is supported through generous gifts from the Charles and Lisa Simonyi Fund for Arts and Sciences, and the Washington Research Foundation. S.S.~Allam and D.L.~Tucker acknowledge that this manuscript has been authored by Fermi Research Alliance, LLC under Contract No. DE-AC02-07CH11359 with the U.S. Department of Energy, Office of Science, Office of High Energy Physics.\\
{\bf SDSS-IV, DR15} Funding for the Sloan Digital Sky Survey IV has been provided by the Alfred P. Sloan Foundation, the U.S. Department of Energy Office of Science, and the Participating Institutions. SDSS acknowledges support and resources from the Center for High-Performance Computing at the University of Utah. SDSS is managed by the Astrophysical Research Consortium for the Participating Institutions of the SDSS Collaboration including the Brazilian Participation Group, the Carnegie Institution for Science, Carnegie Mellon University, Center for Astrophysics at Harvard \& Smithsonian (CfA), the Chilean Participation Group, the French Participation Group, Instituto de Astrof{\'i}sica de Canarias, The Johns Hopkins University, Kavli Institute for the Physics and Mathematics of the Universe (IPMU) / University of Tokyo, the Korean Participation Group, Lawrence Berkeley National Laboratory, Leibniz Institut f{\"u}r Astrophysik Potsdam (AIP), Max-Planck-Institut f{\"u}r Astronomie (MPIA Heidelberg), Max-Planck-Institut f{\"u}r Astrophysik (MPA Garching), Max-Planck-Institut f{\"u}r Extraterrestrische Physik (MPE), National Astronomical Observatories of China, New Mexico State University, New York University, University of Notre Dame, Observat{\'o}rio Nacional / MCTI, The Ohio State University, Pennsylvania State University, Shanghai Astronomical Observatory, United Kingdom Participation Group, Universidad Nacional Autonoma de M{\'e}xico, University of Arizona, University of Colorado Boulder, University of Oxford, University of Portsmouth, University of Utah, University of Virginia, University of Washington, University of Wisconsin, Vanderbilt University, and Yale University. \\
{\bf Gaia EDR3}: This work has made use of data from the European Space Agency (ESA) mission {\it Gaia} (\url{https://www.cosmos.esa.int/gaia}), processed by the {\it Gaia} Data Processing and Analysis Consortium (DPAC,\url{https://www.cosmos.esa.int/web/gaia/dpac/consortium}). Funding for the DPAChas been provided by national institutions, in particular the institutions participating in the {\it Gaia} Multilateral Agreement.\\
{\bf DES DR1}: This project used public archival data from the Dark Energy Survey (DES). Funding for the DES Projects has been provided by the U.S. Department of Energy, the U.S. National Science Foundation, the Ministry of Science and Education of Spain, the Science and Technology FacilitiesCouncil of the United Kingdom, the Higher Education Funding Council for England, the National Center for Supercomputing Applications at the University of Illinois at Urbana-Champaign, the Kavli Institute of Cosmological Physics at the University of Chicago, the Center for Cosmology and Astro-Particle Physics at the Ohio State University, the Mitchell Institute for Fundamental Physics and Astronomy at Texas A\&M University, Financiadora de Estudos e Projetos, Funda{\c c}{\~a}o Carlos Chagas Filho de Amparo {\`a} Pesquisa do Estado do Rio de Janeiro, Conselho Nacional de Desenvolvimento Cient{\'i}fico e Tecnol{\'o}gico and the Minist{\'e}rio da Ci{\^e}ncia, Tecnologia e Inova{\c c}{\~a}o, the Deutsche Forschungsgemeinschaft, and the Collaborating Institutions in the Dark Energy Survey.
The Collaborating Institutions are Argonne National Laboratory, the University of California at Santa Cruz, the University of Cambridge, Centro de Investigaciones Energ{\'e}ticas, Medioambientales y Tecnol{\'o}gicas-Madrid, the University of Chicago, University College London, the DES-Brazil Consortium, the University of Edinburgh, the Eidgen{\"o}ssische Technische Hochschule (ETH) Z{\"u}rich,  Fermi National Accelerator Laboratory, the University of Illinois at Urbana-Champaign, the Institut de Ci{\`e}ncies de l'Espai (IEEC/CSIC), the Institut de F{\'i}sica d'Altes Energies, Lawrence Berkeley National Laboratory, the Ludwig-Maximilians Universit{\"a}t M{\"u}nchen and the associated Excellence Cluster Universe, the University of Michigan, the National Optical Astronomy Observatory, the University of Nottingham, The Ohio State University, the OzDES Membership Consortium, the University of Pennsylvania, the University of Portsmouth, SLAC National Accelerator Laboratory, Stanford University, the University of Sussex, and Texas A\&M University.
Based in part on observations at Cerro Tololo Inter-American Observatory, National Optical Astronomy Observatory, which is operated by the Association of Universities for Research in Astronomy (AURA) under a cooperative agreement with the National Science Foundation.\\
{\bf Pan-STARRS PS1, DR2}: The Pan-STARRS1 Surveys (PS1) and the PS1 public science archive have been made possible through contributions by the Institute for Astronomy, the University of Hawaii, the Pan-STARRS Project Office, the Max-Planck Society and its participating institutes, the Max Planck Institute for Astronomy, Heidelberg and the Max Planck Institute for Extraterrestrial Physics, Garching, The Johns Hopkins University, Durham University, the University of Edinburgh, the Queen's University Belfast, the Harvard-Smithsonian Center for Astrophysics, the Las Cumbres Observatory Global Telescope Network Incorporated, the National Central University of Taiwan, the Space Telescope Science Institute, the National Aeronautics and Space Administration under Grant No. NNX08AR22G issued through the Planetary Science Division of the NASA Science Mission Directorate, the National Science Foundation Grant No. AST-1238877, the University of Maryland, Eotvos Lorand University (ELTE), the Los Alamos National Laboratory, and the Gordon and Betty Moore Foundation.\\
{\bf CFIS} We thank the Canada-France Imaging Survey collaboration for sharing their data with us. This work is based on data obtained as part of the Canada-France Imaging Survey, a CFHT large program of the National Research Council of Canada and the French Centre National de la Recherche Scientifique. Based on observations obtained with MegaPrime/MegaCam, a joint project of CFHT and CEA Saclay, at the Canada-France-Hawaii Telescope (CFHT) which is operated by the National Research Council (NRC) of Canada, the Institut National des Science de l'Univers (INSU) of the Centre National de la Recherche Scientifique (CNRS) of France, and the University of Hawaii.\\
{\bf CDS}:  This research made use of the cross-match service provided by CDS, Strasbourg.





\section*{Data Availability \label{sec:DataAv}}

\noindent The public data described here may be obtained from the following websites: \\
{\bf SDSS-IV, DR15}:\\
\url{www.sdss.org}\\ 
\url{https://dr15.sdss.org/sas/dr15/eboss/photoObj/}\\
{\bf Gaia EDR3}:\\
\url{https://www.cosmos.esa.int/web/gaia/earlydr3}\\
{\bf DES DR1}:\\
\url{http://datalab.noao.edu}\\
{\bf MAST}:\\
\url{https://dx.doi.org/10.17909/T9RP4V}

\vspace{3mm}
\noindent The catalogues generated as part of this work may be accessed at the following links:\\
{\bf New SDSS standard star catalogue, version 4.2}:\\
\url{http://faculty.washington.edu/ivezic/sdss/catalogs/stripe82.html}\\
{\bf Old SDSS standard star catalogue, \pO\ (version 2.6)}:\\
\url{http://faculty.washington.edu/ivezic/sdss/catalogs/stripe82.html}\\


\section*{Software Citations \label{sec:soft}}

\begin{itemize}
\item Numpy \citep{harris2020array}
\item Matplotlib \citep{Hunter:2007}
\item Scipy \citep{2020SciPy-NMeth}
\item Astropy \citep{astropy-1, astropy-2}
\item Pandas \citep{pandas}
\item AstroML \citep{2012cidu.conf...47V}.
\end{itemize}

\newpage

\bibliographystyle{mnras}
\interlinepenalty=10000
\bibliography{S82SSC}







\bsp	
\label{lastpage}
\end{document}